\documentclass[aps,prd,reprint,,superscriptaddress]{revtex4-2}

\usepackage[
colorlinks=true,
filecolor=black,
anchorcolor=blue,
linkcolor=blue,
citecolor=cyan,
urlcolor=cyan,
linktocpage=true,
plainpages=false,
breaklinks=true,
pdfstartview=FitH
]{hyperref}

\usepackage[utf8x]{inputenc}
\DeclareUnicodeCharacter{2212}{\textendash}
\usepackage{comment}
\usepackage{graphicx}
\usepackage{amssymb}
\usepackage{amsmath}
\usepackage{amsthm}

\usepackage{hyperref}
\usepackage{multirow}
\usepackage{bigints}
\usepackage{subfigure}
\usepackage{dcolumn}
\usepackage{bm}
\usepackage{color}
\hypersetup{colorlinks=true,linkcolor=blue,citecolor=cyan,urlcolor=cyan,bookmarks=true}

\usepackage{chngcntr}
\counterwithin{equation}{section}

\begin{document}

\allowdisplaybreaks

\title{Gravitational Lensing of Gravitational Wave for Generalized Navarro-Frenk-White Profile and Einasto Profile}

\author{Hanyu Jiang}
\affiliation{National Astronomical Observatories, Chinese Academy of Sciences, Beijing 100101, China}
\affiliation{School of Astronomy and Space Science, University of Chinese Academy of Sciences, Beijing 100049, China}

\author{Xiao Guo}\email[]{guoxiao@nao.cas.cn}
\affiliation{School of Physics and Technology, Wuhan University, Wuhan 430072, China }

\author{Youjun Lu}\email[]{luyj@nao.cas.cn}
\affiliation{School of Astronomy and Space Science, University of Chinese Academy of Sciences, Beijing 100049, China}
\affiliation{National Astronomical Observatories, Chinese Academy of Sciences, Beijing 100101, China}

\author{Yun-Long Zhang}\email[]{zhangyunlong@nao.cas.cn}
\affiliation{National Astronomical Observatories, Chinese Academy of Sciences, Beijing 100101, China}

\affiliation{School of Fundamental Physics and Mathematical Sciences, Hangzhou Institute for Advanced Study, University of Chinese Academy of Sciences, Hangzhou 310024, China}



\begin{abstract}
The density profiles of Dark matter (DM) halos carry imprints of the DM nature and may be constrained through the lensing effects on gravitational waves (GWs) arising from the halo gravitational potential. In this paper, we investigate GW lensing by two representative types of halo density profiles, i.e., the generalized Navarro-Frenk-White (gNFW) density profile and the Einasto density profile. Using the gravitational lensing equation, we first examine the parameter-space distribution and the imaging characteristics of both profiles under strong lensing, partitioning the parameter space into distinct regions according to the Morse index. We then conduct a detailed analysis of the modulus $\big|F\big|$ and phase $\mathrm{Arg}(F)$ of the amplification factor $F\big(w,y\big)$ at low frequency regime. Our results show that, for a fixed lens mass , increasing the gNFW slope $\gamma$ leads to a larger amplitude, more rapid oscillation, and more distinct wave-packet morphology in the multiple-image regime. Compared with the gNFW case, the Einasto case (with $\alpha=0.16$, $y<0.6$, and the same $M_{200}$) produces a stronger lensing effect. Notably, the evolution of $F\big(w,y\big)$ with frequency for the Einasto profile differs from that of the gNFW case, making its behavior particularly distinctive.
\end{abstract}

\maketitle
\tableofcontents

\section{Introduction}
\label{Sec:Introduction}

Dark matter (DM) as a key component of the Lambda cold dark matter ($\Lambda$CDM) cosmological model plays a crucial role in the formation and evolution of galaxies and large scale structures. In the $\mathrm{\Lambda CDM}$ cosmological model, cold dark matter accounts for approximately 26.4\% of the cosmic critical density and 84\% of the total matter content \cite{10.1051/0004-6361/201833910,10.1146/annurev-astro-081710-102514}. Since DM is lack of emitting EM radiation in any band, and only interacts with baryonic matter through gravitation, it remains extremely difficult to detect. Consequently, our understanding of DM is still limited, and its fundamental nature remains unknown. 

With the detection of the first gravitational wave (GW) event GW150914 \cite{10.1103/PhysRevLett.116.061102}, the era of GW astronomy is coming. GWs provide a new messenger to detect dark universe, especially the nature of DM \cite{2020ScPC....3....7B}. The influence of DM on GW could be detectable for future GW detectors \cite{2025PhRvD.112f3055L}. Besides traditional electromagnetic (EM) lensing, lensed GW is also a crucial method to detect DM. Lensed GWs combine two strengths of GW and gravitational lensing, and is a unique probe to reveal the nature of DM \cite{2018PhRvD..98j4029D, 2019PhRvL.122d1103J, 2020MNRAS.495.2002L, 2021MNRAS.502L..16C, 2022A&A...659L...5C, 2021PhRvD.104f3001C, 2022PhRvD.106b3018G, 2025PhRvL.135k1402J}. The GWs lensed by DMs can accurately detect or constrain density profile of DM halos, thus they can help us understand the nature of DM. 

As a fundamental component of the universe, DM may assemble into halos via the growth of initial perturbations. A standard analytic model for cold dark matter halos is the Navarro–Frenk–White (NFW) profile \cite{10.1086/177173,10.1086/304888}, which has been widely used in cosmological studies because it effectively captures the halo structures found in simulations without invoking complex physics. However, a significant limitation of the NFW profile is its central singularity, where the density formally diverges at radius $r=0$. This ``cuspy'' inner slope is in tension with various observations. For instance, rotation curves of low mass and dwarf galaxies often suggest a nearly constant density core~\cite{2018PhR...730....1T}. This so called ``core–cusp'' problem has motivated the development of modified profiles and the inclusion of additional physical processes to alleviate the central cusp.

While the ``core–cusp'' problem suggests that the central density distribution of DM halos should be shallow, numerical simulation studies have presented a different perspective. \citet{10.1086/311333, 10.1086/312287} point out that the inner slope of the DM profile is steeper than NFW, and they investigate more general scenarios. Refs.~\cite{10.1086/319136, 10.1086/321437,astro-ph/0001288} study a generalized NFW-type (gNFW) profile, the research indicates that the slope at the center of DM halos should be greater than 1. As baryonic matter affects the slope of the density profile, we adopt the gNFW profile to describe the mass distribution of the lens, which theoretically allows for the inclusion of partial effects from baryonic matter \cite{10.1093/mnras/283.3.L72,10.1093/mnras/stz1890,10.1086/377489}. Moreover, The numerical cosmological simulations in \citet{10.1038/s41586-020-2642-9} show that the Einasto profile can accurately describe the density profile of DM halos over a mass range of 20 orders of magnitude, and it provides a better fit than the NFW profile, indicating its universality across different cosmic scales. Therefore, it is necessary to investigate the lensing effect of GWs with the Einasto profile.

Unlike EM wave observations, which probe short wavelengths, GW observations operate at wavelengths that are much longer than those of EM waves. For high- and mid-frequency GWs, geometric optics approximation remains valid for describing their propagation and detection. However, at low frequencies, GW wavelengths can become comparable to or exceed the corresponding Schwarzschild radius of massive celestial objects/systems, such as galaxies or galaxy clusters, where wave-optics effects become relevant. When GWs passage through these celestial objects/systems, it will induce GW lensing effect accompanied by non-negligible wave optics effects \cite{10.1086/377430}. Therefore, diffraction effect needs to be considered when we use  lensed low frequency GWs to probe the nature of DM. 

In this paper, we intend to investigate the lensing effect of two types of generalized DM density profile models: the gNFW profile and the Einasto profile. Both of them exhibit excellent fit across the mass range from dwarf galaxies to galaxy clusters, and therefore they are applicable to all cases. We primarily focus on the lensing effect induced by DM halos, while also incorporating partial effect from baryonic matter (such as galaxies).

The paper is organized as follows. In Sec.~\ref{Sec:Profiles}, we present the model of the gravitational lensing system, introduce the gNFW profile and the Einasto profile used to describe the mass distribution on the lens plane, and specify the configuration parameters of the lensing system. In Sec.~\ref{Sec:Parameter Space}, we carried out a detailed analysis of the imaging properties of gravitational lensing based on the gNFW profile and the Einasto profile. In Sec.~\ref{Sec:F}, building on the analysis of image properties, we further investigate how the amplification factor $F$ evolves with the source position $y$. The conclusions are summarized in Sec.~\ref{Sec:Conclusion}.

Throughout this work, the lens mass distribution is restricted to the axisymmetric case. We further assume that the image position is always positive ($\big|x\big|$), while the sign of the source position $y$ denotes whether the source and the image are on the same side of the origin in the lens plane: $y>0$ for the same side and $y<0$ for the opposite side. We adopt SI with $c_{0}=3\times10^{8}\mathrm{m\,s^{-1}}$ and $G=6.67\times10^{-11}\mathrm{m^{3}kg^{-1}s^{-2}}$.

\section{Profiles of Lens Mass}
\label{Sec:Profiles}

\begin{figure}[!htb]
        \centering
        \includegraphics[width=0.45\textwidth]{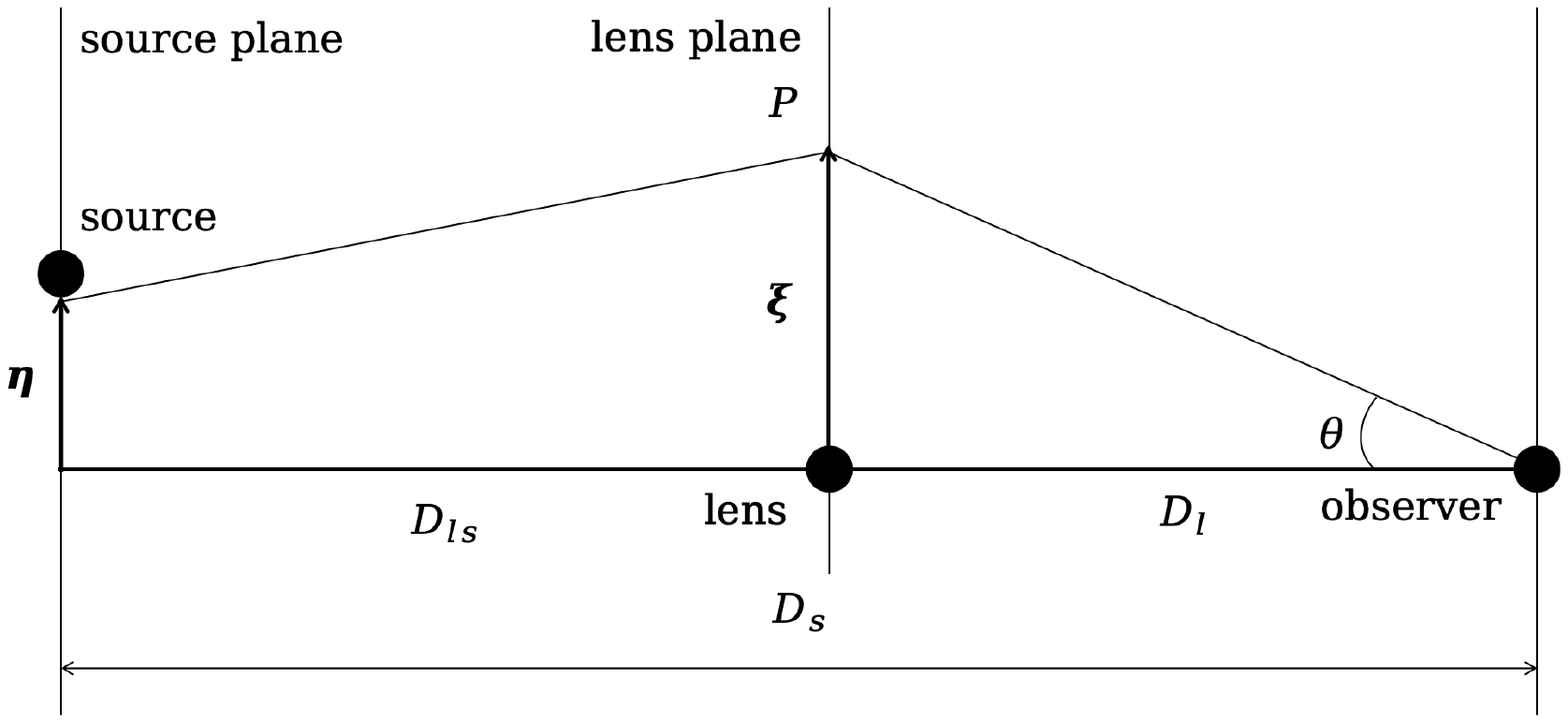}
        \caption{Schematic diagram of the geometrical configuration of the gravitational lensing system (the figure is adapted from \cite{10.1103/PhysRevD.102.124076}). $D_{\mathrm{s}}$ and $D_{\mathrm{l}}$ denote the angular diameter distances from the observer to the source and lens, respectively, and $D_{\mathrm{ls}}$ represents the angular diameter distances between the lens and source. $\boldsymbol{\eta}$ and $\boldsymbol{\xi}$ denote the position vectors of the source on the source plane and image on the lens plane, respectively.}
        \label{GRlens}
    \end{figure}

First, we need to fix the configuration of the gravitational lens system in order to simulate observable lensing phenomena (see Fig.~\ref{GRlens}).

In general, if the density profile $\rho$ on the lens plane takes the following form
    \begin{equation}
        \rho=\rho_{\mathrm{s}}\hat{\rho}\Big(\frac{r}{r_{\mathrm{s}}}\Big),
        \label{rho}
    \end{equation}
here $\rho_{\mathrm{s}}$ and $r_{\mathrm{s}}$ denote the characteristic density and characteristic scale radius, respectively. Then the surface density at a distance $R$ from the center is
    \begin{equation}
        \Sigma\big(R\big)=2\rho_{\mathrm{s}}r_{\mathrm{s}}\int_{0}^{\infty}\hat{\rho}\Big(\sqrt{\Big(\frac{R}{r_{\mathrm{s}}}\Big)^{2}+s^{2}}\Big)ds,
        \label{surface density}
    \end{equation}
during the calculation, we perform the variable substitution $s=z/r_{\mathrm{s}}$, $z$ is the distance from the lens plane. Let $\kappa=\Sigma/\Sigma_{\mathrm{crit}}$, then we obtain the important parameter $\kappa_{\mathrm{s}}=\rho_{\mathrm{s}}r_{\mathrm{s}}/\Sigma_{\mathrm{crit}}$, where $\Sigma_{\mathrm{crit}}=\frac{c_{0}^{2}}{4\pi G} \frac{D_{\mathrm{s}}}{D_{\mathrm{l}}D_{\mathrm{ls}}}$ is the critical surface density. The angular diameter distance is given by
    \begin{equation}
        \begin{aligned}
            D_{A}=\frac{c_{0}}{1+z_{2}}\int_{z_{1}}^{z_{2}}\frac{dz}{H\big(z\big)},
        \end{aligned}
        \label{DA}
    \end{equation}
where $c_{0}$ denotes the speed of light in vacuum. Taking into account the fractional energy densities of the various components in the present day universe \cite{10.1051/0004-6361/201833910,10.1146/annurev-astro-081710-102514}, and in order to simplify the calculations, we adopt a flat $\mathrm{\Lambda CDM}$ cosmology with fiducial parameters $\Omega_{m}=0.3$, $\Omega_{\Lambda}=0.7$, $h\equiv\ H_{0}/\big(100\mathrm{km s^{-1}Mpc^{-1}}\big)=0.7$, thus
    \begin{equation}
        \begin{aligned}
            H\big(z\big)=H_{0}\sqrt{\Omega_{\Lambda}+\Omega_{m}\big(1+z\big)^{3}}.
        \end{aligned}
        \label{H}
    \end{equation}
As noted in \cite{astro-ph/9602053,10.1046/j.1365-8711.2003.06276.x,10.1086/323961,10.1111/j.1365-2966.2012.21983.x,10.1086/322314}, whether a lens system can produce observable gravitational lensing signatures is quantified by the dimensionless surface density parameter $\kappa_{\mathrm{s}}$.
And we state the conclusion: in observational practice, for a lensing system to produce a detectable gravitational lensing effect, it is generally required that $\kappa_{\mathrm{s}}>0.1$ \cite{10.1051/0004-6361/201321618,10.1051/0004-6361:20020226}.

\subsection{gNFW profile}
\label{subSec:gNFW}

We consider a gNFW-type density profile of the form \cite{10.1086/319136,10.1086/321437,astro-ph/0001288},
    \begin{equation}
        \begin{aligned}
            \rho =\frac{\rho_{\mathrm{s}}}{\big(\frac{r}{r_{\mathrm{s}}} \big)^{\gamma }\big(1+\frac{r}{r_{\mathrm{s}}} \big)^{3-\gamma }}.
        \end{aligned}
        \label{generalized NFW Model}
    \end{equation}
For realistic physical considerations, we take $0<\gamma<2$, the central cusp generalizes to $\rho \propto r^{-\gamma }$. When $\gamma=1$ the model reduces to the familiar NFW profile. Here $\rho_{\mathrm{s}}$ and $r_{\mathrm{s}}$ can be written in closed form in terms of a halo mass $M_{\Delta }$ (enclosed within the radius $r_{\Delta }$ at overdensity $\Delta \rho_{\mathrm{crit}}$) and the concentration $c\equiv r_{\Delta }/r_{\mathrm{s}}$ (it should be distinguished from $c_{0}$) as \cite{10.1086/321437,10.1086/323961}
    \begin{equation}
        \begin{aligned}
&\rho_{\mathrm{s}}=\delta_{c}\rho_{\mathrm{crit}},\quad
            \delta_{c}=\frac{\Delta }{3}\frac{c^{3}}{\mathrm{I}\big(c,\gamma\big)},\\
&\mathrm{I}\big(c,\gamma\big)\equiv\int_{0}^{c}x^{2-\gamma}\big(1+x\big)^{\gamma-3}dx,\\
        \end{aligned}
        \label{rhos}
    \end{equation}
and
    \begin{equation}
        \begin{aligned}\qquad\qquad
r_{\mathrm{s}}&=\Big(\frac{M_{\Delta}}{4\pi\rho_{\mathrm{s}}\mathrm{I}\big(c,\gamma\big)}\Big)^\frac{1}{3} 
=\frac{1}{c}\Big(\frac{3M_{\Delta}}{4\pi\Delta\rho_{\mathrm{crit}}}\Big)^\frac{1}{3}.
        \end{aligned}
        \label{rs}
    \end{equation}
Here, $\delta_{c}$ denotes the density perturbation factor, $\mathrm{I}\big(c,\gamma\big)$ is the dimensionless normalization, in our work we choose $\Delta=200$. For the special case $\gamma=1$ one recovers the usual NFW normalization $\mathrm{I}\big(c,1\big)=\mathrm{ln}\big(1+c\big)-c/\big(1+c\big)$\cite{10.1086/321437}. $\rho_{\mathrm{crit}}\equiv3H^{2}\big(z\big)/8\pi G$ is the critical density of the universe at redshift $z$.

Following \cite{10.1111/j.1745-3933.2008.00537.x}, the parameter $M_{200}$ in Eq.~(\ref{rs}) represents the mass of lensing galaxy halo. Based on current observational data \cite{10.1111/j.1745-3933.2008.00537.x,10.1146/annurev-astro-091918-104453,10.1111/j.1365-2966.2008.13348.x,10.1111/j.1365-2966.2012.21984.x,2021Univ....7..139L,10.1146/annurev-astro-081710-102514,10.1111/j.1365-2966.2012.21623.x,10.1088/0004-637X/770/1/57,10.1103/RevModPhys.77.207}, $M_{200}$ is typically in the range of $10^{-4}M_{\mathrm{pivot}}\le M_{200}\le10^{3}M_{\mathrm{pivot}}$, where $M_{\mathrm{pivot}}=2\times 10^{12}h^{-1}{M_{\odot}}$ is the median halo mass and ${M_{\odot}}$ denotes the solar mass. For the concentration parameter $c=c_{200}$, we adopt the empirical relation given in \cite{10.1111/j.1745-3933.2008.00537.x},
    \begin{equation}
        \begin{aligned}
            c_{200}=A_{200}\Big(\frac{M_{200}}{M_{\mathrm{pivot}}}\Big)^{B_{200}}\big(1+z\big)^{C_{200}}.
        \end{aligned}
        \label{c200}
    \end{equation}
Here, we choose $A_{200}=5.71$, $B_{200}=-0.084$, $C_{200}=-0.47$, this prescription provides a good fit to observational data over the redshift interval $0\le z\le2$. It is worth noting that we adopt a simplified treatment here, incorporating the influence of baryonic matter solely through the parameter $\gamma$ of the gNFW profile (See \cite{10.1093/mnras/stw1707, 10.1093/mnras/stu1284}), and neglecting more complex secondary effects as well as potential substructures.

\begin{figure*}
        \centering 
        \subfigure[$\gamma=0.2$]{
        \label{kssub1}
        \includegraphics[width=0.5\textwidth]{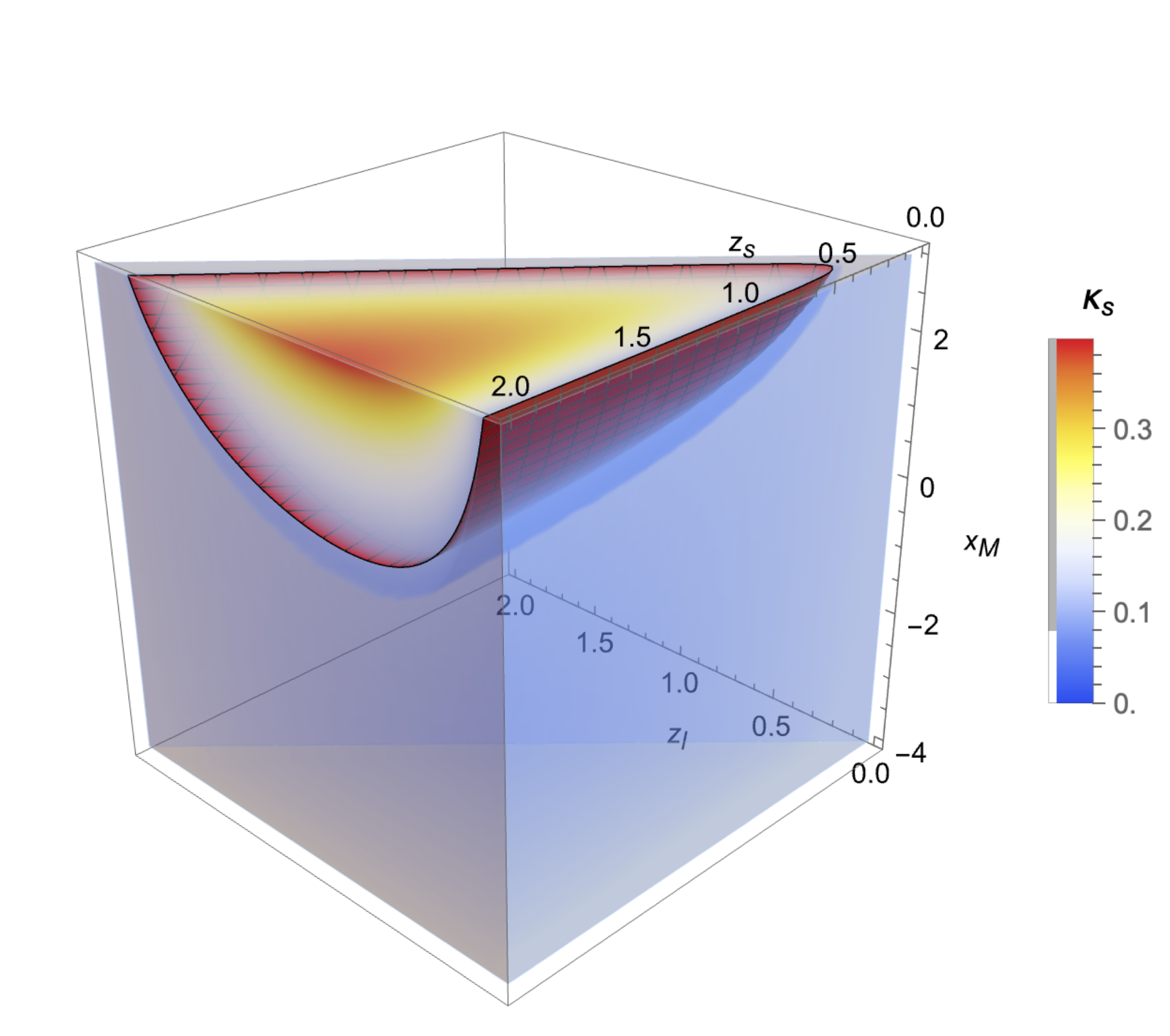}}\subfigure[$\gamma=0.6$]{
        \label{kssub2}
        \includegraphics[width=0.5\textwidth]{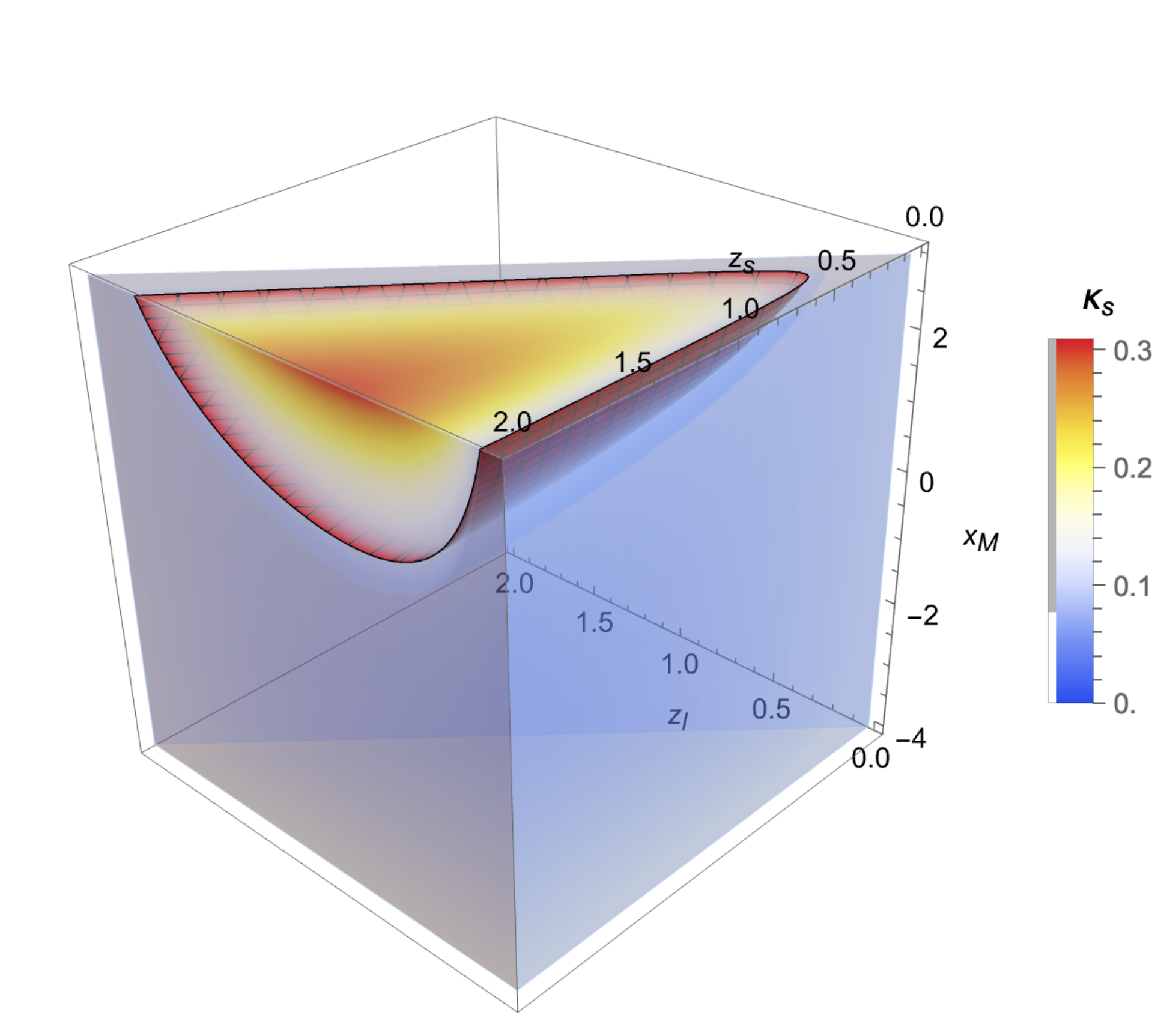}}
        \subfigure[$\gamma=1.4$]{
        \label{kssub3}
        \includegraphics[width=0.5\textwidth]{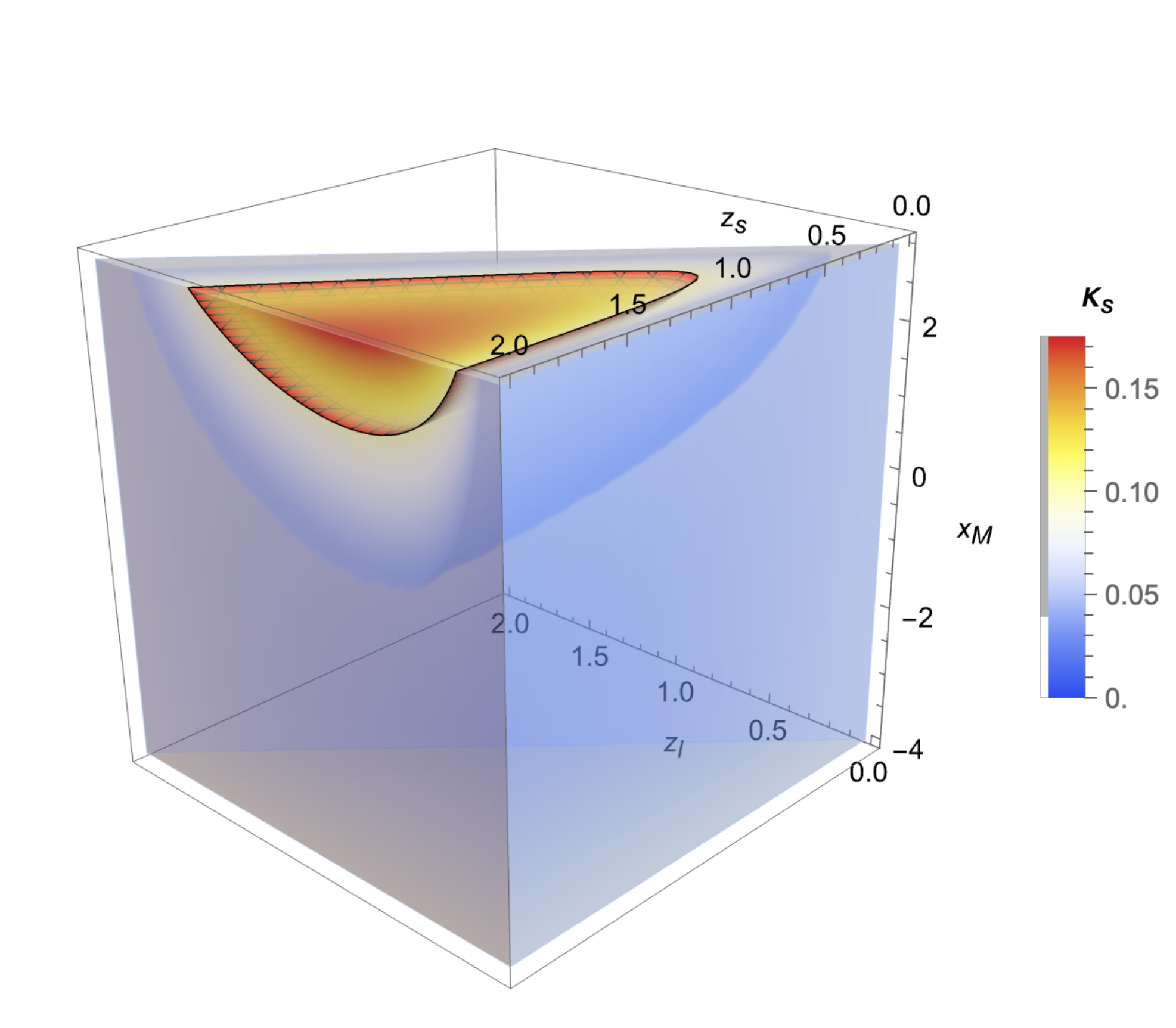}}\subfigure[$\gamma=1.8$]{
        \label{kssub4}
        \includegraphics[width=0.5\textwidth]{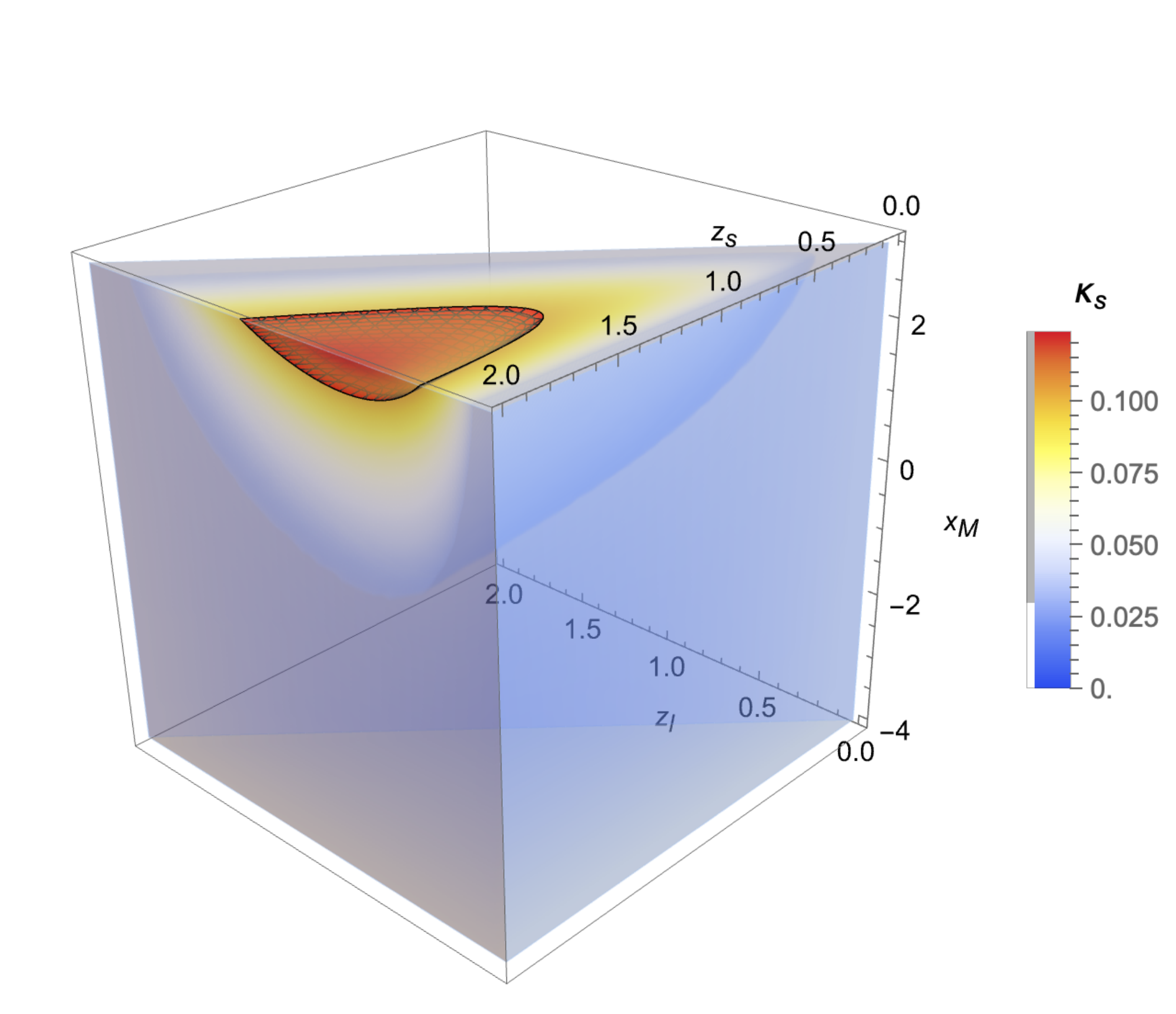}}
        \caption{The distribution of $\kappa_s$ of the gNFW profile, where $x_{\mathrm{M}}=\mathrm{log}\big(M_{200}/M_{\mathrm{pivot}}\big)$. The red envelope surface represents $\kappa_{\mathrm{s}}=0.1$. It is evident that as $\gamma$ increases, the overall value of $\kappa_{\mathrm{s}}$ decreases, leading to fewer lens configurations that satisfy condition $\kappa_{\mathrm{s}}>0.1$. Consequently, the requirements for producing an observable gravitational lensing effect become increasingly stringent.}
        \label{ks}
    \end{figure*}

After reviewing the parameters of the gNFW profile, we can determine the parameters ($z_{\mathrm{l}}$, $z_{\mathrm{s}}$, $\mathrm{log}\big(M_{200}/M_{\mathrm{pivot}}\big)$) of the gravitational lensing system, where $z_{\mathrm{s}}$ and $z_{\mathrm{l}}$ represent the redshifts of the sources and the lens, respectively. The distribution of $\kappa_s$ in the parameter space $\big(z_{\mathrm{l}},z_{\mathrm{s}},\mathrm{log}\big(M_{200}/M_{\mathrm{pivot}}\big)\big)$ of the gravitational lens system with different $\gamma$ is shown in Fig.~\ref{ks}.

From Fig.~\ref{ks}, we find that for larger $\gamma$, the condition required to satisfy $\kappa_{\mathrm{s}}>0.1$ becomes more stringent. Among these, the most crucial factor is the mass of the lens, only sufficiently massive lenses are capable of producing observable gravitational lensing phenomena. Otherwise, even if lensing occurs in reality, it would remain undetectable with current observational capabilities. In addition, when the lens is located approximately midway between the source and the observer within the gravitational lensing system, the value of $\kappa_{\mathrm{s}}$ can be effectively enhanced.

\subsection{Einasto profile}\label{subSec:Einasto}

Then, we consider another important profiles: the Einasto profiles, whose density profile is given by \cite{10.1051/0004-6361/201219539, 10.1051/0004-6361/201118543}
    \begin{equation}
        \begin{aligned}
            \rho\big(r\big)=\rho_{-2}\mathrm{exp}\Big\{-\frac{2}{\alpha}\Big[\Big(\frac{r}{r_{-2}}\Big)^{\alpha}-1\Big]\Big\},
        \end{aligned}
        \label{Einasto profile}
    \end{equation}
where $\rho_{-2}$ and $r_{-2}$ are the density and radius at which $\rho\big(r\big)\propto r^{-2}$. For convenience, we assume that in the Einasto profile $\rho_{-2}=\rho_{\mathrm{s}}$ and $r_{-2}=r_{\mathrm{s}}$, which satisfy
    \begin{equation}
        \begin{aligned}
            \rho_{\mathrm{s}}&=\delta_{c}\rho_{\mathrm{crit}},\\
            \delta_{c}&=\frac{\Delta}{3}\frac{c^{3}}{\mathrm{ln}\big(1+c\big)-c/\big(1+c\big)},\\
        \end{aligned}
        \label{rho-2}
    \end{equation}
and
    \begin{equation}
        \begin{aligned}
            r_{\mathrm{s}}&=\Big\{\frac{M_{\Delta}}{4\pi\rho_{\mathrm{s}}\big[\mathrm{ln}\big(1+c\big)-c/\big(1+c\big)\big]}\Big\}^\frac{1}{3},\\
            &=\frac{1}{c}\Big(\frac{3M_{\Delta}}{4\pi\Delta\rho_{\mathrm{crit}}}\Big)^\frac{1}{3}.
        \end{aligned}
        \label{r-2}
    \end{equation}
For the concentration parameter $c_{200}$ of the Einasto profile, we can also use Eq.~(\ref{c200}), because \cite{10.1111/j.1745-3933.2008.00537.x} adopts the value of $A_{200}=6.4$, $B_{200}=-0.108$, $C_{200}=-0.62$. However, recent results from \cite{10.1038/s41586-020-2642-9} provide improved findings. \cite{10.1038/s41586-020-2642-9} simulates the cosmic structure of DM halos over 20 orders of magnitude in halo mass and finds that the halo density profiles are universal across the entire mass range, with both the NFW and Einasto profiles providing good descriptions of halo properties. Fig.~3 in \cite{10.1038/s41586-020-2642-9} presents a more general relation of $c_{200}$ over a mass range of 20 orders of magnitude. We therefore adopt the results as the reference for the values of $c_{200}$ used in this work, and the empirical expression for $c_{200}$ is \cite{10.1038/s41586-020-2642-9}
    \begin{equation}
        \begin{aligned}
            c_{200}=\sum_{i=1}^{6}c_{i}\Big[\mathrm{ln}\Big(2\times10^{12}\frac{M_{200}}{M_{\mathrm{pivot}}}\Big)\Big]^{i-1},
        \end{aligned}
        \label{c200Einasto}
    \end{equation}
where $c_{i}=\big[27.112,−0.381,−1.853\times10^{−3},−4.141\times10^{−4},−4.334\times10^{−6},3.208\times10^{−7}\big]$ for $i\in\left\{1,...,6\right\}$. Finally, it should be noted that some references use the parameter $n=1/\alpha$ to denote Eq.~(\ref{Einasto profile}), but its physical meaning is identical and does not affect the analysis.

\begin{figure}[!htb]
        \centering
        \includegraphics[width=0.5\textwidth]{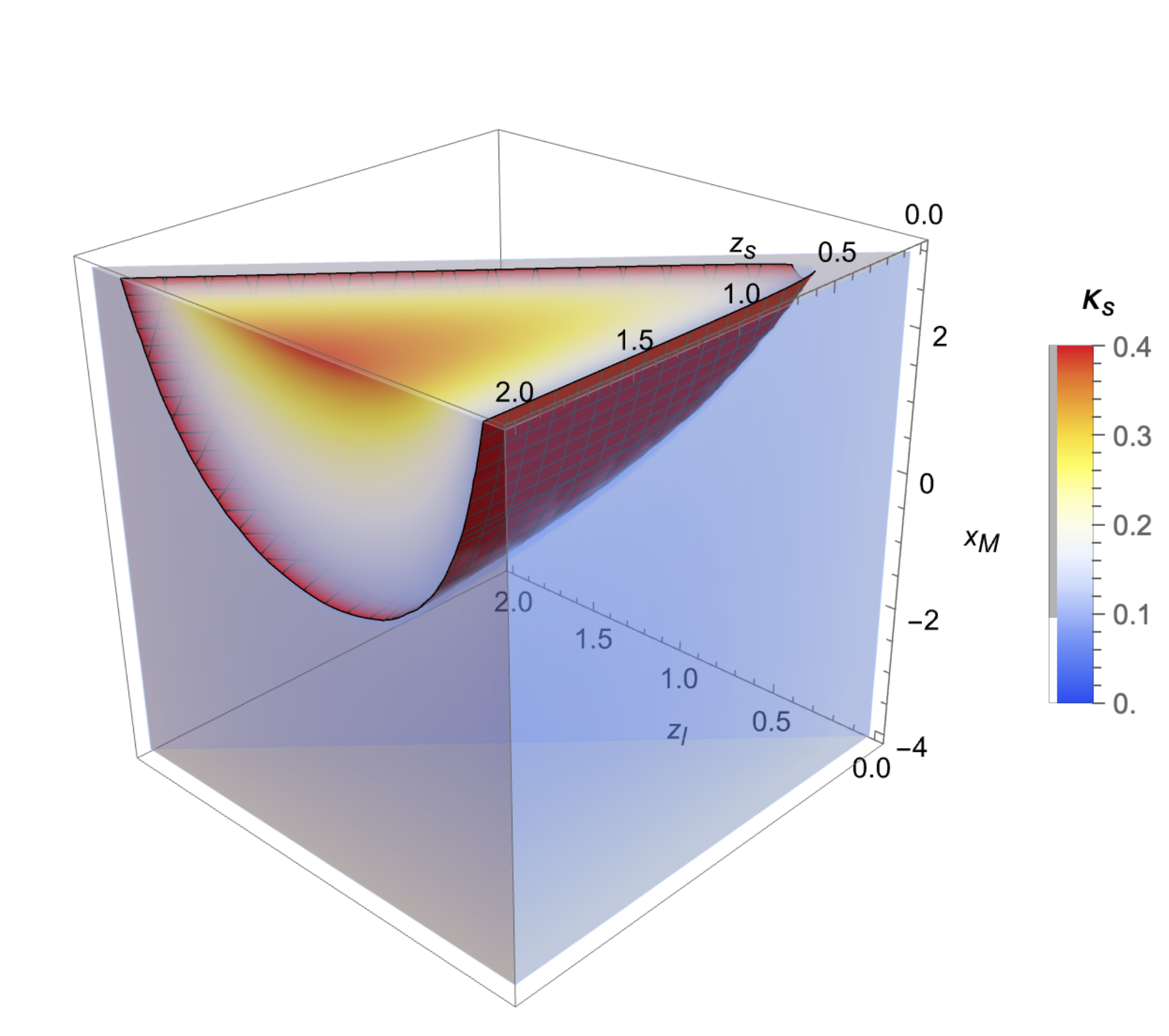}
        \caption{The distribution of $\kappa_s$ of the Einasto profile, where $x_{\mathrm{M}}=\mathrm{log}\big(M_{200}/M_{\mathrm{pivot}}\big)$. The red envelope surface represents $\kappa_{\mathrm{s}}=0.1$. Compared to the gNFW profile, $\kappa_{\mathrm{s}}>0.1$ is more easily satisfied with Einasto profile.}
        \label{ksEinasto}
    \end{figure}

Similar to the case of gNFW profile, we present the distribution of $\kappa_s$ in the parameter space $\big(z_{\mathrm{l}},z_{\mathrm{s}},\mathrm{log}\big(M_{200}/M_{\mathrm{pivot}}\big)\big)$ for the Einasto profile (see Fig.~\ref{ksEinasto}).

Finally, after comparing the $\kappa_s$ distribution of the two density profiles, we adopt configuration $\big(z_{\mathrm{l}}=0.5$, $z_{\mathrm{s}}=1.5$, $\mathrm{log}\big(M_{200}/M_{\mathrm{pivot}}\big)=3\big)$ as the representative setup for the gravitational lensing system considered in this work.

\section{Parameter Space of Lensing}
\label{Sec:Parameter Space}

\subsection{Lens equation}
\label{subSec:T and critical condition}

Refer to Fig.~\ref{GRlens}, we introduce $\xi_{0}$ as the normalization constant of the length in the lens plane, so it is useful to express the time delay function $T\big(\boldsymbol{x},\boldsymbol{y})$ in terms of dimensionless quantities, which is given by
    \begin{equation}
        \begin{aligned}
            T\big(\boldsymbol{x},\boldsymbol{y})=\frac{1}{2}|\boldsymbol{x}-\boldsymbol{y}|^{2}-\psi\big(\boldsymbol{x}\big)+\phi_{\mathrm{m}}\big(\boldsymbol{y}\big),
        \end{aligned}
        \label{T}
    \end{equation}
where
    \begin{equation}
        \begin{aligned}
            \boldsymbol{x}=\frac{\boldsymbol{\xi}}{\xi_{0}},\qquad \boldsymbol{y}=\frac{D_{\mathrm{l}}}{\xi_{0}D_{\mathrm{s}}}\boldsymbol{\eta},
        \end{aligned}
        \label{dimensionless parameters}
    \end{equation}
and $\phi_{\mathrm{m}}\big(\boldsymbol{y}\big)$ makes $T_{\mathrm{min}}\big(\boldsymbol{x},\boldsymbol{y})=0$, i.e. $\phi_{\mathrm{m}}\big(\boldsymbol{y}\big)= -\big(\frac{1}{2}|\boldsymbol{x}-\boldsymbol{y}|^{2}-\psi\big(\boldsymbol{x}\big)\big)_{\mathrm{min}}$. Since this work considers only the axisymmetric case, $\boldsymbol{x}$ and $\boldsymbol{y}$ can be reduced to scalars.

In this work, we choose characteristic scale radius $r_{\mathrm{s}}$ as the unit scale for convenience, $\psi\big(x\big)$ is the lens potential. In general, for both the gNFW profile and the Einasto profile, $\psi\big(x\big)$ does not admit a simple analytic expression and can only be expressed in integral form,
    \begin{equation}
        \begin{aligned}
            \psi\big(x\big)=4\kappa_{\mathrm{s}}\int_{0}^{x}t\hat{\kappa}\big(t\big)\mathrm{ln}\frac{x}{t}dt.
        \end{aligned}
        \label{potential}
    \end{equation}
For the gNFW profile, $\hat{\kappa}\big(x\big)$ has the expression \cite{2001astro.ph..2341K}
    \begin{equation}
        \begin{aligned}
            \hat{\kappa}\big(x\big)&=x^{1-\gamma}\Big[\big(1+x\big)^{\gamma-3}\\
            &\quad+\big(3-\gamma\big)\int_{0}^{1}\big(y+x\big)^{\gamma-4}\Big(1-\sqrt{1-y^2}\Big)dy\Big].
        \end{aligned}
        \label{kbargNFW}
    \end{equation}
While for the Einasto profile, $\hat{\kappa}\big(x\big)$ is
    \begin{equation}
        \begin{aligned}
            \hat{\kappa}\big(x\big)=\int_{0}^{\infty}\mathrm{exp}\Big\{-\frac{2}{\alpha}\Big[\Big(x^{2}+s^{2}\Big)^{\alpha/2}-1\Big]\Big\}ds.
        \end{aligned}
        \label{kbarEinasto}
    \end{equation}

\begin{figure}[!htb]
        \centering
        \includegraphics[width=0.45\textwidth]{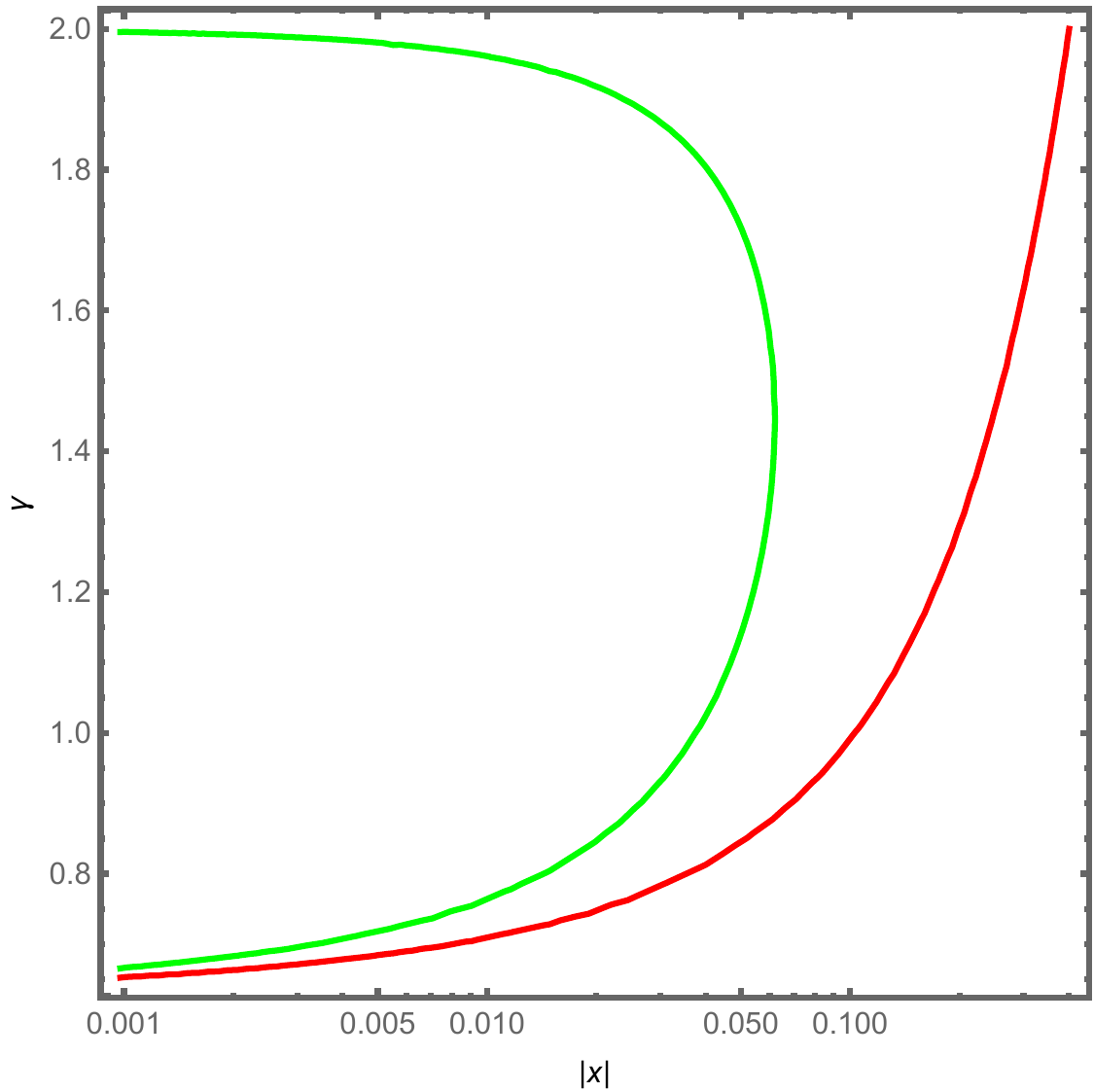}
        \caption{Two types of critical condition for gNFW profile, $M_{200}=2.86\times 10^{15}{M_{\odot}}$. Red: the tangential critical condition ($y=0$, i.e. Einstein ring); Green: the radial critical condition.}
        \label{critical condition}
    \end{figure}

\begin{figure}[!htb]
        \centering
        \includegraphics[width=0.45\textwidth]{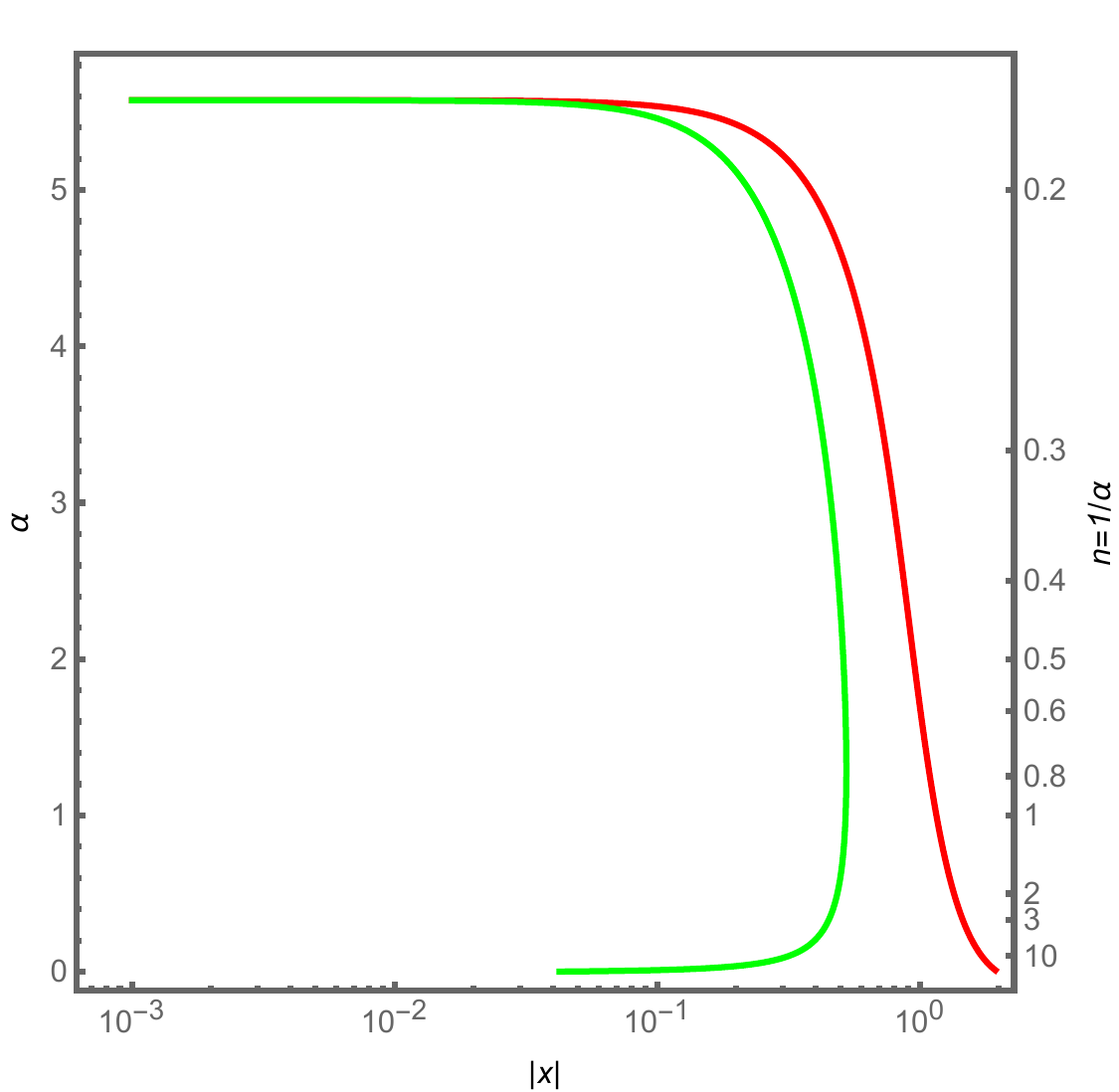}
        \caption{Same as Fig.~\ref{critical condition}, but for Einasto profile.}
        \label{critical condition Einasto}
    \end{figure}

According to Fermat’s principle, GWs follow paths that make $T\big(x,y)$ stationary, leading to the condition
    \begin{equation}
        \begin{aligned}
            y&=x-{\psi}'\big(x\big),\\
            {\psi}'\big(x\big)&=\alpha\big(x\big),\\
            \alpha\big(x\big)&=4\kappa_{\mathrm{s}}\frac{\int_{0}^{x}t\hat{\kappa}\big(t\big)dt}{x}.
        \end{aligned}
        \label{lightfunction}
    \end{equation}
The above equation corresponds to the tangential critical condition. In particular, when $y=0$, the lensing image forms an Einstein ring (see the red curves in Figs.~\ref{critical condition} and \ref{critical condition Einasto}).

\begin{figure}[!htb]
        \centering
        \includegraphics[width=0.45\textwidth]{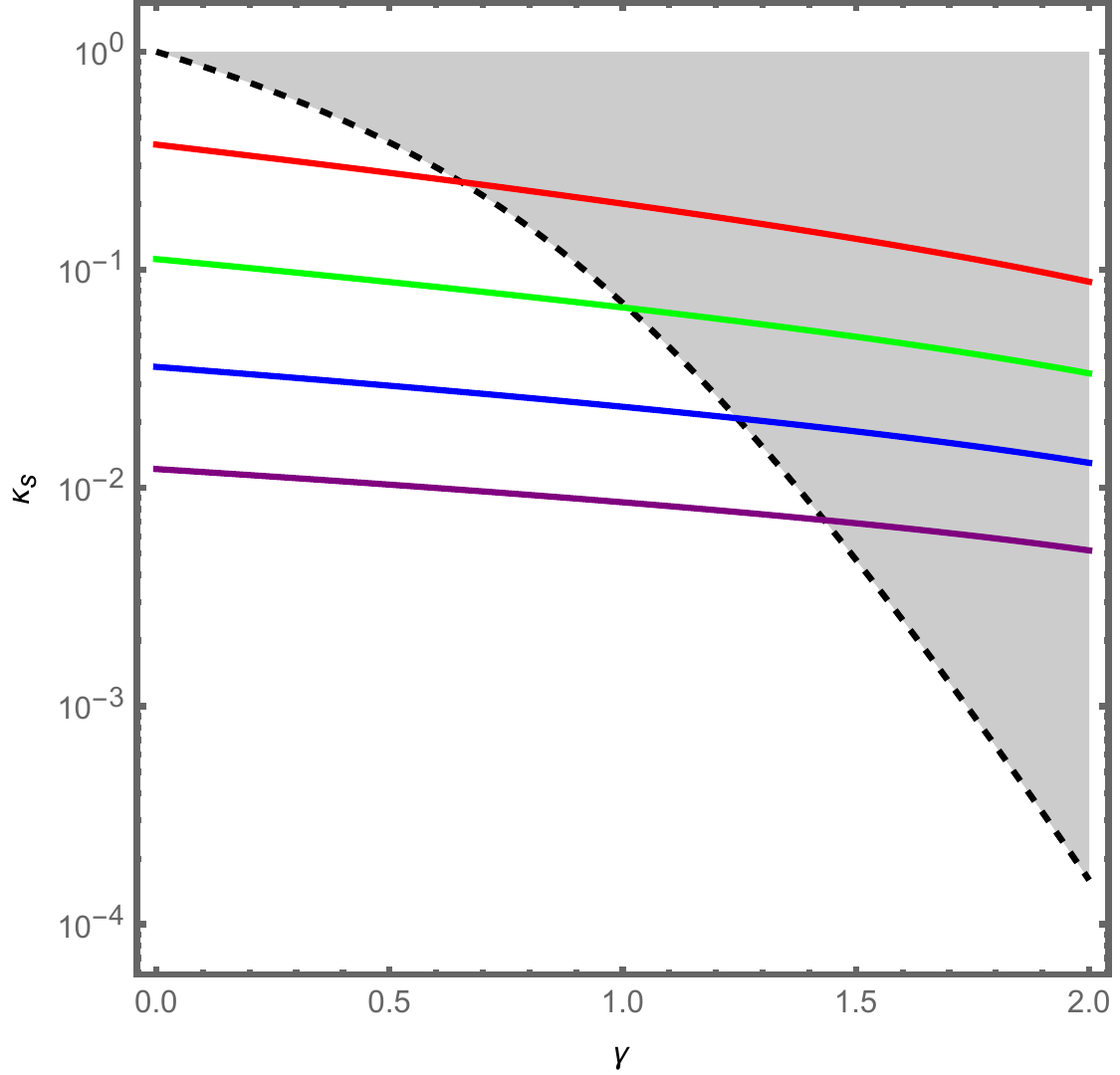}
        \caption{$\kappa_{s}$ for the gNFW profile. The gray region in the figure indicates the range allowed for the formation of Einstein rings. Dashed: the minimum $\kappa_{s}$ required for the gNFW profile to produce Einstein rings; Red: $M_{200}=2.86\times 10^{15}{M_{\odot}}$; Green: $M_{200}=2.86\times 10^{13}{M_{\odot}}$; Blue: $M_{200}=2.86\times 10^{11}{M_{\odot}}$; Purple: $M_{200}=2.86\times 10^{9}{M_{\odot}}$.
        }
        \label{kappasgamma}
    \end{figure}

\begin{figure}[!htb]
        \centering
        \includegraphics[width=0.5\textwidth]{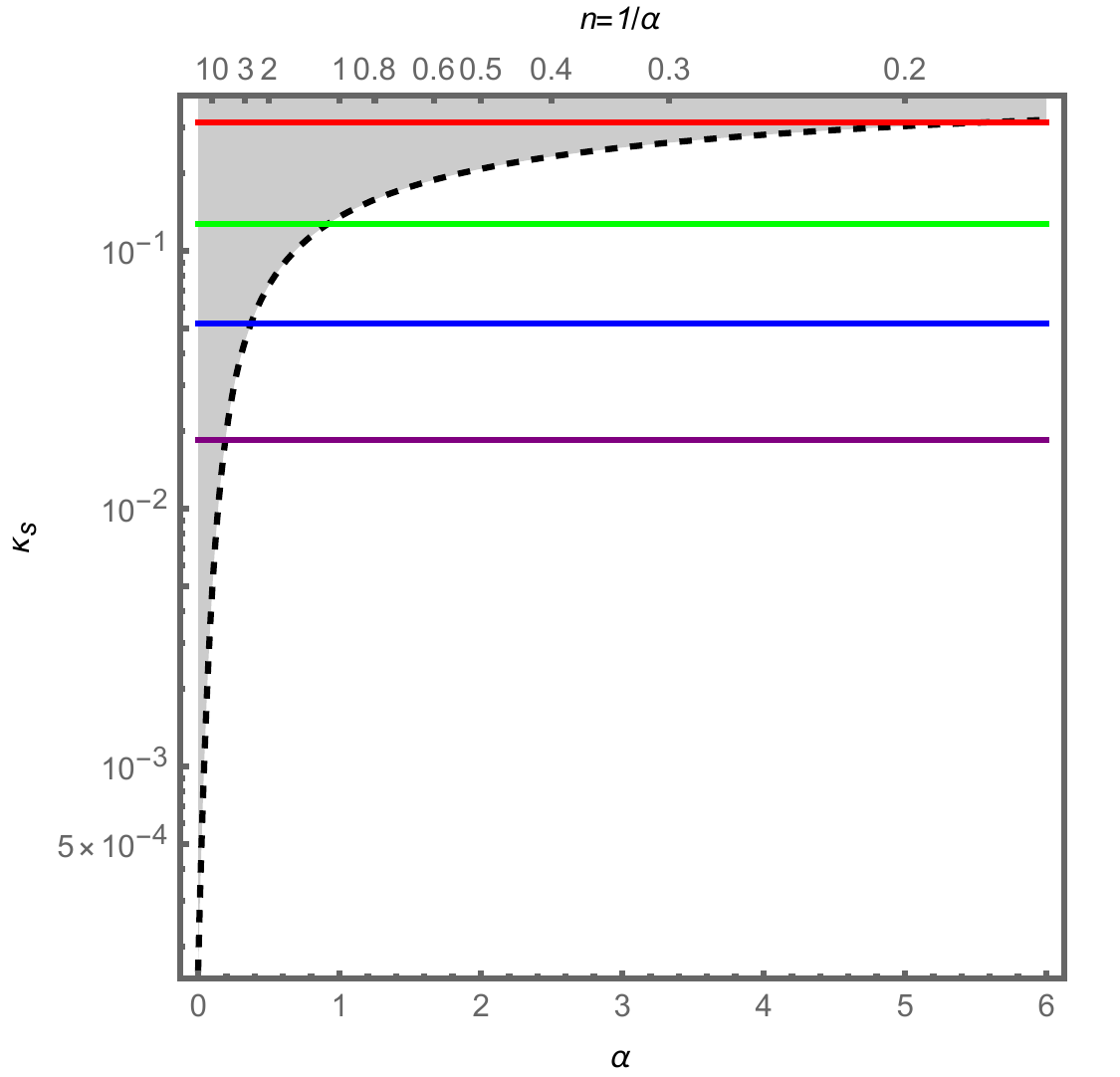}
        \caption{Same as Fig.~\ref{kappasgamma}, but for Einasto profile.
        }
        \label{kappasalpha}
    \end{figure}

Based on the gravitational lens system configuration used in this study, we set a lower limit for the Einstein ring radius of $x_{\mathrm{Emin}}=10^{-3}$, corresponding to the angular radius of $\theta_{\mathrm{Emin}}=x_{\mathrm{Emin}}r_{\mathrm{s}}/D_{\mathrm{l}}\approx0.15''$.
Under this criterion, we determine the minimum $\kappa_{s}$ required for two profiles to produce Einstein rings (shown as the black dashed line in Figs.~\ref{kappasgamma} and \ref{kappasalpha}), and calculate the $\kappa_{s}$ for different lens masses with two profiles (shown as the colored lines in Figs.~\ref{kappasgamma} and \ref{kappasalpha}).

According to the literature \cite{10.1086/589989,10.1088/0004-637X/748/2/129,10.3847/1538-4357/aa9794,10.1051/0004-6361/202451341,10.1111/j.1365-2966.2012.21041.x}, 
the Einstein ring produced by galaxys generally lies in the range $\mathcal{O}\big (0.1''-1''\big)$, while in the case of strong lensing by galaxy clusters, the Einstein ring radius distribution extends to larger angular scales (about $\mathcal{O}\big (10''\big)$), with some observations reaching up to $70''$ \cite{10.1111/j.1365-2966.2012.21041.x}. For the resolving ability of current telescopes, for example, the Hubble Space Telescope (HST) achieves an angular resolution of approximately $0.05''$, while the James Webb Space Telescope (JWST) provides a resolution of about $0.07''$ at a wavelength of $2\mathrm{\mu m}$, so the overall structure of the Einstein ring can be clearly resolved. This makes the criterion $x_{\mathrm{Emin}}=10^{-3}$ physically meaningful, models satisfying $x_{\mathrm{Emin}}=10^{-3}$ are therefore capable of producing observable gravitational lensing phenomena in practice.

Meanwhile, the radial critical condition is given by
    \begin{equation}
        \begin{aligned}
            {y}'&=1-{\psi}''\big(x\big)=0,\\
            {\psi}''\big(x\big)&=4\kappa_{\mathrm{s}}\Big[\hat{\kappa}\big(x\big)-\frac{\int_{0}^{x}t\hat{\kappa}\big(t\big)dt}{x^{2}}\Big].
        \end{aligned}
        \label{radial}
    \end{equation}
If the above equation has a positive real solution $x_{\mathrm{crit}}$, then combining it with Eq.~(\ref{lightfunction}) yields
    \begin{equation}
        \begin{aligned}
            y_{\mathrm{crit}}&=2x_{\mathrm{crit}}\big(1-2\kappa_{\mathrm{s}}\hat{\kappa}(x_{\mathrm{crit}})\big).
        \end{aligned}
        \label{ycrit}
    \end{equation}
The variation of $x_{\mathrm{crit}}$ with the model parameters ($\gamma$ and $\alpha$) are shown by the green curves in Fig.~\ref{critical condition} and Fig.~\ref{critical condition Einasto}. It can be seen that $x_{\mathrm{crit}}$ has an upper bound. For gNFW mode $x_{\mathrm{max}}=0.062$, $\gamma\big(x_{\mathrm{max}}\big)=1.446$; for Einasto profile $x_{\mathrm{max}}=0.525$, $\alpha\big(x_{\mathrm{max}}\big)=1.282$. We find that when $0<x\le x_{\mathrm{max}}$, $y_{\mathrm{crit}}<0$ for all cases. This result is of great significance, as together with Eq.~(\ref{lightfunction}) and Eq.~(\ref{radial}), it determines the nature of the image.

\begin{figure}[!htb]
        \centering
        \includegraphics[width=0.45\textwidth]{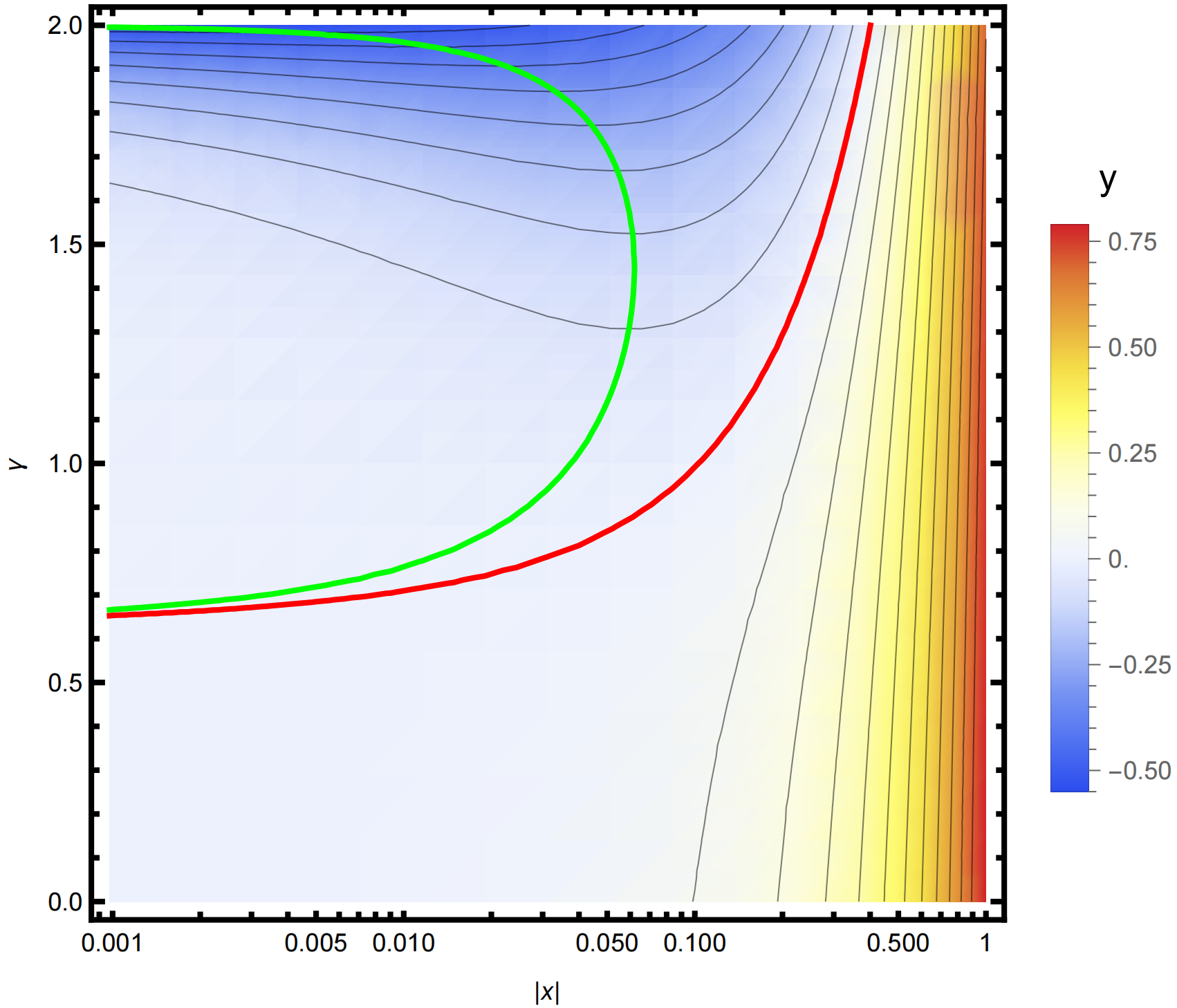}
        \caption{Density map for the gNFW profile, with $M_{200}=2.86\times 10^{15}{M_{\odot}}$. The physical meanings of the red and green curves are the same as those in Fig.~\ref{critical condition}, while the black curves represent contour lines. From this figure, it can be seen that for a given $\left|y\right|$, there is always one image on the same side as the source, and possibly up to two additional images on the opposite side.}
        \label{imagine}
    \end{figure}

\begin{figure}[!htb]
        \centering
        \includegraphics[width=0.45\textwidth]{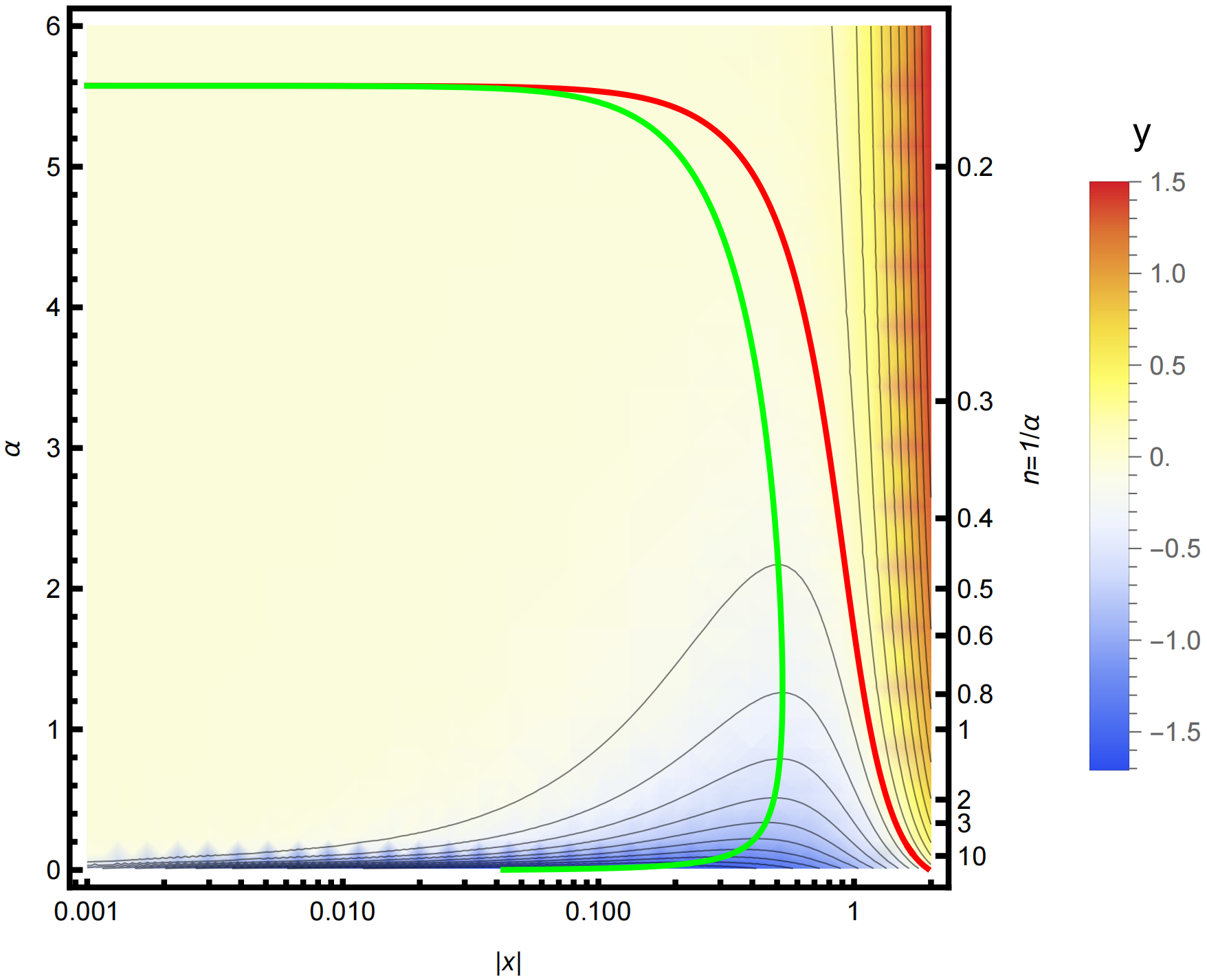}
        \caption{Same as Fig.~\ref{imagine}, but for Einasto profile.}
        \label{imagineEinasto}
    \end{figure}

Finally, we present the density map satisfying Eq.~(\ref{lightfunction}). For given model parameters, Figs.~\ref{imagine} and \ref{imagineEinasto} allow a quick determination of the number of images and whether the images and the source lie on the same or opposite sides of the lens plane center. In these figures, the image coordinates are defined to be always positive. Therefore, $y>0$ indicates that the image and the source lie on the same side of the lens plane center, while $y<0$ denotes that they are on opposite sides.

\subsection{Magnification $\mu$ and the Morse Index}\label{subSec:magnification}

\begin{figure}[!htb]
        \centering
        \includegraphics[width=0.45\textwidth]{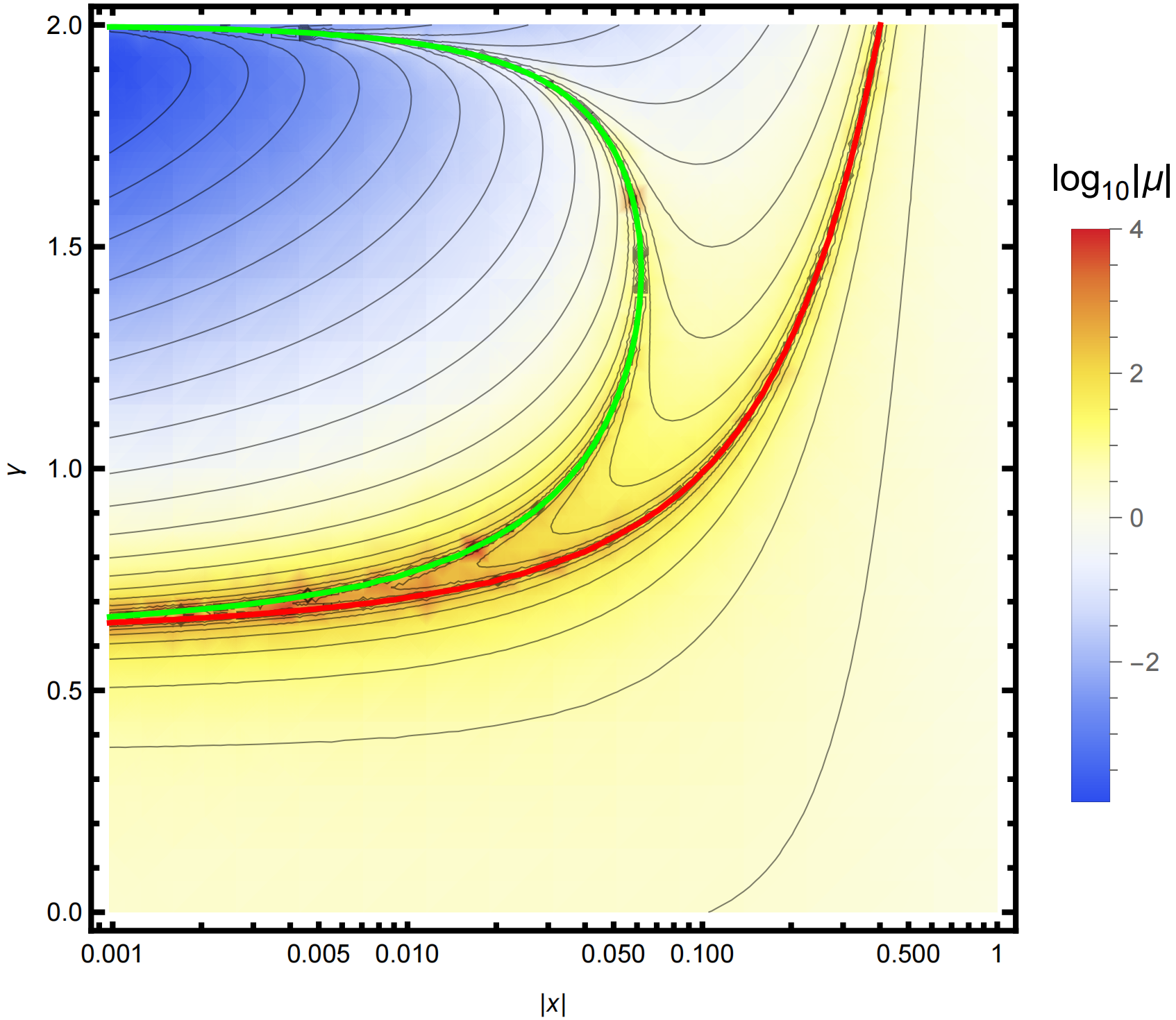}
        \caption{Density map of $\mathrm{log_{10}|\mu|}$ for gNFW profile, with $M_{200}=2.86\times 10^{15}{M_{\odot}}$. The physical meanings of the red and green curves are the same as those in Fig.~\ref{critical condition}, while the black curves represent contour lines. From the figure, it can be seen that when the imaging configuration satisfies either of the critical conditions, the magnification $\mu$ in Eq.~(\ref{miu}) diverges, resulting in exceptionally bright lensing images.}
        \label{magnification}
    \end{figure}

\begin{figure}[!htb]
        \centering
        \includegraphics[width=0.45\textwidth]{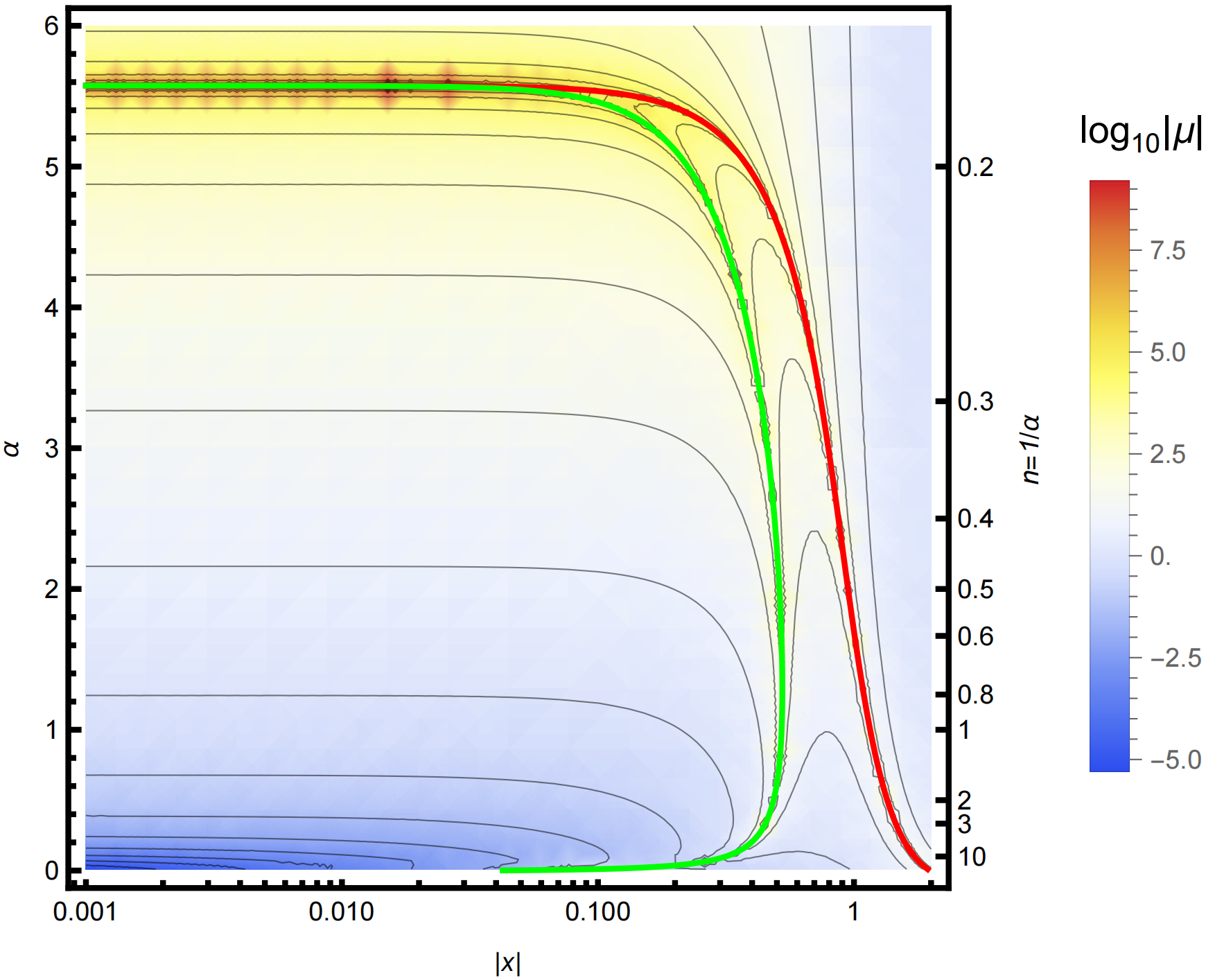}
        \caption{Same as Fig.~\ref{magnification}, but for Einasto profile.}
        \label{magnificationEinasto}
    \end{figure}

In the axisymmetric case, the Hessian matrix of Eq.~(\ref{T}) takes a simple diagonal form when expressed in polar coordinates,
    \begin{equation}
        \begin{aligned}
            A&=\mathrm{Hess}(T)=I-\mathrm{Hess}(\psi),\\
            &=\begin{pmatrix}
                1-\psi''\big(r\big)&0\\
                0&1-\frac{\psi'\big(r\big)}{r}
              \end{pmatrix}.
        \end{aligned}
        \label{Hessian}
    \end{equation}
$A$ has two eigenvalues, $\lambda_{\mathrm{r}}=1-\psi''\big(r\big)$ and $\lambda_{\mathrm{t}}=1-\frac{\psi'\big(r\big)}{r}$, the magnification $\mu$ is given by the inverse of the determinant of $A$,
    \begin{equation}
        \begin{aligned}
            \mu=\frac{1}{\left|A\right|}=\frac{1}{\lambda_{\mathrm{r}}\lambda_{\mathrm{t}}}=\frac{1}{\Big(1-\psi''\big(r\big)\Big)\Big(1-\frac{\psi'\big(r\big)}{r}\Big)}.
        \end{aligned}
        \label{miu}
    \end{equation}
The sign of the magnification $\mu$ merely indicates whether the image is mirrored, and in this work we only focus on the magnitude of the magnification and do not consider image parity. Consequently, we obtain the density map of $\mu$ with different conditions, as shown in Fig.~\ref{magnification} and Fig.~\ref{magnificationEinasto}. From these figures, it is clear that the magnification $\mu$ diverges when either of the two critical conditions is satisfied. Some studies \cite{10.1007/978-3-662-03758-4,10.1007/BF00654034,10.1086/164389} have pointed out that even in the geometrical optics limit, wave optics effects must be taken into account when the magnification diverges. However, this complication is beyond the scope of the present work.

Next, we proceed to examine the Morse index $n_{\mathrm{M}}$ of each image. The Morse index of an image is defined as the number of negative eigenvalues of the Hessian matrix in Eq.~(\ref{Hessian}), $n_{\mathrm{M}}=0,1,2$ when image position is a minimum, saddle, or maximum point of $T\big(x,y\big)$, respectively. Combining with Eq.~(\ref{miu}), the Morse index $n_{\mathrm{M}}$ can be easily inferred from Fig.~\ref{magnification}: $n_{\mathrm{M}}=2$, on the left side of the radial critical curve (green); $n_{\mathrm{M}}=1$, in the region between the radial critical curve (green) and the tangential critical curve $y=0$ (red); $n_{\mathrm{M}}=0$, on the right side of the tangential critical curve $y=0$ (red).

By combining the figures of Eqs.~(\ref{lightfunction}) and (\ref{miu}), we find that the magnification of the image on the same side as the source ($n_{\mathrm{M}}=0$) is generally higher than that of the images on the opposite side ($n_{\mathrm{M}}=1$ or $2$). Among the opposite side images, the one located closer to the center of the lens plane ($n_{\mathrm{M}}=2$) typically has very small magnification, and is therefore difficult to detect in actual observations.

\subsection{Overview of phase diagram}
\label{subSec:phase diagram}

\begin{figure}[!htb]
        \centering
        \includegraphics[width=0.45\textwidth]{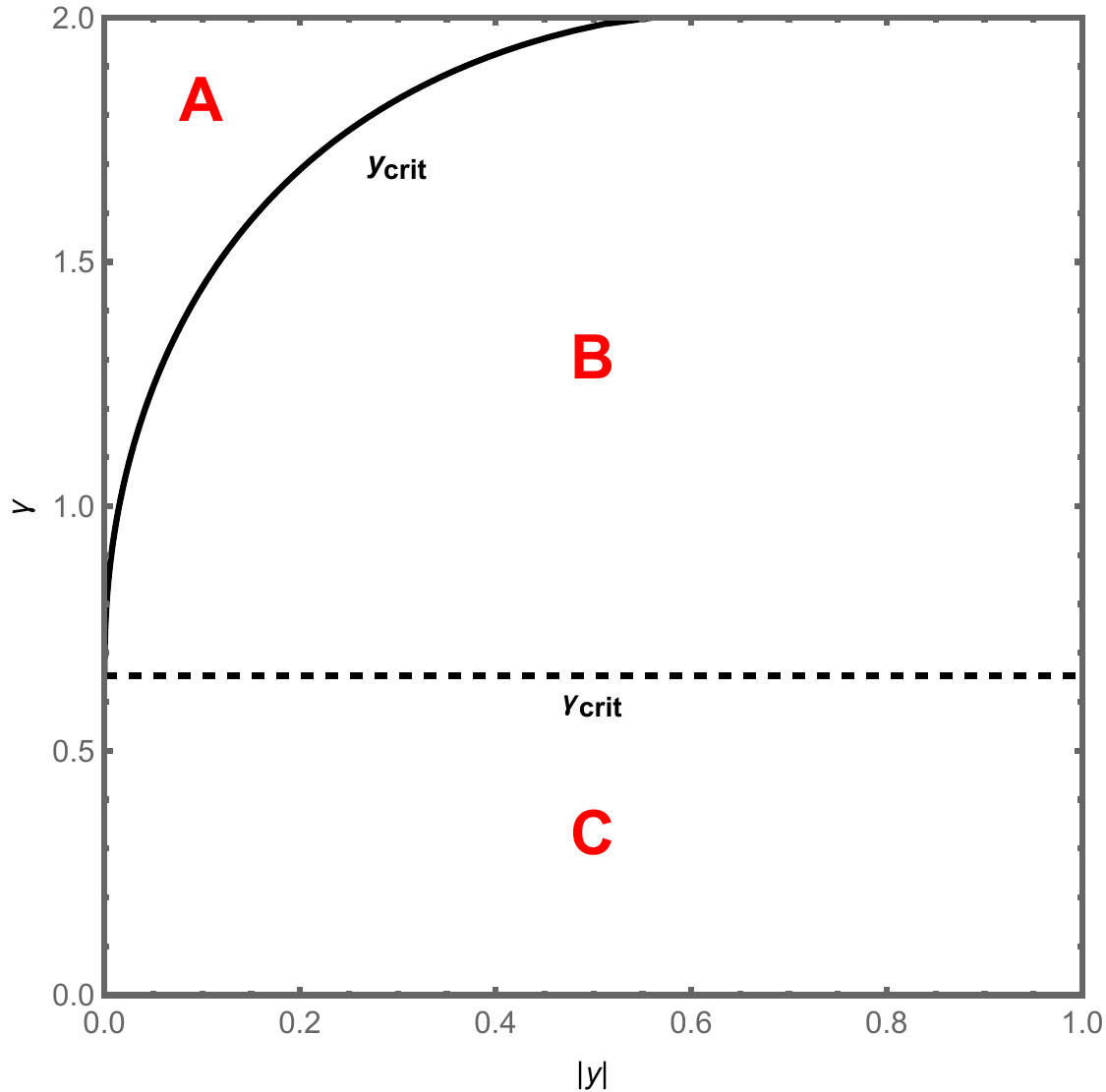}
        \caption{$\big(\big|y\big|,\gamma\big)$ phase diagram for gNFW profile, $M_{200}=2.86\times 10^{15}{M_{\odot}}$. The diagram is divided by two critical curves ($\gamma_{\mathrm{crit}}$ and $y_{\mathrm{crit}}$) into three regions: A, B and C. $\gamma_{\mathrm{crit}}$ denotes the minimum value of $\gamma$ required for the formation of an Einstein ring, while $y_{\mathrm{crit}}$ represents the curve $y_{\mathrm{crit}}=y_{\mathrm{crit}}\big(\gamma\big)$ of Eq.~(\ref{ycrit}) when Eq.~(\ref{radial}) has a solution. A: three images with Morse indices $n_{\mathrm{M}}=0,1,2$; B, C: one image with $n_{\mathrm{M}}=0$.}
        \label{yGamma}
    \end{figure}

\begin{figure}[!htb]
        \centering
        \includegraphics[width=0.45\textwidth]{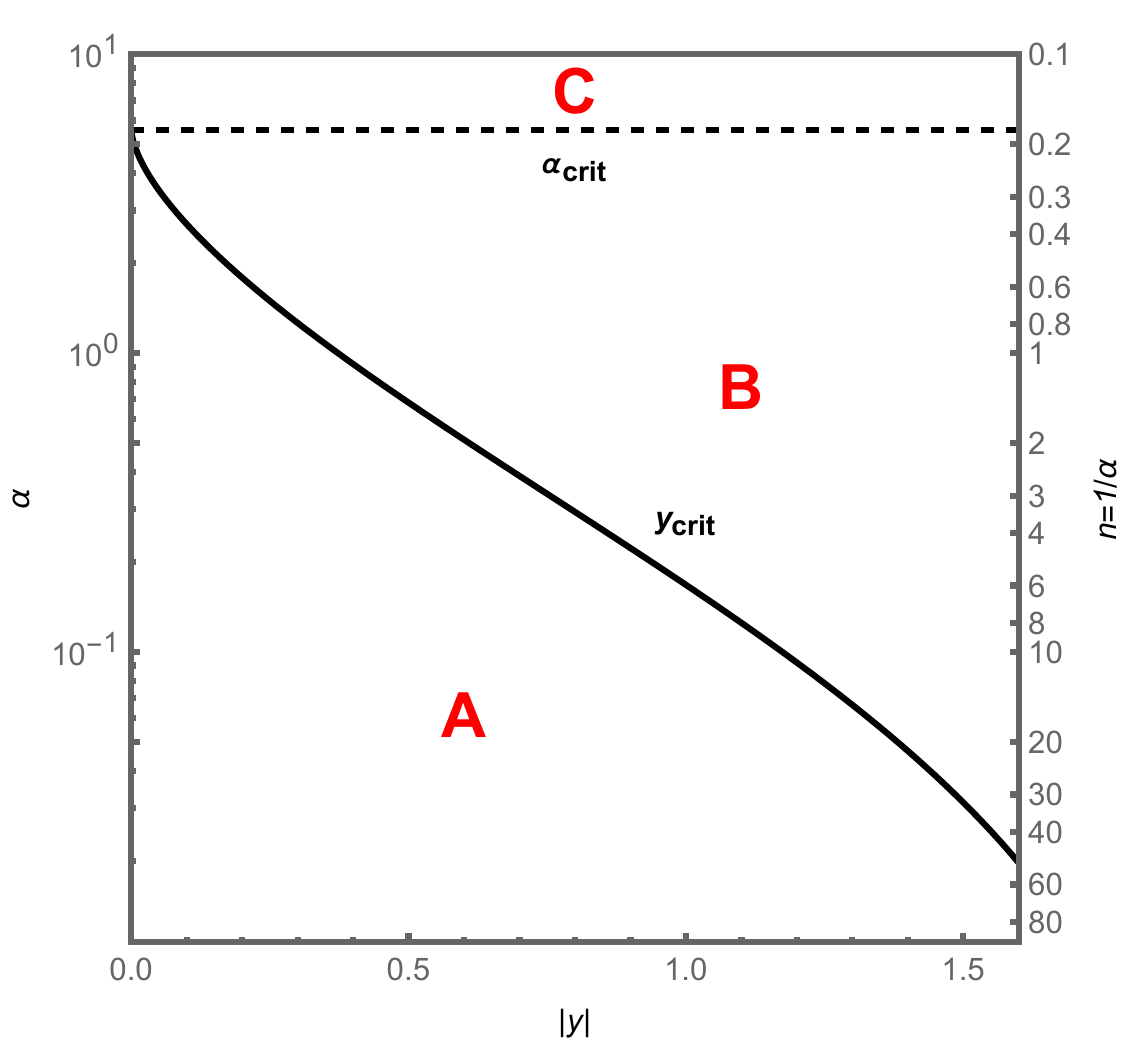}
        \caption{Same as Fig.~\ref{yGamma}, but for Einasto profile.}
        \label{ynEinasto}
    \end{figure}

In this section, we summarize the imaging characteristics of the gravitational lensing system. For a given lens configuration with $z_{\mathrm{l}}=0.5,z_{\mathrm{s}}=1.5,\mathrm{log}\big(M_{200}/M_{\mathrm{pivot}}\big)=3$, the imaging properties of gNFW profile and Einasto profile can be identified within the parameter space $\big(\big|y\big|,\gamma\big)$ and $\big(\big|y\big|,\alpha\big)$ (see Fig.~\ref{yGamma} and Fig.~\ref{ynEinasto}). It is worth noting that the curves $\gamma_{\mathrm{crit}}$ and $y_{\mathrm{crit}}$ in Fig.~\ref{yGamma} indeed intersect at $y=0$, the proof is straightforward and is provided in Appendix \ref{App:Proof}. Under axisymmetric conditions, this conclusion is independent of single parameter density models and holds universally.

\section{Amplification Factor $F$}\label{Sec:F}

Based on Sec.~\ref{Sec:Parameter Space}, we can now investigate the amplification factor $F$ for a given model. The general form of the amplification factor $F$ is \cite{10.1051/0004-6361:20040212,10.1103/PhysRevD.102.124076}
    \begin{equation}
        \begin{aligned}
F\big(w,\boldsymbol{y}\big)=\frac{w}{2\pi i}\int e^{iwT(\boldsymbol{x},\boldsymbol{y})}d^{2}x.
        \end{aligned}
        \label{F}
    \end{equation}
The parameter $w$ denotes the dimensionless frequency, which characterizes the wave optics or geometrical optics nature of the observed source, 
    \begin{equation}
        \begin{aligned}
            w=\frac{D_{\mathrm{s}}\xi_{0}^{2}}{D_{\mathrm{l}}D_{\mathrm{ls}}c_{0}}\big(1+z_{\mathrm{l}}\big)2\pi f_{\mathrm{obs}}.
        \end{aligned}
        \label{w}
    \end{equation}
Here $f_{\mathrm{obs}}$ denotes the observed GW frequency, we set the parameters $\big(z_{\mathrm{l}}=0.5$, $z_{\mathrm{s}}=1.5$, $\mathrm{log}\big(M_{200}/M_{\mathrm{pivot}}\big)=3\big)$ and $\xi_{0}=r_{\mathrm{s}}$. Under the axisymmetric assumption adopted in this work, Eq.~(\ref{F}) can be reduced to a one dimensional integral along the radial direction \cite{10.1051/0004-6361:20040212, 10.1103/PhysRevD.102.124076},
    \begin{equation}
        \begin{aligned}
            F\big(w,y\big)&=-iwe^{iw\big(y^{2}/2+\phi_{\mathrm{m}}(y)\big)}\\
            &\quad\times\int_{0}^{\infty}xJ_{0}\big(wxy\big)e^{iw\big(x^{2}/2-\psi(x)\big)}dx.
        \end{aligned}
        \label{Fone}
    \end{equation}
Here $J_{n}\big(x\big)$ denotes the $n$th-order Bessel function, which oscillates with $J_{n}\big(x\big)\to0$ as $x\to\infty$. The integral in the above expression involves an oscillatory kernel, when $x$ is large, the integrand exhibits rapid oscillations. In the Appendix \ref{App:Numerical Methods}, we give three numerical methods for efficiently computing Eq.~(\ref{Fone}) and compare their results.

In this section, we focus on the variation of the amplification factor $F\big(w,y\big)$ (the modulus $\big|F\big|$ and the phase $\mathrm{Arg}(F)$) with the source position $y$. Our results show that, as $y$ increases, the number of gravitational lensed images decreases from three to one. This transition in image multiplicity has a significant impact on $F\big(w,y\big)$, with particularly pronounced changes occurring around $y\sim y_{\mathrm{crit}}$.

\subsection{gNFW profile}\label{subSec:gNFW profile}

\begin{figure}[!htb]
        \centering
        \includegraphics[width=0.45\textwidth]{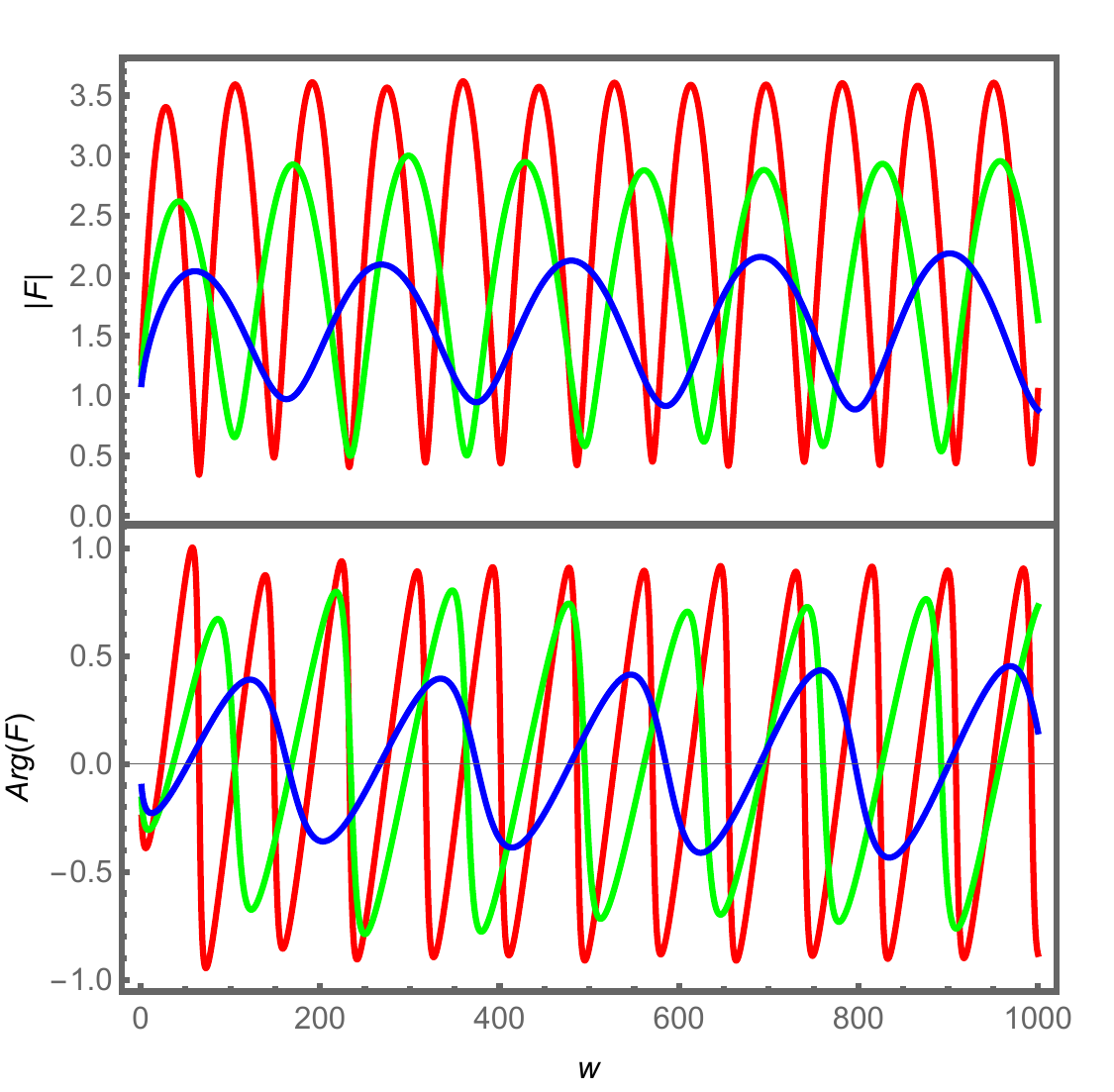}
        \caption{$F\big(w,y\big)$ for gNFW profile with $\gamma=1.89$ and $y=0.1$. Red: $M_{200}=2.86\times 10^{15}{M_{\odot}}$; Green: $M_{200}=2.86\times 10^{14}{M_{\odot}}$; Blue: $M_{200}=2.86\times 10^{13}{M_{\odot}}$.}
        \label{gNFWxM}
    \end{figure}

We first investigate the effect of lens mass $M_{200}$ on $F\big(w,y\big)$, Fig.~\ref{gNFWxM} illustrates the variation of $F\big(w,y\big)$ with different lens mass $M_{200}$ when $\gamma=1.89$ and $y=0.1$. It indicates that the smaller the lens mass, the weaker the gravitational effect it can produce, and the oscillation frequency of $F\big(w,y\big)$ also decreases as expected.
    
Then we investigate the effect of $\gamma$ on $F\big(w,y\big)$, here we choose $M_{200}=2.86\times 10^{15}{M_{\odot}}$ and $\kappa_{s}\big(\gamma\big)$ is shown in Fig.~\ref{kappasgamma} (red line) and Table.~\ref{tablekappasgamma}, which demonstrate that for the gNFW profile, a massive lens always produces a strong gravitational effect.

    \begin{table}[]
        \begin{tabular}{c|cc}
        \hline
       Profile & $\gamma$ or $\alpha$   & $\kappa_{s}$   \\ \hline
       \multirow{5}{*}{gNFW} 
       & $\gamma=$0.2  & 0.3338  \\
       & $\gamma=$0.6 & 0.2616  \\
        & $\gamma=$1.0  & 0.2010  \\
        & $\gamma=$1.6  & 0.1279  \\     
        & $\gamma=$2.0  & 0.08819 \\ \hline
       Einasto & $\alpha=$0.16 & 0.315                    \\ \hline
        \end{tabular}
        \caption{Values of $\kappa_{s}$ for gNFW and Einasto profile ($\alpha=0.16$), $M_{200}=2.86\times 10^{15}{M_{\odot}}$.}
        \label{tablekappasgamma}
    \end{table}
    
\begin{figure}[!htb]
        \centering
        \includegraphics[width=0.45\textwidth]{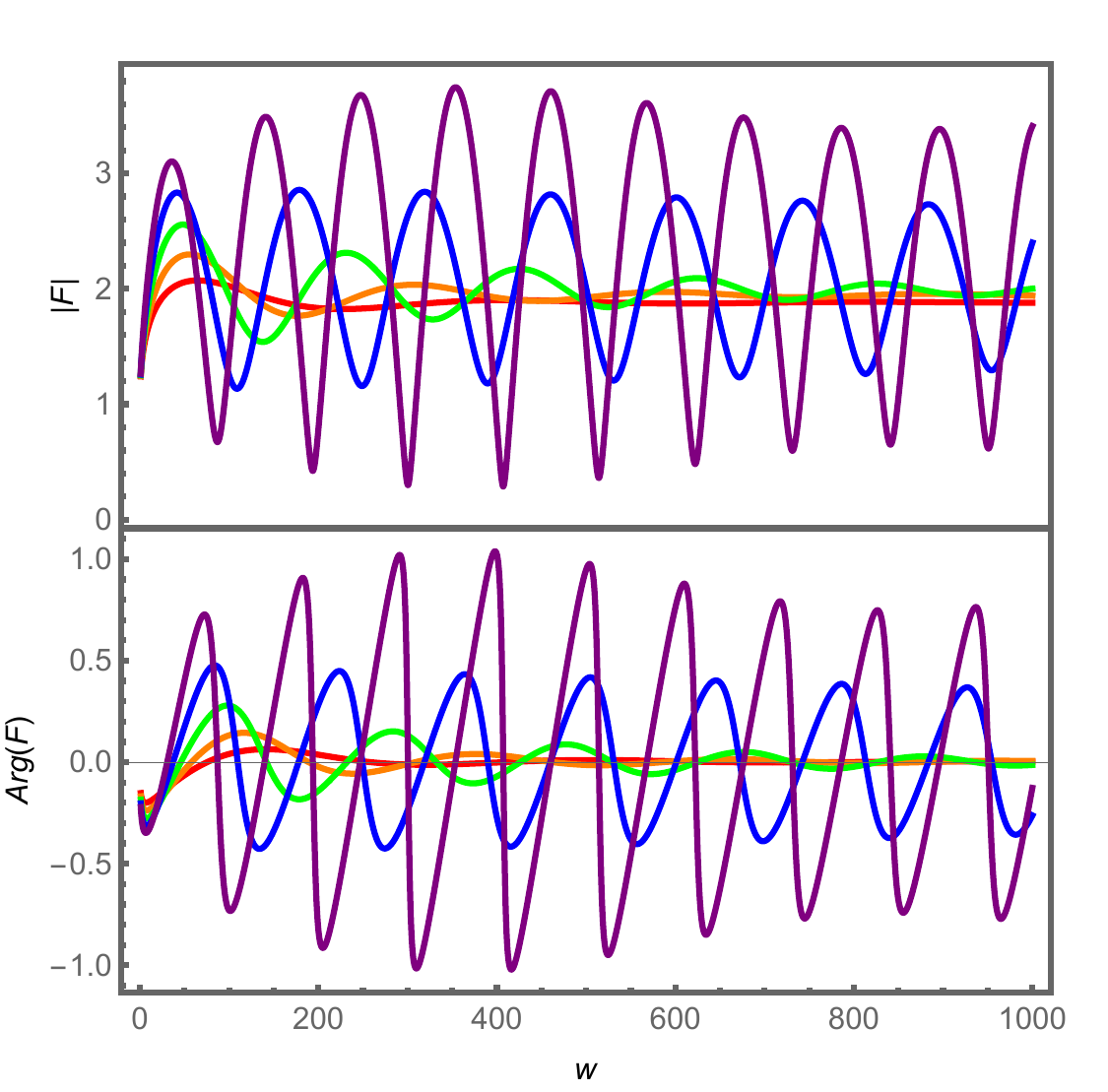}
        \caption{$F\big(w,y\big)$ for gNFW profile with $y=0.1$. Red: $\gamma=0.8$; Orange: $\gamma=1$ (NFW profile); Green: $\gamma=1.2$; Blue: $\gamma=1.4$; Purple: $\gamma=1.6$.}
        \label{gNFWFgamma}
    \end{figure}

Fig.~\ref{gNFWFgamma} illustrates the variation of $F\big(w,y\big)$ when $y=0.1$, we observe that $\gamma$ has a significant impact on both the amplitude and oscillation frequency of $\big|F\big|$ and $\mathrm{Arg}(F)$, an increase in $\gamma$ leads to a larger amplitude and oscillation frequency, as well as a clearer wave packet profile for multiple-image regime. It is evident from Fig.~\ref{yGamma} that for $y=0.1$, the system exhibits three images when $\gamma>1.4$ and one image when $\gamma<1.4$, while $\gamma\sim 1.4$ represents a near-critical case.

Now, we show the evolution of $F\big(w,y\big)$ with frequency across different regimes, here we choose $\gamma=1.89$, the lens mass is $M_{200}=2.86\times 10^{15}{M_{\odot}}$.

\begin{figure}[!htb]
        \centering 
        \subfigure[$w$: $1\sim1000$]{
        \label{gNFWF3imagesub1}
        \includegraphics[width=0.45\textwidth]{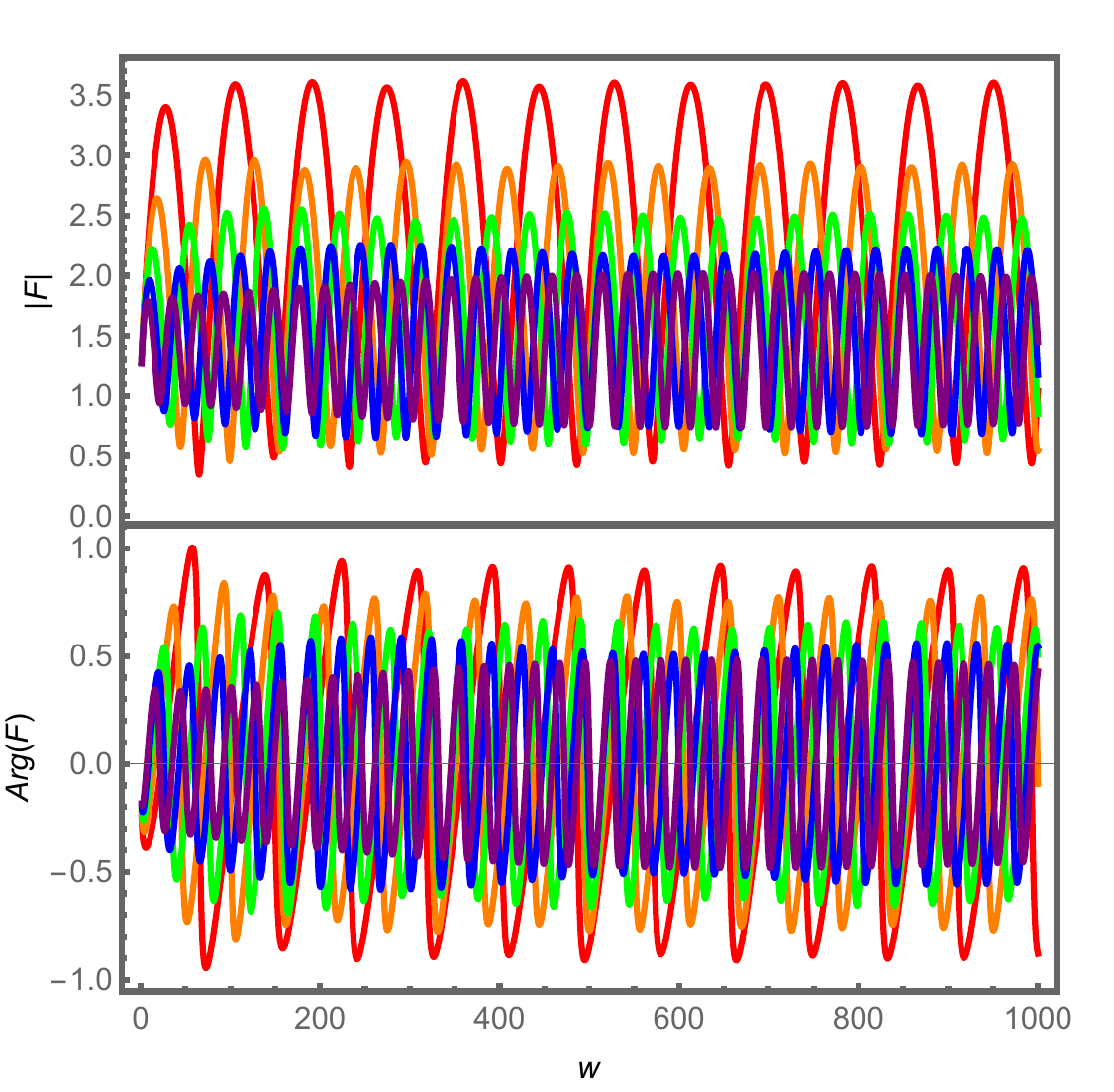}}
        \subfigure[$w$: $900\sim1000$]{
        \label{gNFWF3imagesub2}
        \includegraphics[width=0.45\textwidth]{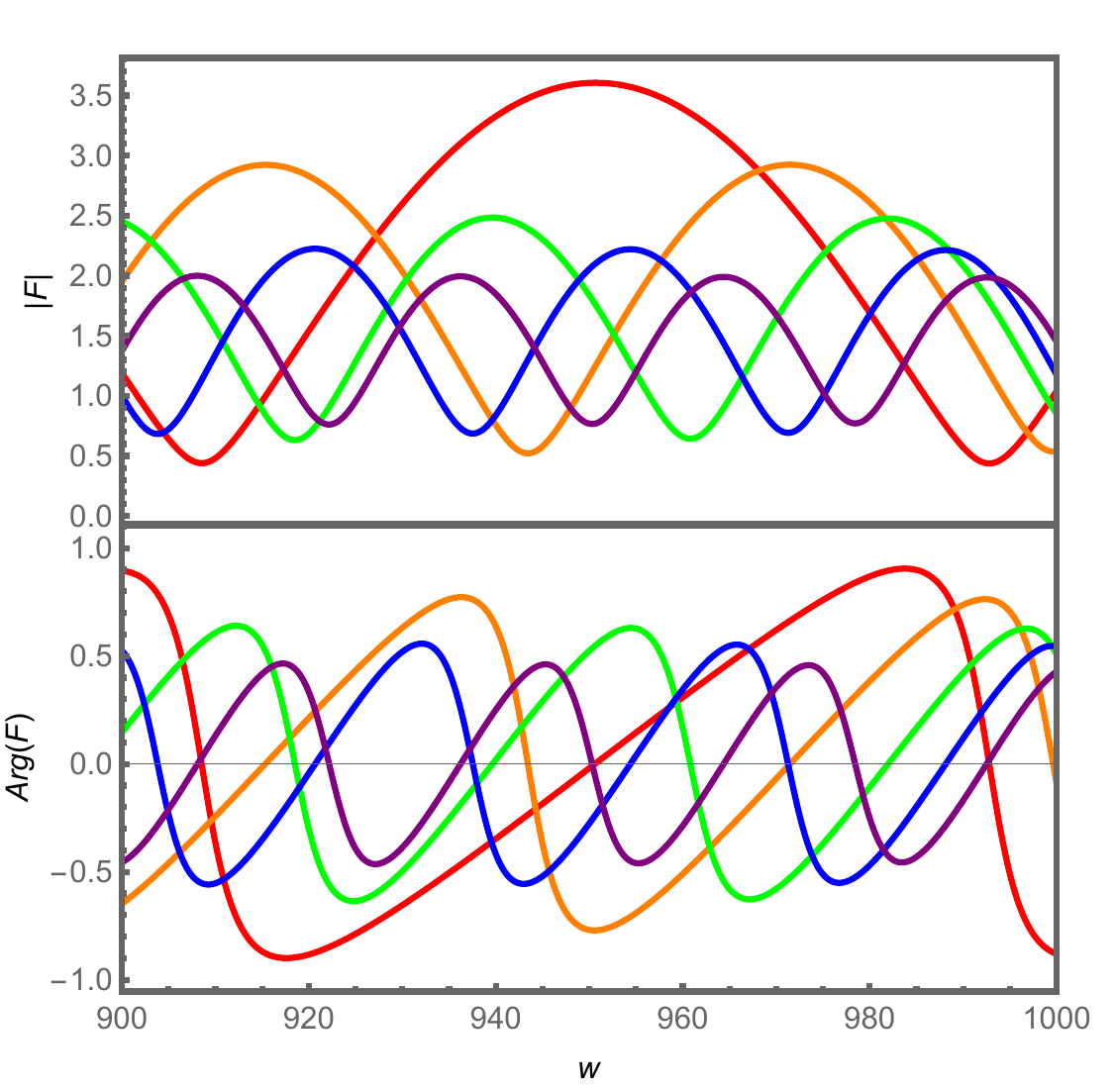}}
        \caption{Multiple-image regime, $F\big(w,y\big)$ for gNFW profile ($\gamma=1.89$). Red: $y=0.1$; Orange: $y=0.15$; Green: $y=0.2$; Blue: $y=0.25$; Purple: $y=0.3$.}
        \label{gNFWF3image}
    \end{figure}

{\it Multiple-image regime.---} For $y<0.9y_{\mathrm{crit}}$, corresponding to the region A in Fig.~\ref{yGamma}, the system is well away from the critical line $y_{\mathrm{crit}}$. As shown in Fig.~\ref{gNFWF3image}, the oscillation frequencies of both $\big|F\big|$ and $\mathrm{Arg}(F)$ increase with increasing $y$, whereas the oscillation frequency of the wave packet decreases accordingly, and the average value of $\big|F\big|$ also diminishes. So in the multiple-image regime, $F\big(w,y\big)$ can sustain stable oscillations, and the envelope is symmetric about the mean value. Notably, the critial parameter $y_{\mathrm{crit}}$ for the gNFW profile is smaller than that for the Einasto profile. This results a relatively lower oscillation frequency of $F\big(w,y\big)$ throughout the parameter range, which consequently renders the envelopes of $\big|F\big|$ and $\mathrm{Arg}(F)$ poorly resolved.

\begin{figure}[!htb]
        \centering
        \includegraphics[width=0.45\textwidth]{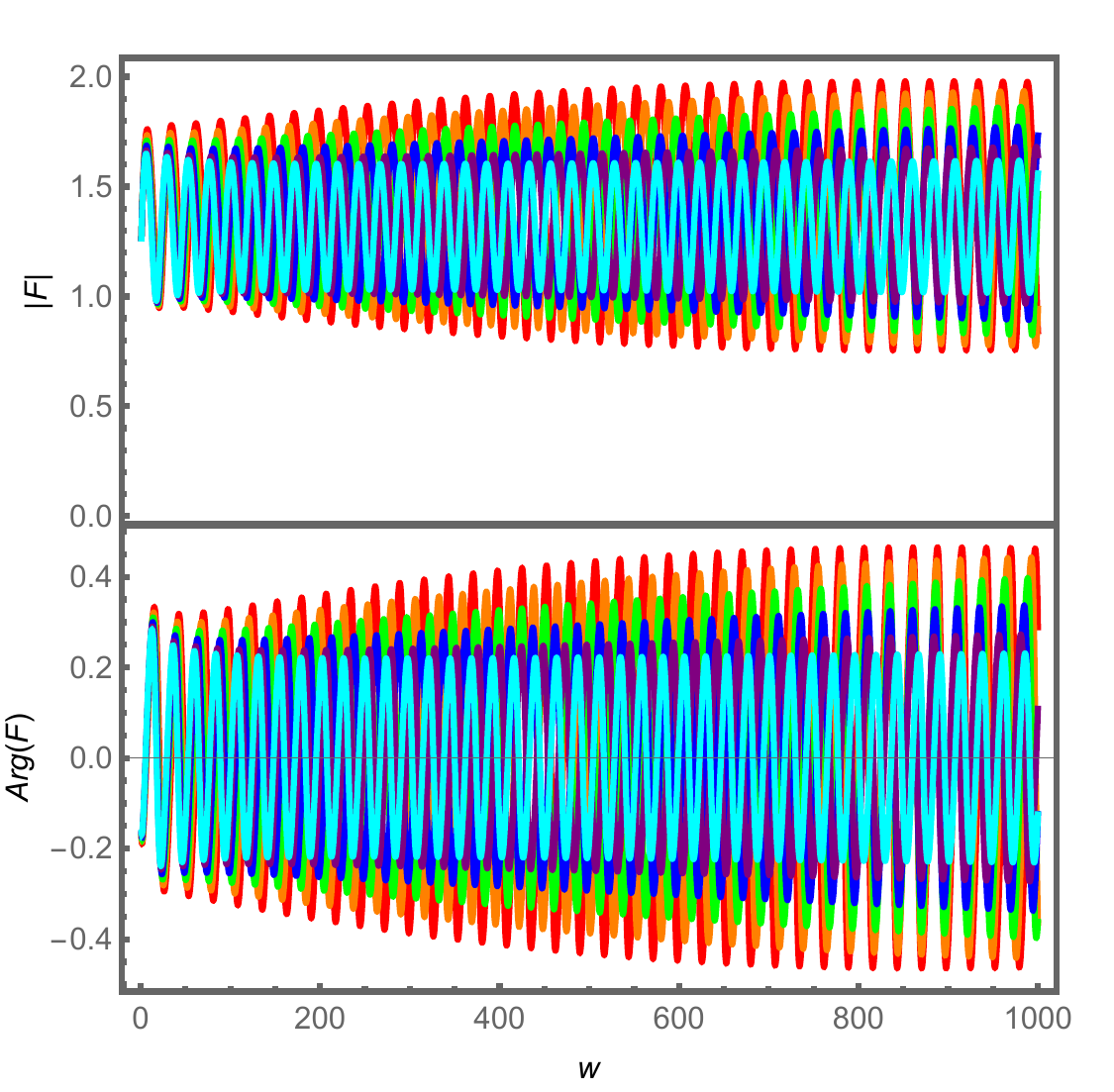}
        \caption{Transition regime, $F\big(w,y\big)$ for gNFW profile with $\gamma=1.89$. Red: $y=0.31$; Orange: $y=0.32$; Green: $y=0.33$; Blue: $y=0.34$; Purple: $y=0.35$; Cyan: $y=y_{\mathrm{crit}}=0.35690236876$.}
        \label{gNFWFtrans}
    \end{figure}

{\it Transition regime.---} For $0.9y_{\mathrm{crit}}\le y\le y_{\mathrm{crit}}$, corresponding to region A in Fig.~\ref{yGamma}, the system lies in the vicinity of the critical line $y_{\mathrm{crit}}$. As shown in Fig.~\ref{gNFWFtrans}, for both $\big|F\big|$ and $\mathrm{Arg}(F)$, the envelopes start to deform while remaining symmetric. With increasing $y$, the envelopes gradually contract, with the region $w<200$ contracting at a faster rate. Consequently, the envelopes exhibit a pronounced necking.

\begin{figure}[!htb]
        \centering
        \includegraphics[width=0.45\textwidth]{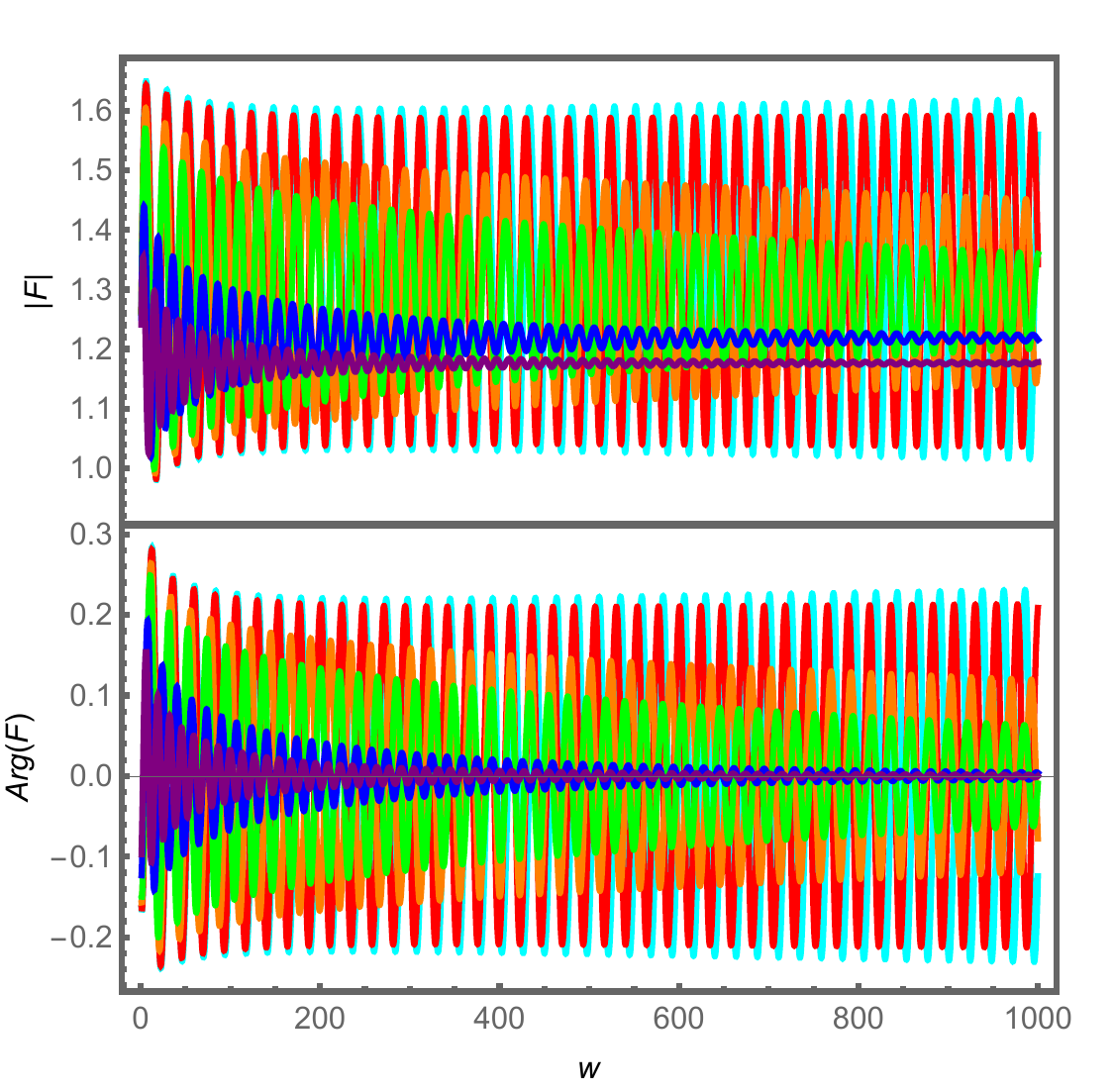}
        \caption{Single-image regime, $F\big(w,y\big)$ for gNFW profile with $\gamma=1.89$. Cyan: $y=y_{\mathrm{crit}}=0.35690236876$; Red: $y=0.36$; Orange: $y=0.38$; Green: $y=0.4$; Blue: $y=0.5$; Purple: $y=0.6$.}
        \label{gNFWF1image}
    \end{figure}

{\it Single image regime.---} $y>y_{\mathrm{crit}}$, corresponding to region B in Fig.~\ref{yGamma}. As shown in Fig.~\ref{gNFWF1image}, for both $\big|F\big|$ and $\mathrm{Arg}(F)$, the entire envelope begins to contract rapidly as $y$ increases, while the average value of $\big|F\big|$ also diminishes.

\subsection{Einasto profile}\label{subSec:FEinasto}
\begin{figure}[!htb]
        \centering
        \includegraphics[width=0.5\textwidth]{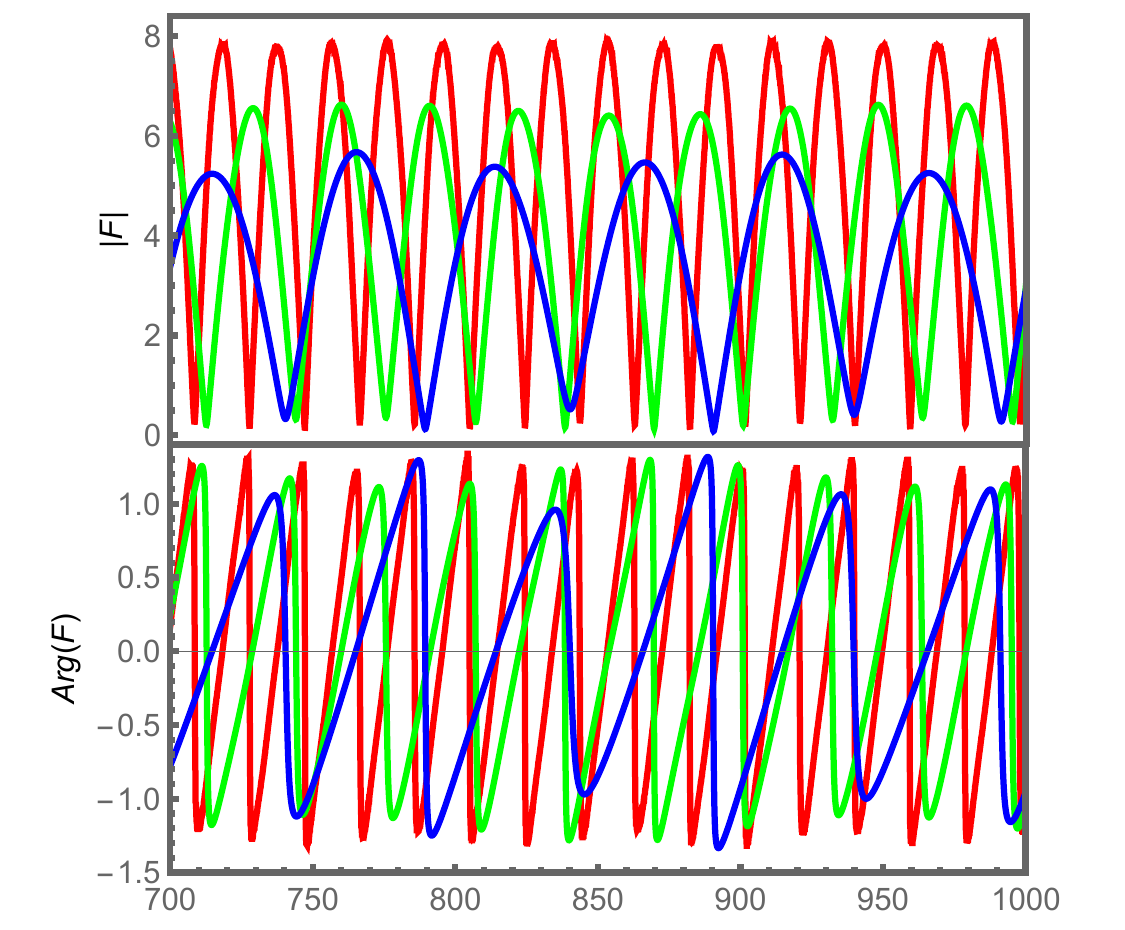}
        \caption{$F\big(w,y\big)$ for Einasto profile with $\alpha=0.16$ and $y=0.1$ ($700\le w\le1000$). Red: $M_{200}=2.86\times 10^{15}{M_{\odot}}$; Green: $M_{200}=2.86\times 10^{14}{M_{\odot}}$; Blue: $M_{200}=2.86\times 10^{13}{M_{\odot}}$.}
        \label{EinastoxM}
    \end{figure}

The effect of lens mass on the Einasto profile ($\alpha=0.16$ and $y=0.1$) is similar to that on the gNFW profile, as shown in Fig.~\ref{EinastoxM}. Similarly, the boundary between the diffraction effects of $F\big(w,y\big)$ and the geometric optics approximation for lenses of different masses resembles that of the gNFW case. Given that the oscillation frequency of $F\big(w,y\big)$ with the Einasto profile is much higher than that with the gNFW profile, and that the oscillation in the multiple-image regime are periodic, so only a partial interval is shown in Fig.~\ref{EinastoxM} for clarity.

Following \cite{10.1038/s41586-020-2642-9}, we adopt $\alpha=0.16$ for detailed analysis, we choose lens mass $M_{200}=2.86\times 10^{15}{M_{\odot}}$ and $\kappa_{s}=0.315$. By comparing $\kappa_{s}$ with the two profiles, we find that the lensing effect produced by the Einasto profile is generally stronger than that of the gNFW profile (for $\gamma>0.3$). Unlike the case with gNFW profile, for Einasto profile, the transition regime exhibits considerable complexity, with $F\big(w,y\big)$ undergoing drastic changes.

\begin{figure}[!htb]
        \centering 
        \subfigure[$w$: $1\sim1000$]{
        \label{EinastoNatureF3imagesub1}
        \includegraphics[width=0.45\textwidth]{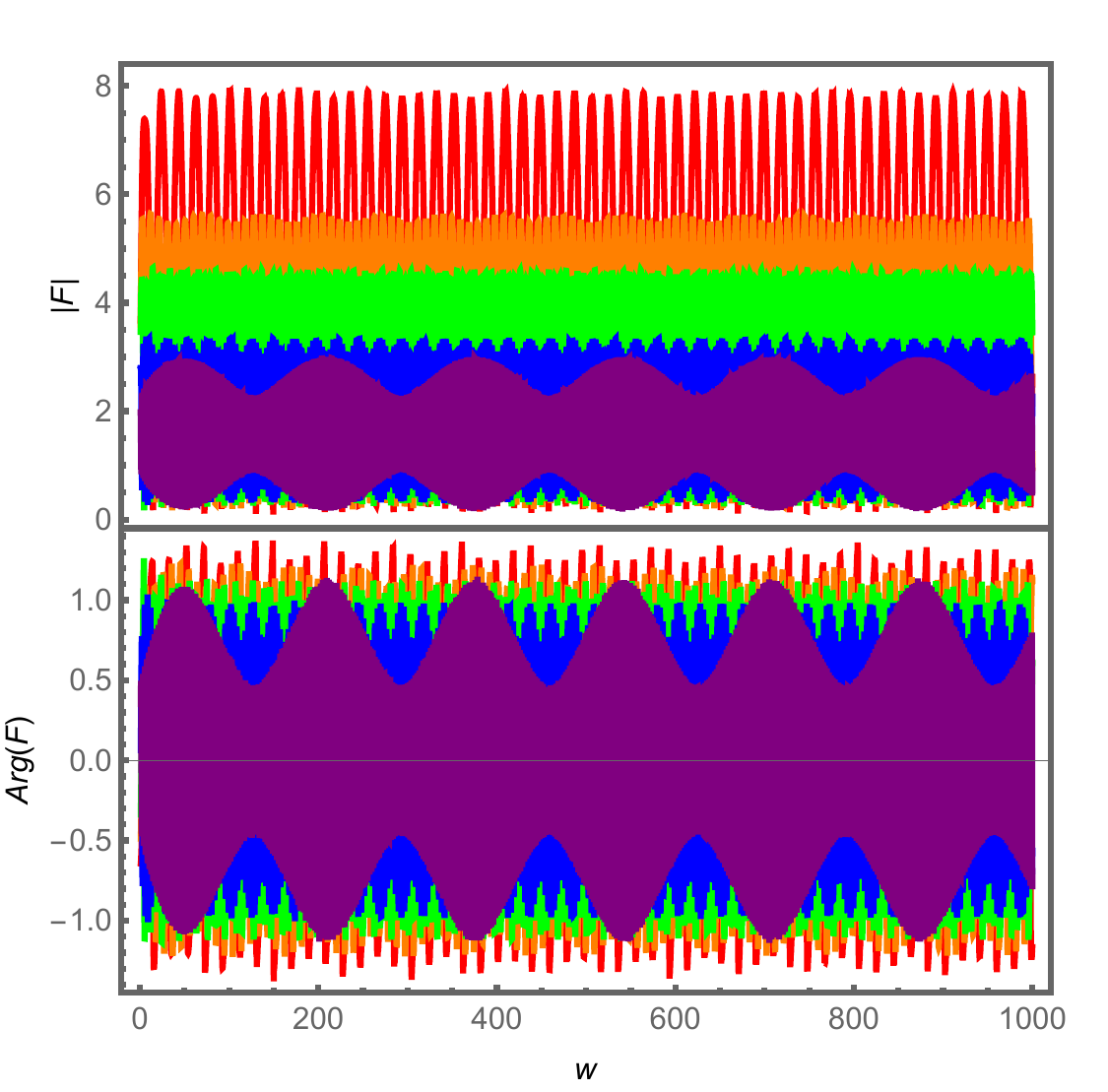}}
        \subfigure[$w$: $900\sim1000$]{
        \label{EinastoNatureF3imagesub2}
        \includegraphics[width=0.45\textwidth]{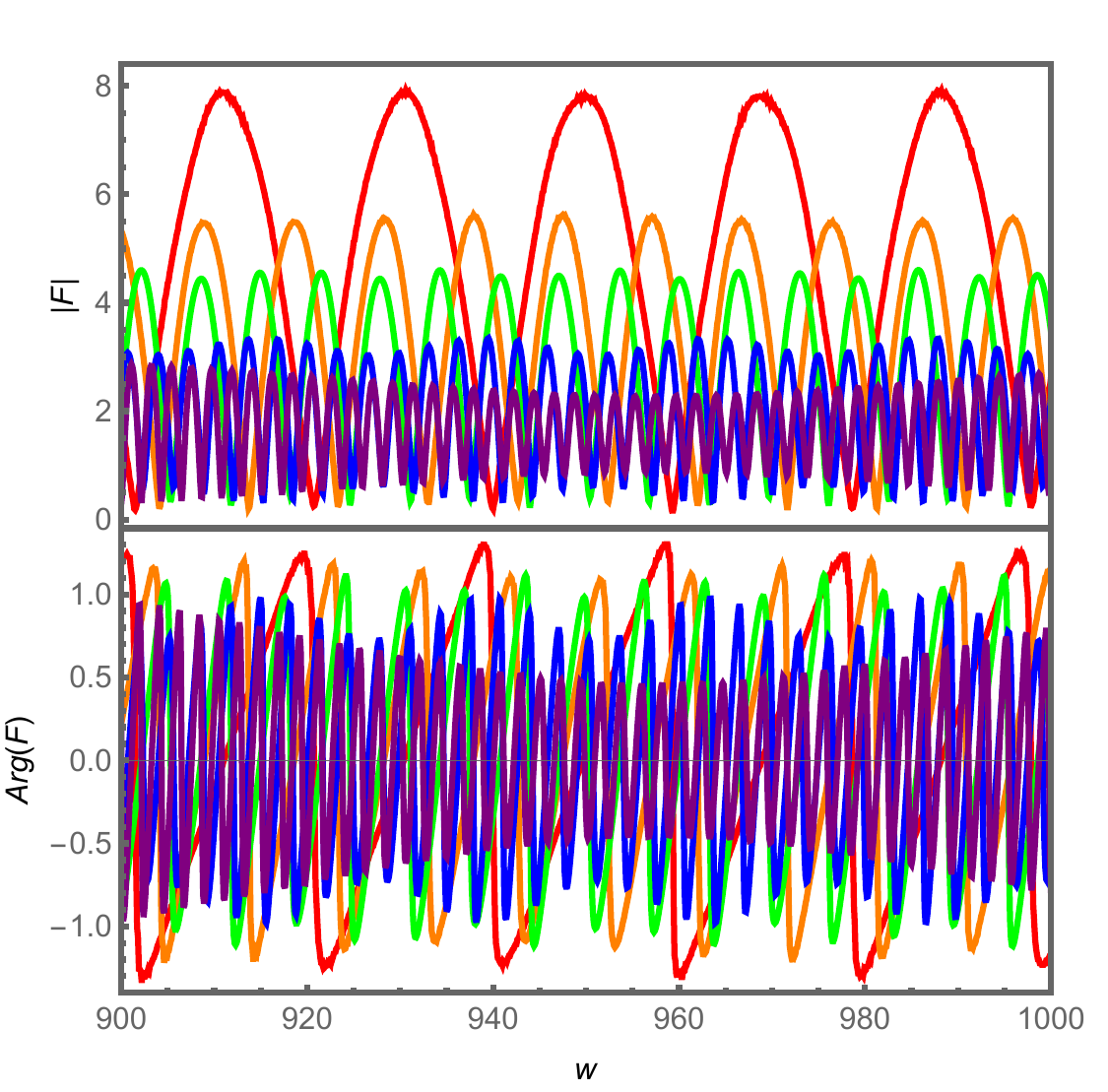}}
        \caption{Multiple-image regime, $F\big(w,y\big)$ for Einasto profile ($\alpha=0.16$). Red: $y=0.1$; Orange: $y=0.2$; Green: $y=0.3$; Blue: $y=0.6$; Purple: $y=0.9$.}
        \label{EinastoNatureF3image}
    \end{figure}

{\it Multiple-image regime.---} For $y<0.95y_{\mathrm{crit}}$, corresponding to the region A in Fig.~\ref{ynEinasto}, the system is well away from the critical line $y_{\mathrm{crit}}$. As shown in Fig.~\ref{EinastoNatureF3image}, similar to the gNFW profile, both $\big|F\big|$ and $\mathrm{Arg}(F)$ exhibit the rapidly oscillating with stable amplitude, while the oscillation frequency of the wave packet decreases with increasing $y$, and the envelopes maintain good symmetry about the mean value.

{\it Transition regime.---} For $0.95y_{\mathrm{crit}}\le y\le 1.05y_{\mathrm{crit}}$, corresponding to the system lies in the vicinity of the critical line $y_{\mathrm{crit}}$ in Fig.~\ref{ynEinasto}. In this regime, the situation differs entirely from the gNFW profile, $F\big(w,y\big)$ undergoes dramatic variations, which can be divided into three regimes.

$\big(i\big)$: For $\big|F\big|$, starting from the multiple-image regime, the envelope continues to stretch, the oscillation frequency of the wave packet decreases, and the lower envelope exhibits pronounced deformation, breaking the axisymmetric between the upper and lower envelopes. As $y$ increases, the oscillatory behavior of the lower envelope progressively converges to that of the upper envelope; For $\mathrm{Arg}(F)$, it exhibits a transition behavior that is entirely distinct from that of $\big|F\big|$, featuring particularly rich morphological variations. In this regime, as $y$ increases, while the oscillation frequency of wave packet continues to decrease, the widest portion of the wave packet undergoes significant morphological transformation, ``fins" emerge, the width of which is equal to that of the arched portion of $\big|F\big|$ lower envelope. Fig.~\ref{EinastoNatureFtransi} presents a comparison of $F\big(w,y\big)$ for different $y$, detailed descriptions are provided in Appendix \ref{App:FTrans}.

$\big(ii\big)$: From Fig.~\ref{EinastoNatureFtransii} in Appendix \ref{App:FTrans}, it can be seen that for $\big|F\big|$, based on regime $\big(i\big)$, as $y$ increases, the lower envelope keeps evolving, and in the critical case ($y=y_{\mathrm{crit}}$) the upper and lower envelopes regain their symmetry about the mean value; For $\mathrm{Arg}(F)$, the ``fin" structure appears at the lowest point of $\big|F\big|$ lower envelope. As $y$ gradually approaches $y_\mathrm{crit}$, the smooth portion of $\mathrm{Arg}(F)$ envelopes progressively replace the ``fins" and symmetry is subsequently restored.

\begin{figure}[!htb]
        \centering
        \includegraphics[width=0.45\textwidth]{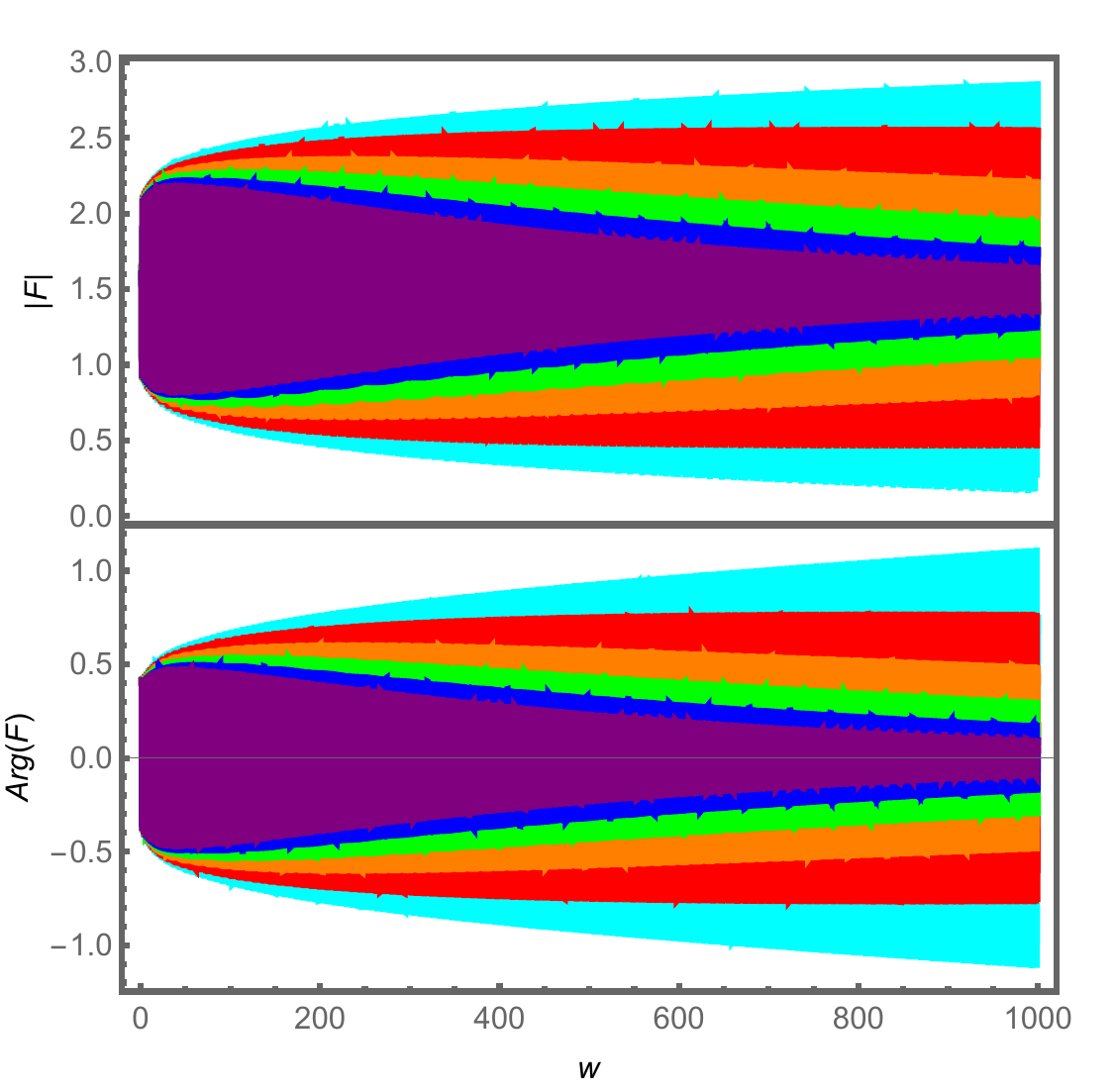}
        \caption{Transition regime $\big(iii\big)$, $F\big(w,y\big)$ for Einasto profile ($\alpha=0.16$). Cyan: $y=y_{\mathrm{crit}}=1.0161268$; Red: $y=1.02$; Orange: $y=1.025$; Green: $y=1.03$; Blue: $y=1.035$; Purple: $y=1.04$.}
        \label{EinastoNatureFtrans3}
    \end{figure}

$\big(iii\big)$: After crossing the critical case, the number of images decreases from three to one. For both $\big|F\big|$ and $\mathrm{Arg}(F)$, as $y$ increases further, the envelopes begin to contract, and the contraction rate accelerating. Throughout this process, the upper and lower envelopes remain axisymmetric (see Fig.~\ref{EinastoNatureFtrans3}).

\begin{figure}[!htb]
        \centering
        \includegraphics[width=0.45\textwidth]{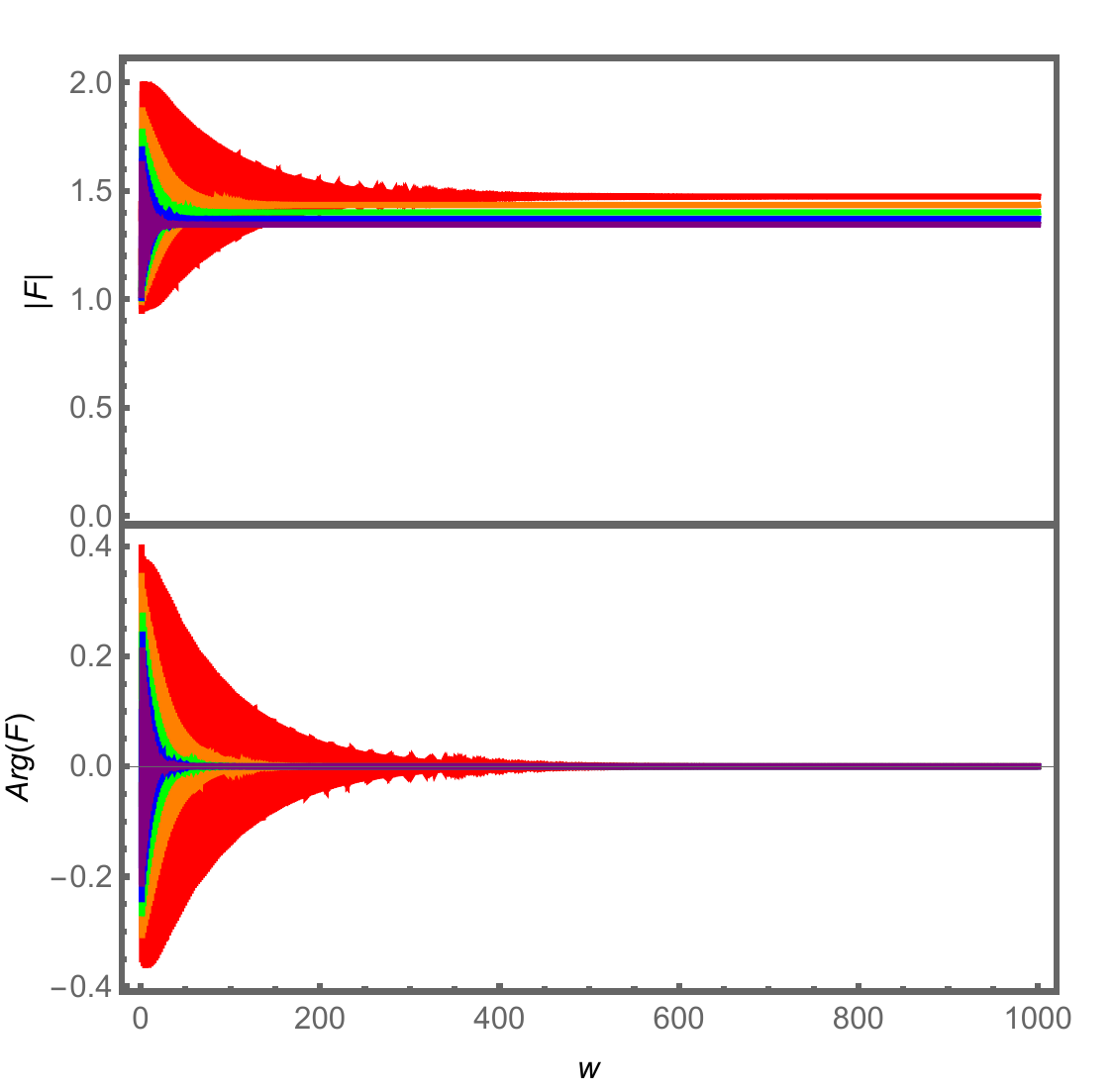}
        \caption{Single-image regime, $F\big(w,y\big)$ for Einasto profile ($\alpha=0.16$). Red: $y=1.1$; Orange: $y=1.2$; Green: $y=1.3$; Blue: $y=1.4$; Purple: $y=1.5$.}
        \label{EinastoNatureF1image}
    \end{figure}

{\it Single-image regime.---} $y>1.05y_{\mathrm{crit}}$, corresponding to region B in Fig.~\ref{ynEinasto}. As $w$ increases, both $\big|F\big|$ and $\mathrm{Arg}(F)$ rapidly converge and ultimately exhibit small amplitude oscillations around a constant value (see Fig.~\ref{EinastoNatureF1image}), consistent with the evolution in the single-image regime of gNFW profile.

Through a comprehensive comparative study, we find that while the evolution trends of $F\big(w,y\big)$ with the two profiles are roughly similar, but the detailed evolutionary processes differ significantly. These differences include the amplitude and oscillation frequency of $F\big(w,y\big)$, the oscillation frequency of wave packets during the multiple-image regime, and the morphological evolution of $F\big(w,y\big)$ during the transition regime. In addition, both the amplitude and the mean value of $\big|F\big|$ with the Einasto profile ($y<0.6$ and $\alpha=0.16$) are generally larger than those with the gNFW profile (for any $y$ and $0<\gamma<2$).

\section{Conclusion}\label{Sec:Conclusion}
With the rapid advancement of GW astronomy, wave-optics effects in GW lensing have become increasingly important. Unlike conventional optical lensing, GWs with extremely low frequencies have wavelengths sufficiently longer to produce prominent diffraction effects, making a key distinction from the conventional optical lensing. In this paper, we examine the gravitational lensing properties produced by the gNFW profile and the Einasto profile, and determine the frequency range of gravitational waves where diffraction effects become relevant. To ensure that our models capture realistic strong lensing configurations, we impose the condition of the dimensionless surface density parameter $\kappa_{\mathrm{s}}>0.1$. Guided by this constraint, we fix the lens configuration as $\big(z_{\mathrm{l}}=0.5, z_{\mathrm{s}}=1.5, M_{200}=2.86\times 10^{15}{M_{\odot}}\big)$.

Starting from the dimensionless time delay function $T$, we derive the imaging equation Eq.~(\ref{lightfunction}) from Fermat’s principle and examine two types of critical condition, i.e. tangential and radial critical condition. The former produces characteristic Einstein rings ($y=0$) and the latter governs image multiplicity and properties. The density maps of Eq.~(\ref{lightfunction}) (Fig.~\ref{imagine}, Fig.~\ref{imagineEinasto}) and Eq.~(\ref{miu}) (Fig.~\ref{magnification}, Fig.~\ref{magnificationEinasto}) show that these critical cases partition the phase diagram into three regions (Fig.~\ref{yGamma}, Fig.~\ref{ynEinasto}), each associated with distinct imaging behavior (characterized by the Morse index $n_{\mathrm{M}}$) and substantially different magnifications $\mu$. This distinction is important for computing $F_{\mathrm{geo}}$ and magnification correction $\mathrm{d}F_\mathrm{m}$ (see Appendix \ref{App:Q-Geo}). In particular, we perform a detailed analysis of the evolution of the amplification factor $F\big(w,y\big)$ (the modulus $\big|F\big|$ and the phase $\mathrm{Arg}(F)$) in the axisymmetric case. For both profiles, the influence of lens mass is identical: a more massive lens produces a stronger gravitational effect, with larger amplitude and frequency of $F\big(w,y\big)$.

For the gNFW profile, its evolutionary process is relatively simple, and throughout the entire evolutionary process, the upper and lower envelopes of $\big|F\big|$ and $\mathrm{Arg}(F)$ maintain symmetry. We attribute this to the relatively small value of $y_{\mathrm{crit}}$, which causes the oscillation frequency of $F\big(w,y\big)$ to remain consistently low throughout the entire evolutionary process. This directly results in an extremely blurred envelope of $F\big(w,y\big)$, rendering it incapable of displaying meaningful information. With increasing inner slope $\gamma$, oscillation amplitude and frequency of $\big|F\big|$ and $\arg\big(F\big)$ is greater.

For the Einasto profile, the evolution of $F\big(w,y\big)$ is considerably more complex, exhibiting rich and varied evolutionary characteristics in both $\big|F\big|$ and $\mathrm{Arg}(F)$. We show the evolution across three regimes: In multiple-image regime (Fig.~\ref{EinastoNatureF3image}), both $\big|F\big|$ and $\mathrm{Arg}(F)$ exhibit stable oscillations, with the upper and lower envelopes maintaining axial symmetry, and the oscillation frequency of the wave packet decreases as $y$ increases; in transition regime (Figs.~\ref{EinastoNatureFtransi}, \ref{EinastoNatureFtransii}, and \ref{EinastoNatureFtrans3}), $F\big(w,y\big)$ exhibits significant morphological evolution. The wave packet of $\big|F\big|$ is continuously stretched, while the symmetry of its upper and lower envelopes is broken, an arched structure emerges in the lower envelope. Meanwhile, the ``fin" structure appears at the widest part of $\mathrm{Arg}(F)$ wave packet, with a width equal to that of the arched portion of $\big|F\big|$ lower envelope. Subsequently, as $y\to y_{\mathrm{crit}}$, the upper and lower envelopes of both $\big|F\big|$ and $\mathrm{Arg}(F)$ gradually restore their symmetry, and the ``fin" structure of $\mathrm{Arg}(F)$ gradually disappears. When $y>y_{\mathrm{crit}}$, the envelopes of $\big|F\big|$ and $\mathrm{Arg}(F)$ contract rapidly while maintaining symmetry, and simultaneously, the number of images changes from three to one; Single-image regime (Fig.~\ref{EinastoNatureF1image}), as $w$ and $y$ increases, $\big|F\big|$ and $\mathrm{Arg}(F)$ converges rapidly and approach a nearly constant value, accompanied only by minor oscillations.

Finally, our comparison reveals that although gNFW and Einasto profiles are important theoretical models describing DM density distribution, the evolution of $F\big(w,y\big)$ with the two density profiles is almost entirely distinct, with numerous differences in detail. Both the amplitude and the mean value of $\big|F\big|$ with the Einasto profile ($y<0.6$ and $\alpha=0.16$) are generally larger than those with the gNFW profile (for any $y$ and $0<\gamma<2$). This suggests that there are more physical properties of DM yet to be discovered, which warrants further investigation in the future. With the advancement of GW theory research and detection technology, we hope to improve the localization accuracy of GWs in the future and observe lensed GW events.

\acknowledgments 
This work is partly supported by the National Natural Science Foundation of China (under grant Nos. 12503001, 12273050, 12375059, 12533009), the Project of National Astronomical Observatories, Chinese Academy of Sciences (No. E4TG6601), and the Strategic Priority Program of the Chinese Academy of Sciences (grant no. XDB0550300).

\appendix

\section{The Proof of Intersection}
\label{App:Proof}

Proving that the curves $\gamma_{\mathrm{crit}}$ and $y_{\mathrm{crit}}$ in Fig.~\ref{yGamma} intersect at $y=0$ is equivalent to proving that the two critical curves in Fig.~\ref{imagine} intersect at $x=0$. For the tangential critical condition ($y=0$), as $x\to 0_{+}$, Eq.~(\ref{lightfunction}) reduces to
    \begin{equation}
        \begin{aligned}
            \lim_{x\to0_{+}}\frac{\int_{0}^{x}t\hat{\kappa}\big(t,\gamma\big)dt}{x^{2}}&=\lim_{x\to0_{+}}\frac{1}{4\kappa_{\mathrm{s}}\big(\gamma\big)},\\
            \frac{\hat{\kappa}\big(0_{+},\gamma_{1}\big)}{2}&=\frac{1}{4\kappa_{\mathrm{s}}\big(\gamma_{1}\big)}.
        \end{aligned}
        \label{lightfunctionx0}
    \end{equation}
It should be noted that $\gamma$ in the above equation is not an independent variable, it depends on $x$ and the limit is $\lim_{x\to0_{+}}\gamma=\gamma_{1}$. Similarly, as $x\to0_{+}$, the radial critical condition Eq.~(\ref{radial}) reduces to
    \begin{equation}
        \begin{aligned}
            \lim_{x\to0_{+}}\Big[\hat{\kappa}\big(x,\gamma\big)-\frac{\int_{0}^{x}t\hat{\kappa}\big(t,\gamma\big)dt}{x^{2}}\Big]&=\lim_{x\to0_{+}}\frac{1}{4\kappa_{\mathrm{s}}\big(\gamma\big)},\\
            \frac{\hat{\kappa}\big(0_{+},\gamma_{2}\big)}{2}&=\frac{1}{4\kappa_{\mathrm{s}}\big(\gamma_{2}\big)},
        \end{aligned}
        \label{radialx0}
    \end{equation}
$\gamma$ satisfies Eq.~(\ref{radial}) and $\lim_{x\to0_{+}}\gamma=\gamma_{2}$.

By comparing Eq.~(\ref{lightfunctionx0}) and Eq.~(\ref{radialx0}), we observe that the equations for $\gamma$ in the two critical cases are identical in the limit $x\to0_{+}$. This shows that the two critical curves in Fig.~\ref{imagine} intersect at $x=0$.

Therefore, the proposition is proved.

\section{Numerical Methods for Evaluating $F$}\label{App:Numerical Methods}

For both models adopted in this work, Eq.~(\ref{Fone}) does not have a simple analytical expression and must be evaluated numerically. Because the integral in Eq.~(\ref{Fone}) contains a highly oscillatory kernel, here we take the gNFW profile as an example and present three efficient numerical methods for evaluating $F$.

\begin{figure}[!htb]
        \centering
        \includegraphics[width=0.40\textwidth]{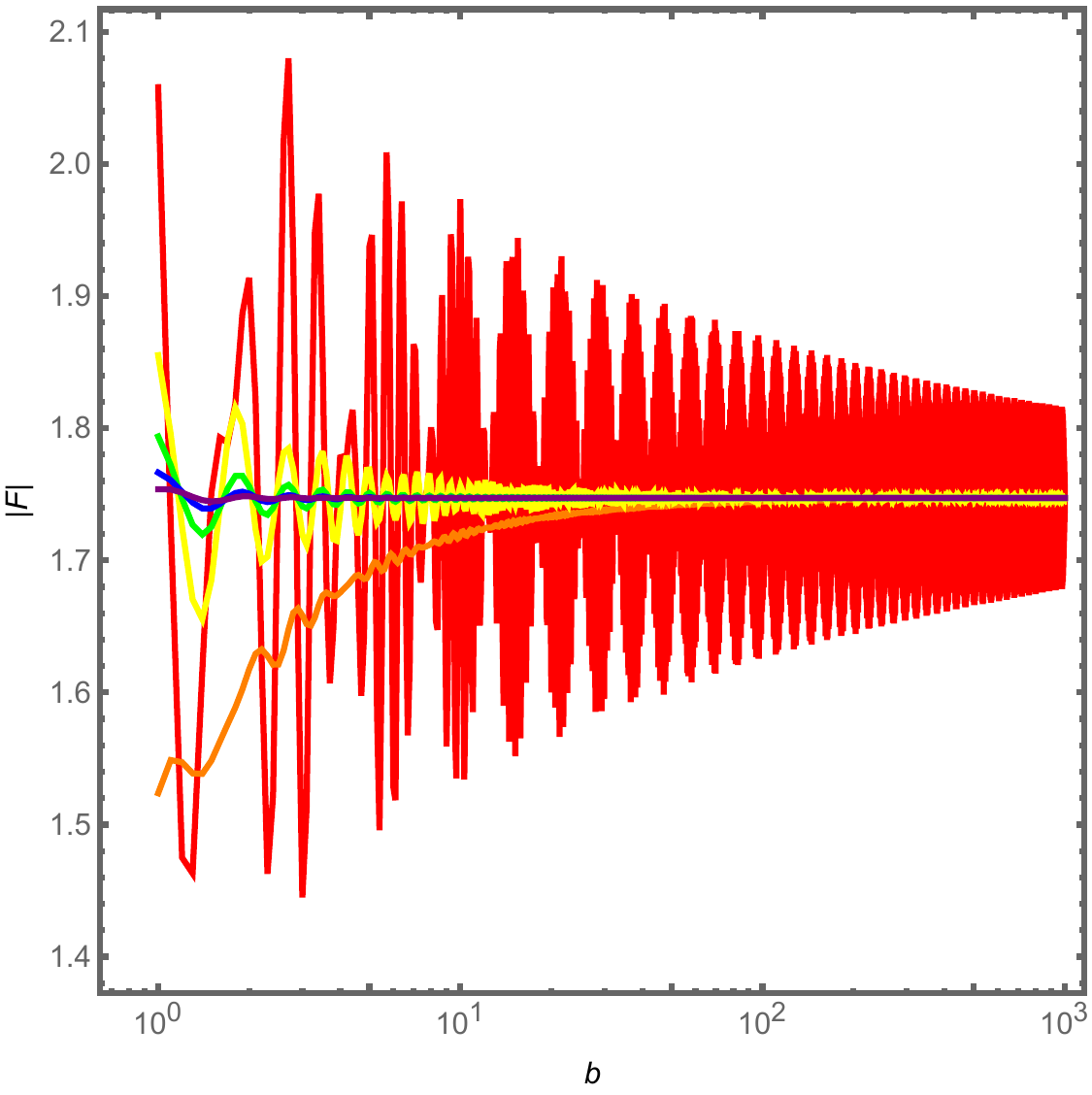}
        \caption{The amplification factor $\big|F\big|$ as a function of the upper limit $b$ of the Eq.~(\ref{Fone}) with $w=10, y=0.3, z_{\mathrm{l}}=0.5, z_{\mathrm{s}}=1.5, \mathrm{log}\big(M_{200}/M_{\mathrm{pivot}}\big)=3$. The red curve is plotted using Levin’s method; the orange curve is plotted using integral mean method; the yellow, green, blue, purple curves are plotted using asymptotic expansion method, with $n_{\mathrm{max}}=1,2,3,4$.}
        \label{Fb}
    \end{figure}

{\it Levin’s method}. By making the transform $z=x^{2}/2$ into the integral of Eq.~(\ref{Fone}), we obtain 
    \begin{equation}
        \begin{aligned}
            f\big(z\big)&=J_{0}\big(wy\sqrt{2z}\big)e^{iw\big(z-\psi(\sqrt{2z})\big)}.
        \end{aligned}
        \label{fz}
    \end{equation}
For integrals with highly oscillatory kernel we adopt the Levin’s method \cite{levin1982procedures}, which is highly efficient for this class of integrals. Fig.~\ref{Fb} shows $F\big(w,y\big)$ with the approximation $\int_{0}^{b}f\big(z\big)dz$ obtained with the Levin’s method (red curve). As the upper integration limit $b$ increases, the numerical result forms a rapidly oscillating wave packet and converges slowly.

{\it Integral mean method}. According to \cite{10.1103/PhysRevD.102.124076}, if the integrand satisfies the Ces$\grave{a}$ro summability condition (hereafter referred to as $\big(C,1\big)$ summable), the integral mean method can be used for the calculation. In this work, Eq.~(\ref{Fone}) represents a rapidly oscillatory improper integral that satisfies the $\big(C,1\big)$ summability condition and can therefore be computed using this method, i.e.
    \begin{equation}
        \begin{aligned}
            &\int_{0}^{\infty}f\big(z\big)dz=\lim_{b\to\infty}I_{C}\big(b\big),\\
            &~~I_{C}\big(b\big)\equiv \int_{0}^{b}f\big(z\big)\Big(1-\frac{z}{b}\Big)dz.
        \end{aligned}
        \label{imm}
    \end{equation}
Compared with direct computation using the Levin’s method, the integral mean method yields results with significantly reduced oscillations and achieves rapid convergence to a constant for sufficiently large $b$ (see Fig.~\ref{Fb} orange curve), demonstrating a marked improvement in performance. Hence, they may be used as an approximation to the Eq.~(\ref{Fone}). For the underlying mathematical principles and detailed explanations, please refer to Appendix B of \cite{10.1103/PhysRevD.102.124076}.

{\it Asymptotic expansion method}. According to \cite{10.1051/0004-6361:20040212,0521431085}, the integral of Eq.~(\ref{fz}) can be regarded as a Fourier integral. The numerical evaluation of such an integral allows the original expression to be rewritten as
    \begin{equation}
        \begin{aligned}
            &\int_{0}^{\infty}g\big(z\big)e^{iwz}dz=\int_{0}^{b}g\big(z\big)e^{iwz}dz+\int_{b}^{\infty}g\big(z\big)e^{iwz}dz,\\
            &=\int_{0}^{b}g\big(z\big)e^{iwz}dz+e^{iwb}\sum_{n=1}^{\infty}\frac{\big(-1\big)^{n}}{\big(iw\big)^{n}}\frac{\partial^{n-1}g(z)}{\partial z^{n-1}}\Big|_{z=b},
        \end{aligned}
        \label{aem}
    \end{equation}
where
    \begin{equation}
        \begin{aligned}
            g\big(z\big)&=J_{0}\big(wy\sqrt{2z}\big)e^{-iw\psi\big(\sqrt{2z}\big)},
        \end{aligned}
        \label{gz}
    \end{equation}
as $z\to\infty$, $g\big(z\big)\to0$. Therefore, an upper integration limit $b$ can be introduced to truncate the integral, while the remaining small contribution can be treated using an asymptotic expansion. 

When the upper integration limit $b$ is sufficiently large, Eq.~(\ref{aem}) converges to a constant. The remaining term beyond this limit is infinitesimal, so considering practical constraints, we introduce an upper cutoff $n_{\mathrm{max}}$ and neglect higher order terms, thus we have
    \begin{equation}
        \begin{aligned}
            &I_{A.E.}\big(b,n_{\mathrm{max}}\big)=\\
            &\int_{0}^{b}g\big(z\big)e^{iwz}dz+e^{iwb}\sum_{n=1}^{n_{\mathrm{max}}}\frac{\big(-1\big)^{n}}{\big(iw\big)^{n}}\frac{\partial^{n-1}g(z)}{\partial z^{n-1}}\Big|_{z=b}.
        \end{aligned}
        \label{aemn}
    \end{equation}
For different upper cutoff $n_{\mathrm{max}}$, the corresponding results of Eq.~(\ref{Fone}) are shown in Fig.~\ref{Fb} as the yellow, green, blue and purple curves. It is evident that the asymptotic expansion method provides more stable numerical results. For the same upper integration limit $b$, the amplification factor $F\big(b\big)$ rapidly converges to a constant as the upper cutoff $n_{\mathrm{max}}$ increases. Even for $b<10$, this method yields highly accurate estimates, while the integral mean method still exhibits noticeable fluctuations in this range and shows no clear advantage.

\section{$F\big(w,y\big)$ in the Transition Regime}\label{App:FTrans}

\begin{figure*}
        \centering 
        \subfigure[$y=0.95$]{
        \label{EinastoNatureFtrans1sub1}
        \includegraphics[width=0.35\textwidth]{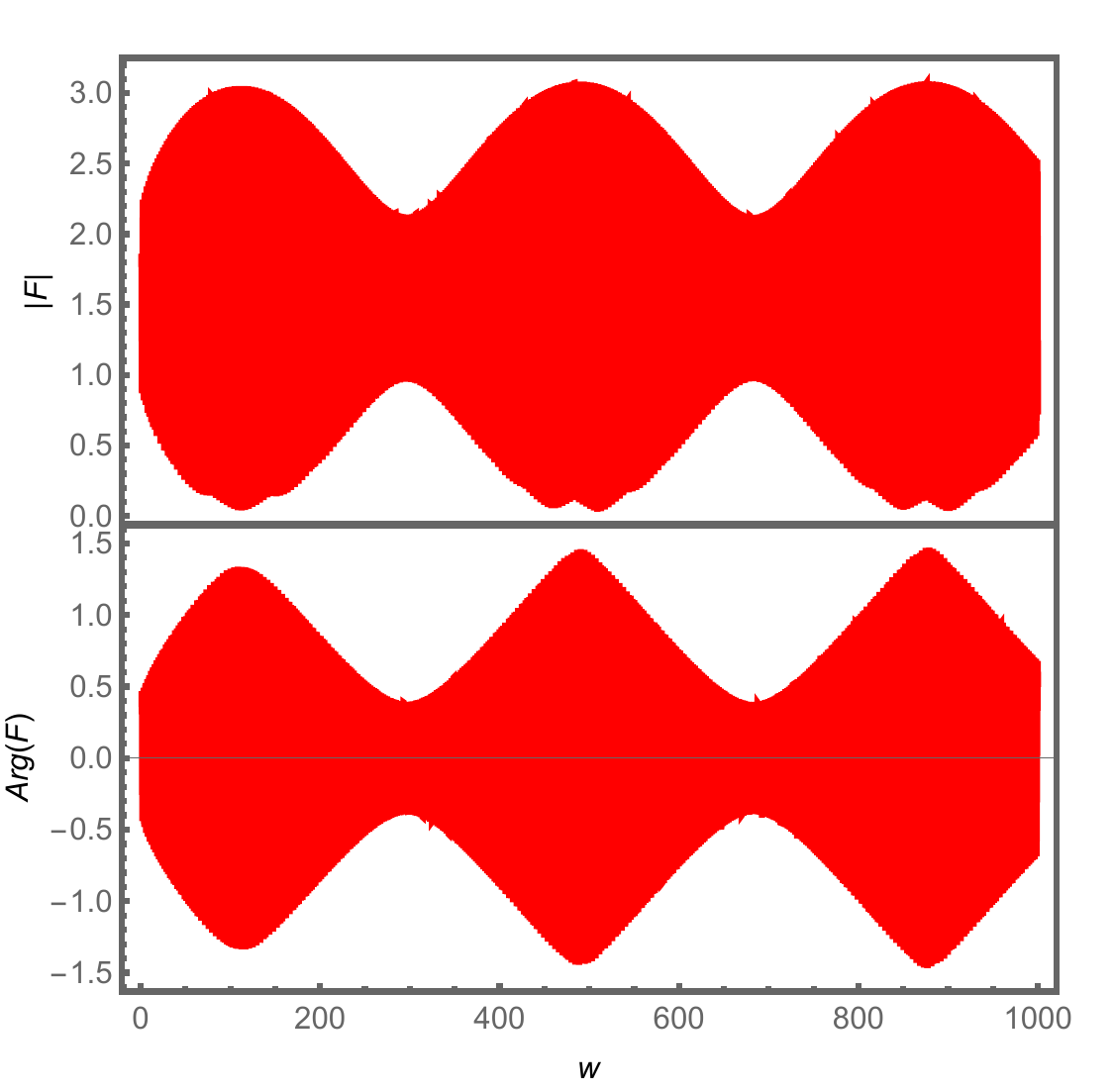}}
        \subfigure[$y=0.97$]{
        \label{EinastoNatureFtrans1sub2}
        \includegraphics[width=0.35\textwidth]{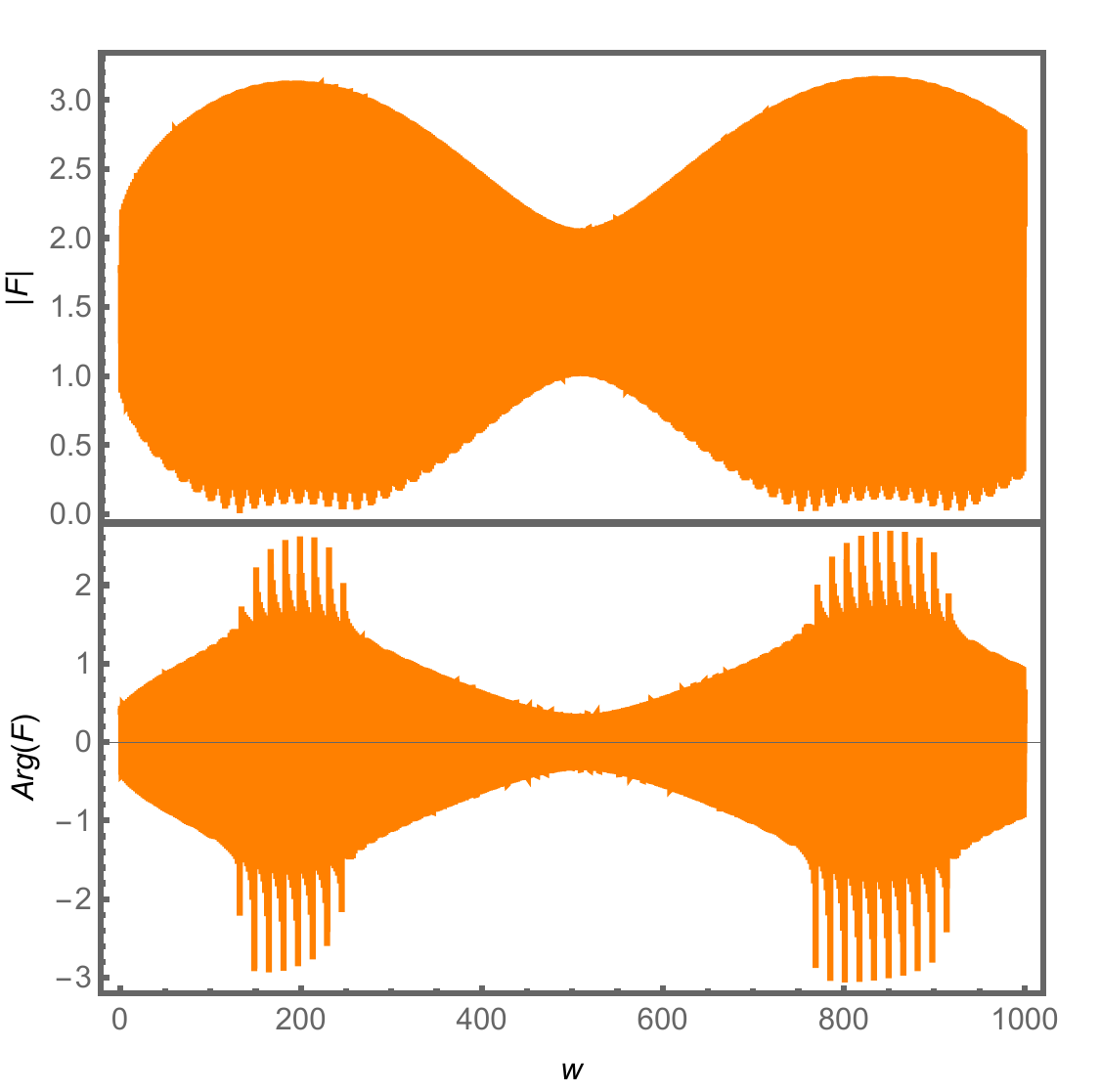}}
        \subfigure[$y=0.98$]{
        \label{EinastoNatureFtrans1sub3}
        \includegraphics[width=0.35\textwidth]{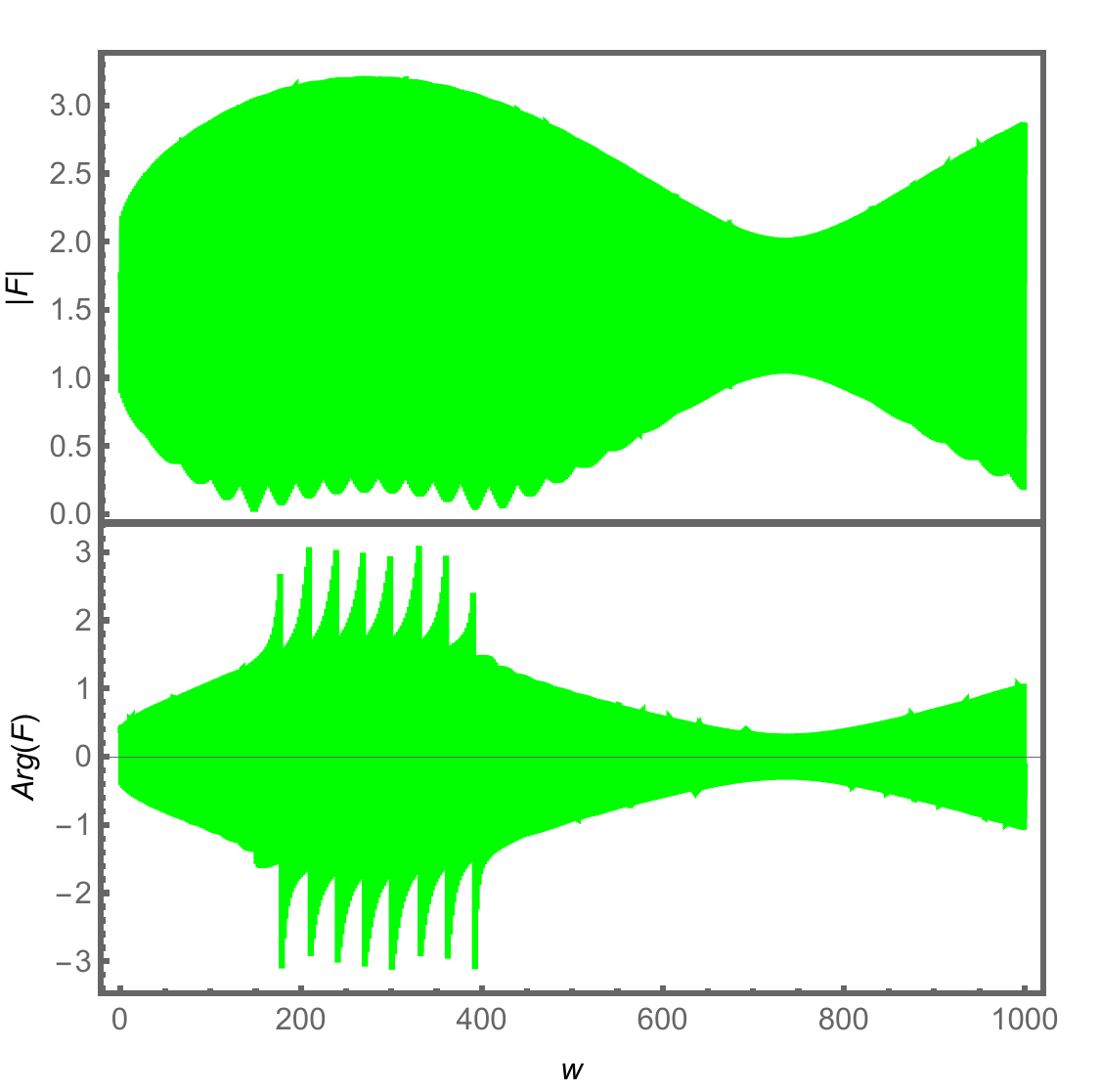}}
        \subfigure[$y=0.99$]{
        \label{EinastoNatureFtrans1sub4}
        \includegraphics[width=0.35\textwidth]{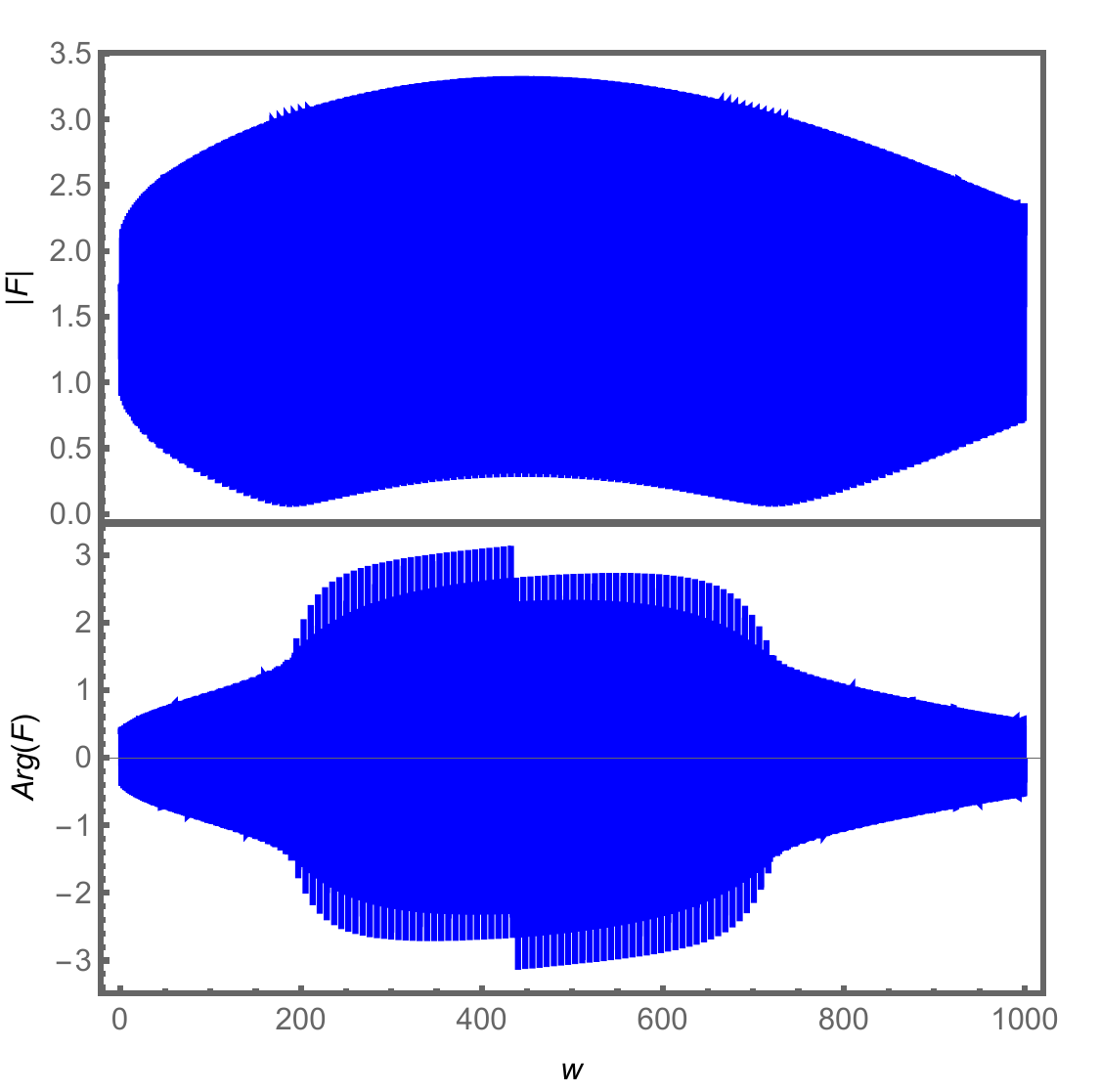}}
        \subfigure[$y=1$]{
        \label{EinastoNatureFtrans1sub5}
        \includegraphics[width=0.35\textwidth]{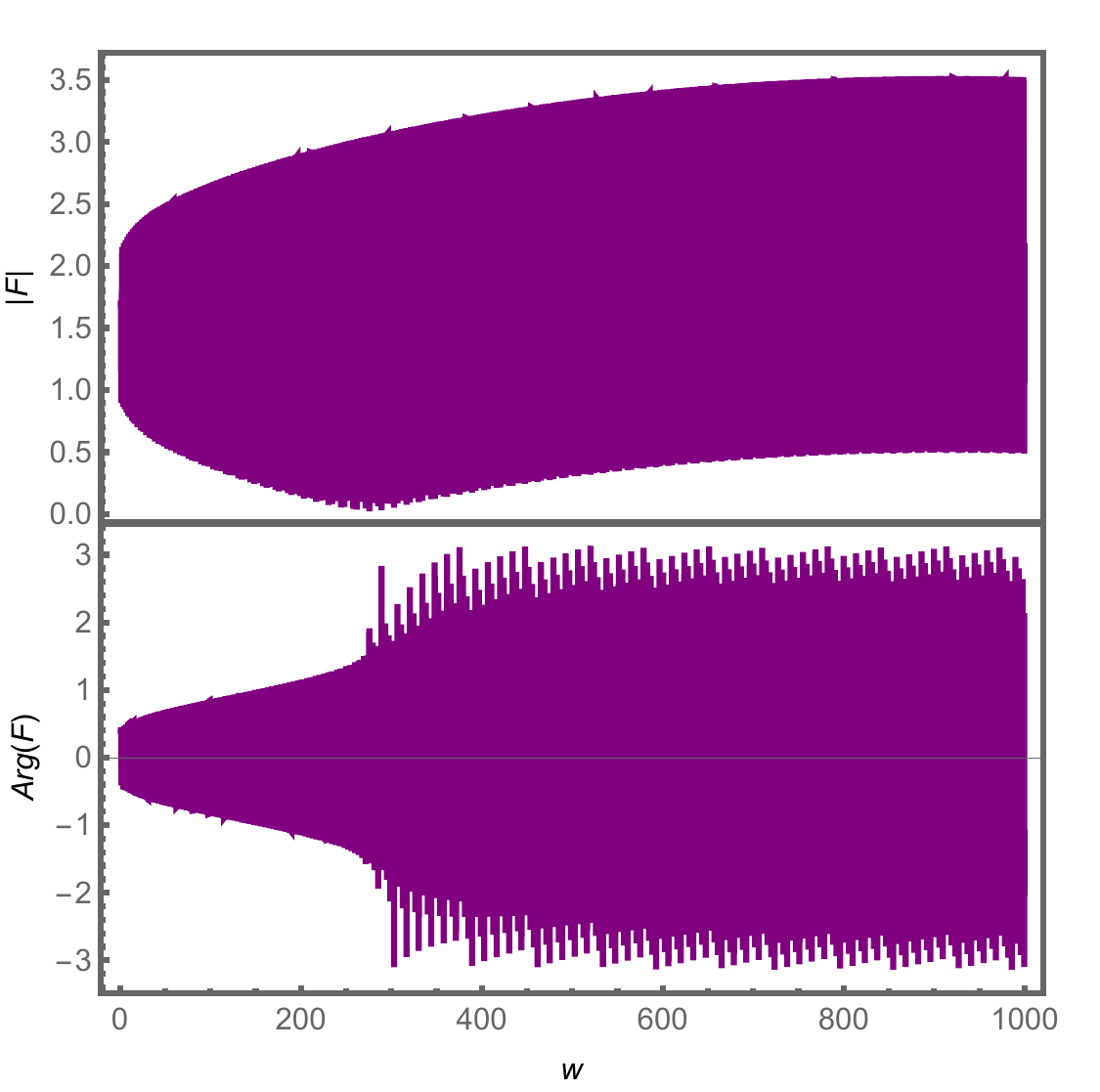}}
        \caption{Transition regime $\big(i\big)$, $F\big(w,y\big)$ for Einasto profile ($\alpha=0.16$). (a) Red: $y=0.95$; (b) Orange: $y=0.97$; (c) Green: $y=0.98$; (d) Blue: $y=0.99$; (e) Purple: $y=1$.
        }
 \label{EinastoNatureFtransi}
 
    \end{figure*}

\begin{figure*}
        \centering 
        \subfigure[$y=1.003$]{
        \label{EinastoNatureFtrans2sub1}
        \includegraphics[width=0.35\textwidth]{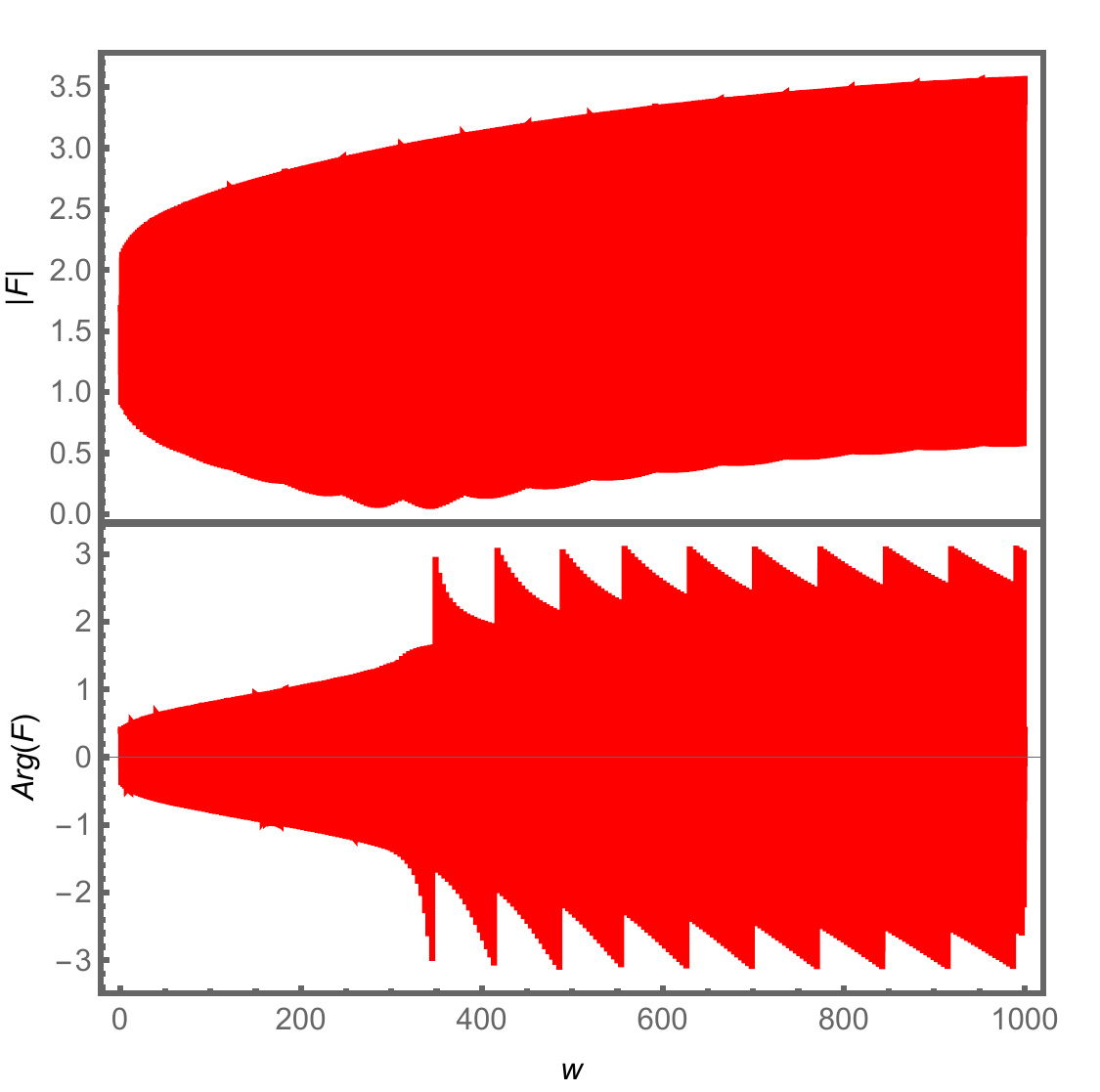}}
        \subfigure[$y=1.005$]{
        \label{EinastoNatureFtrans2sub2}
        \includegraphics[width=0.35\textwidth]{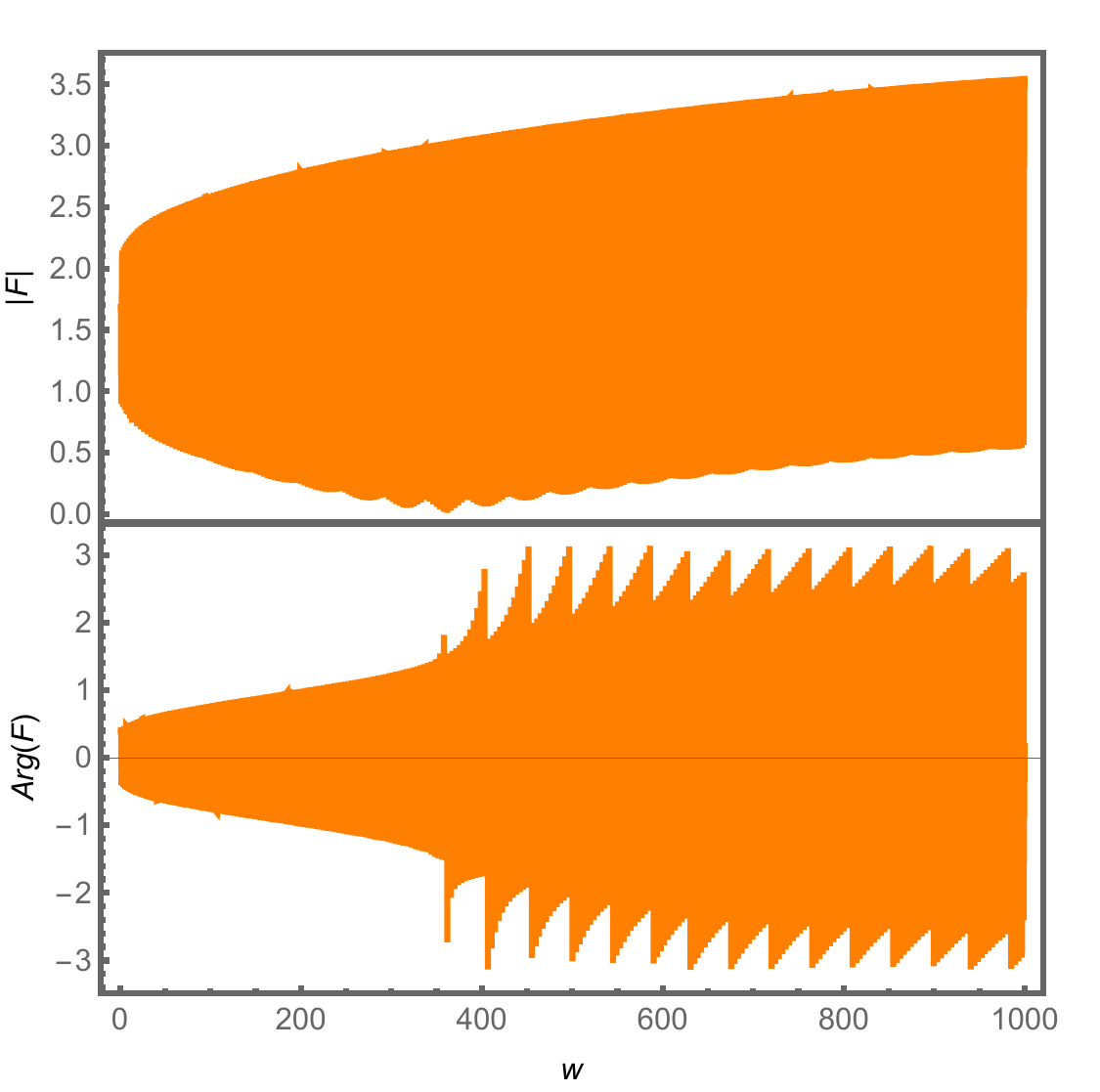}}
        \subfigure[$y=1.008$]{
        \label{EinastoNatureFtrans2sub3}
        \includegraphics[width=0.35\textwidth]{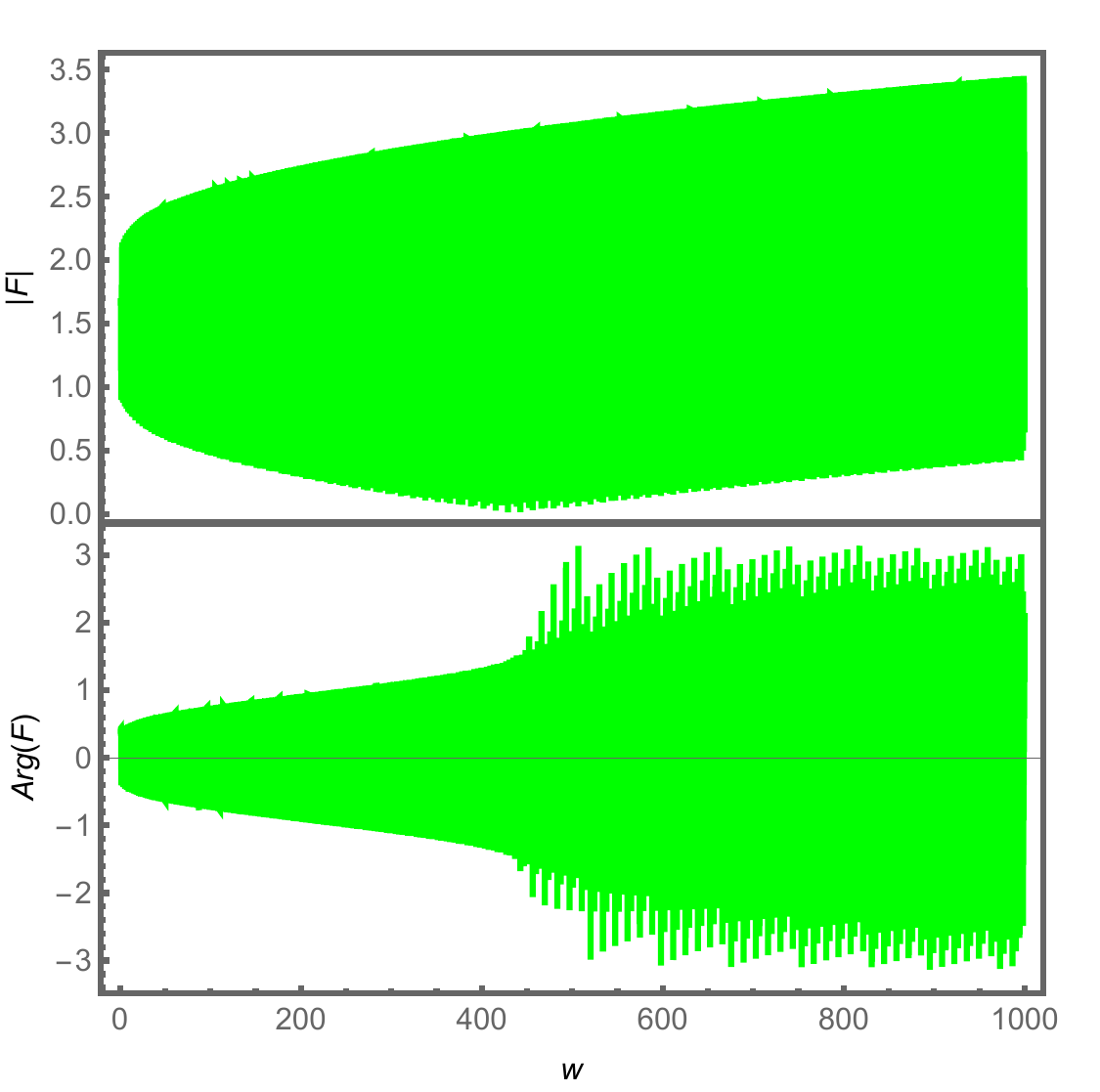}}
        \subfigure[$y=1.01$]{
        \label{EinastoNatureFtrans2sub4}
        \includegraphics[width=0.35\textwidth]{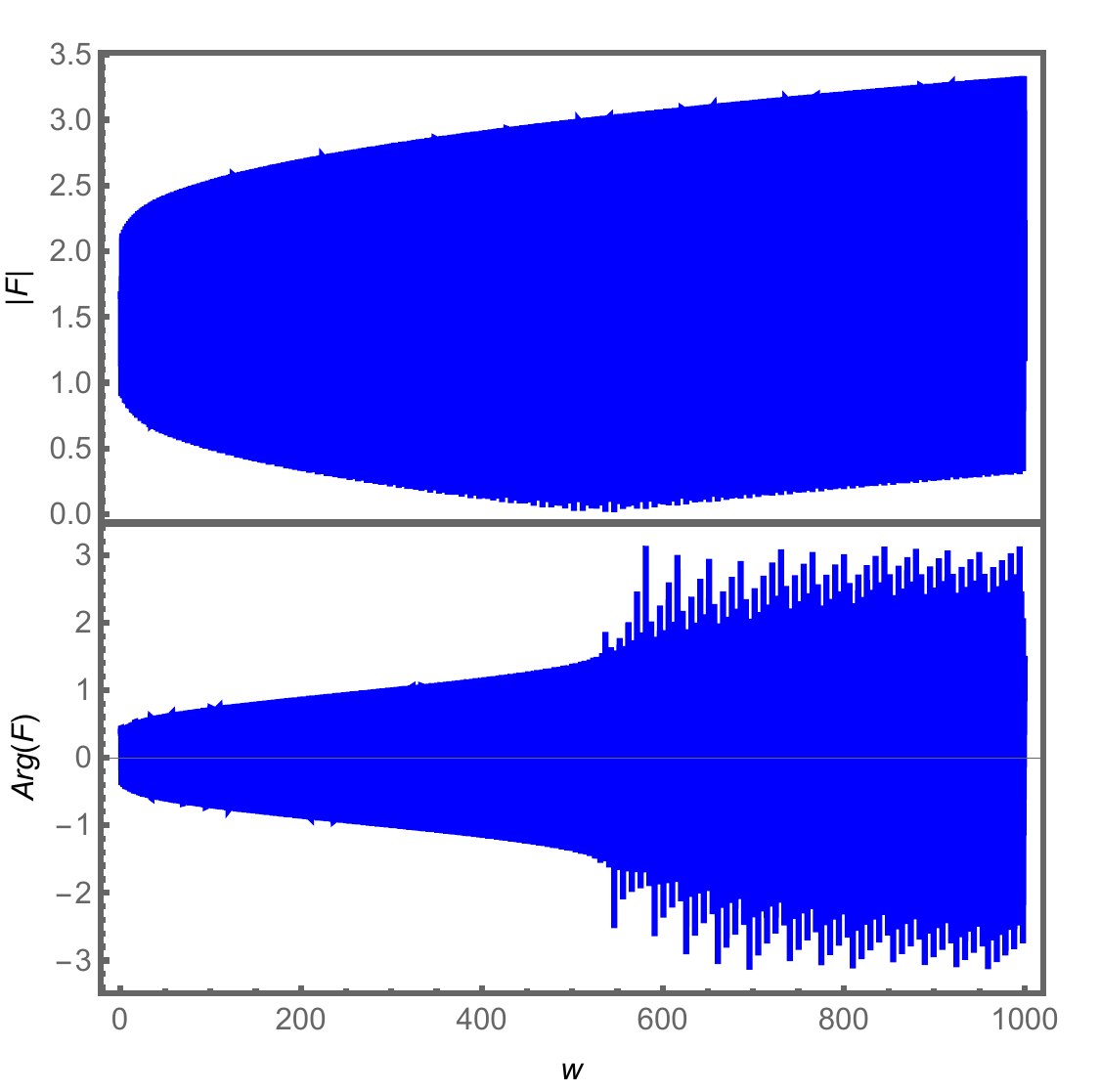}}
        \subfigure[$y=1.015$]{
        \label{EinastoNatureFtrans2sub5}
        \includegraphics[width=0.35\textwidth]{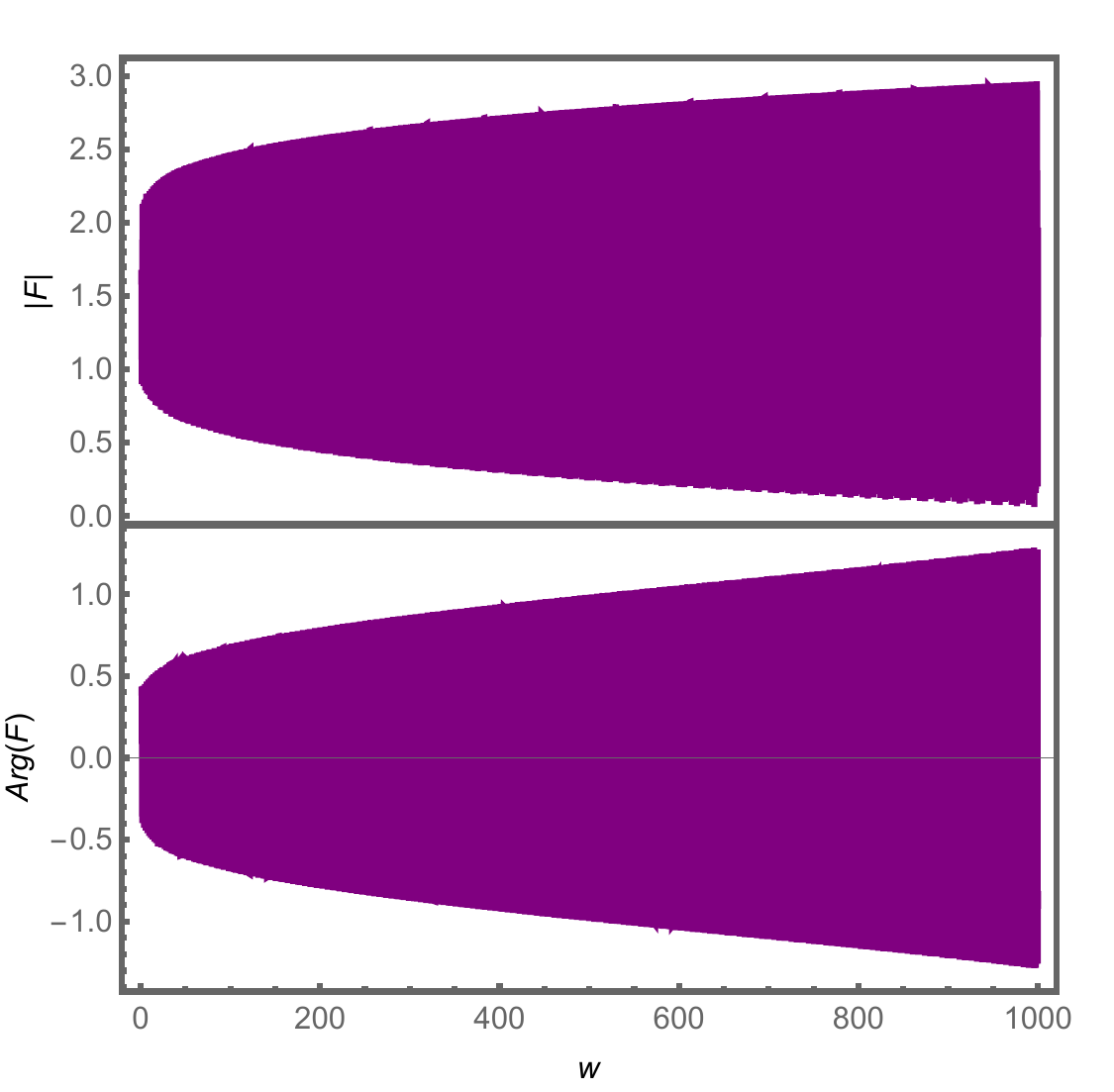}}
        \subfigure[$y=y_{\mathrm{crit}}=1.0161268$]{
        \label{EinastoNatureFtrans2sub1}
        \includegraphics[width=0.35\textwidth]{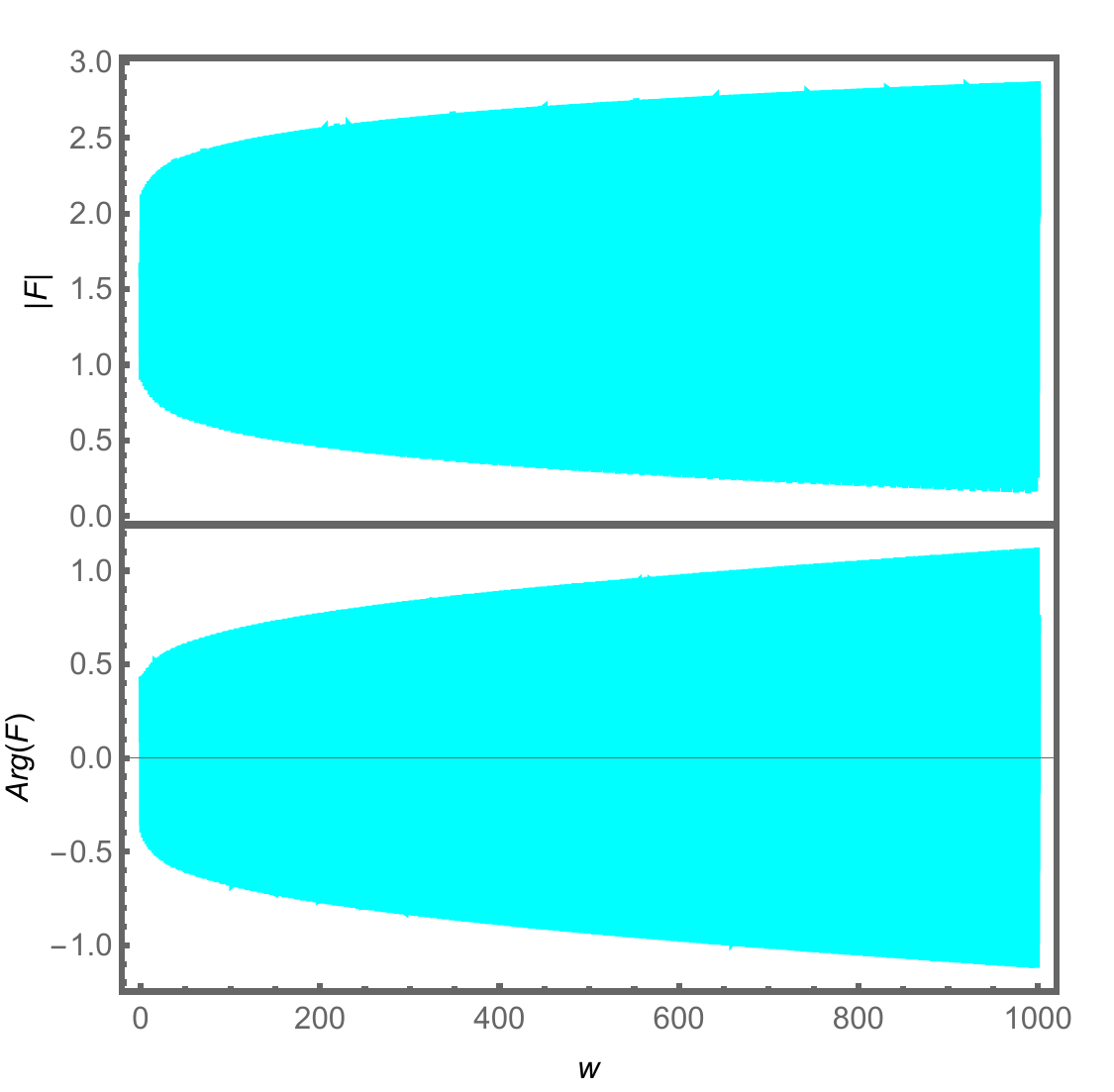}}
        \caption{Transition regime $\big(ii\big)$, $F\big(w,y\big)$ for Einasto profile ($\alpha=0.16$). (a) Red: $y=1.003$; (b) Orange: $y=1.005$; (c) Green: $y=1.008$; (d) Blue: $y=1.01$; (e) Purple: $y=1.015$; (f) Cyan: $y=y_{\mathrm{crit}}=1.0161268$.}
        \label{EinastoNatureFtransii}
    \end{figure*}

For the Einasto profile ($\alpha=0.16$), the evolution of $F\big(w,y\big)$ during the transition regime are shown in Fig.~\ref{EinastoNatureFtransi} and Fig.~\ref{EinastoNatureFtransii}.

\section{(Quasi-)Geometrical Optics Approximation}
\label{App:Q-Geo}

In this section, taking the gNFW profile as an example, we investigate the differences on the amplification factor $F$ with the geometrical and quasi-geometrical optics approximation, and identify the frequency range in which wave effect corrections must be taken into account.

{\it Geometrical optics approximation}. In the geometrical optics limit, the amplification factor $F$ depends only on the properties of the image. Therefore, by solving Eq.~(\ref{lightfunction}) derived from Fermat’s principle and Eq.~(\ref{miu}) to obtain the position $x_{j}$, the magnification $\mu_{j}$ and the Morse index $n_{j}$ of the image, we can compute the $F_{\mathrm{geo}}$ in geometrical optical limit, where the index $j$ runs over all images \cite{10.1143/ptps.133.137,10.1007/978-3-662-03758-4},
    \begin{equation}
        \begin{aligned}
            F_{\mathrm{geo}}\big(w,y\big)=\sum_{j}\left|\mu_{j}\right|^{\frac{1}{2}}e^{iwT_{j}-i\pi n_{j}/2}.
        \end{aligned}
        \label{Fgeo}
    \end{equation}
    
{\it Correction to the image magnification}. According to \cite{10.1051/0004-6361:20040212}, the quasi-geometrical optics approximation differs from the geometrical optics approximation in that the time delay function $T(\vec{x},\vec{y})$ must be expanded to higher orders to account for diffraction effects. Consequently, an additional correction term $\mathrm{d}F_{\mathrm{m}}$ is introduced,
    \begin{equation}
        \begin{aligned}
            \mathrm{d}F_{\mathrm{m}}\equiv\frac{i}{w}\sum_{j}\left|\mu_{j}\right|^{\frac{1}{2}}\Delta_{j}e^{iwT_{j}-i\pi n_{j}/2}.
        \end{aligned}
        \label{dFm}
    \end{equation}
Here, the diffraction effect is contained in $\Delta_{j}$,
    \begin{equation}
        \begin{aligned}
\Delta_{j}=&\frac{1}{16}\Bigg[\frac{1}{2\alpha_{j}^{2}}\psi_{j}^{(4)}+\frac{5}{12\alpha_{j}^{3}}\Big(\psi_{j}^{(3)}\Big)^{2}\\
&\quad+\frac{1}{\alpha_{j}^{2}}\frac{\psi_{j}^{(3)}}{\left|x_{j}\right|}+\frac{\alpha_{j}-\beta_{j}}{\alpha_{j}\beta_{j}}\frac{1}{\left|x_{j}\right|^{2}}\Bigg],
        \end{aligned}
        \label{delta}
    \end{equation}
where
    \begin{equation}
        \begin{aligned}
            \alpha_{j}=\frac{1}{2}\Big(1-\psi_{j}^{(2)}\Big),~~
            \beta_{j}=\frac{1}{2}\Big(1-\frac{\psi_{j}^{(1)}}{\left|x_{j}\right|}\Big),
        \end{aligned}
        \label{alphabeta}
    \end{equation}
and $\psi_{j}^{(n)}\equiv \mathrm{d}^{n}\psi\big(\left|x_{j}\right|\big)/\mathrm{d}x^{n}$ represents the $n$th-order derivative of the lensing potential $\psi$ evaluated at the point of the image $x_{j}$. When $w\gg1$, the correction $\mathrm{d}F_{\mathrm{m}}$ becomes negligible and the approximation reduces to the geometrical optics limit. \cite{10.1051/0004-6361:20040212} further notes that the correction terms need only account for the contributions from stationary points (i.e., points satisfying Eq.~(\ref{lightfunction})), while those from non-stationary points (not satisfying Eq.~(\ref{lightfunction})) can be ignored. Hence, with the quasi-geometrical optics approximation, $F\big(w,y\big)$ in Eq.~(\ref{Fone}) is given by
    \begin{equation}
        \begin{aligned}
            F\big(w,y\big)&=F_{\mathrm{geo}}\big(w,y\big)+\mathrm{d}F_{\mathrm{m}}\\
            &=\sum_{j}\left|\mu_{j}\right|^{\frac{1}{2}}\Big(1+\frac{i}{w}\Delta_{j}\Big)e^{iwT_{j}-i\pi n_{j}/2}+\mathcal{O}\big(w^{-2}\big).
        \end{aligned}
        \label{Fqgeo}
    \end{equation}

{\it Correction to the central cusp of the lens}. In addition to the correction $\mathrm{d}F_{\mathrm{m}}$ discussed above, we also consider the correction arising from the central cusp of the lens (hereafter denoted as $\mathrm{d}F_{\mathrm{c}}$). When the source is located close to the center of the lens plane ($y\to0$), the correction $\mathrm{d}F_{\mathrm{c}}$ becomes significant and should therefore be included in the present analysis. The lensing potential $\psi\big(x\big)$ satisfies $\psi\big(x\big)=\psi_{0}x^{3-\gamma}$, by combining Eq.~(\ref{potential}) and Eq.~(\ref{kbargNFW}), we obtain $\psi_{0}$ as $x\to0$, 
    \begin{equation}
        \begin{aligned}
            \psi_{0}=\frac{4\kappa_{\mathrm{s}}}{\big(3-\gamma\big)^{2}}\Big[1+\big(3-\gamma\big)\int_{0}^{1}y^{\gamma-4}\big(1-\sqrt{1-y^{2}}\big)dy\Big].
        \end{aligned}
        \label{psi0}
    \end{equation}
It should be noted that $x$ in this case is independent of the image position $x_{j}$.

Therefore, $\mathrm{d}F_{\mathrm{c}}$ can be expressed as \cite{10.1051/0004-6361:20040212}
    \begin{equation}
        \begin{aligned}
            \mathrm{d}F_{\mathrm{c}}=-\frac{1}{2}\Big(\frac{y}{2}\Big)^{\gamma-5}\psi_{0}w^{\gamma-3}e^{iw\big(y^{2}/2+\phi_{\mathrm{m}}(y)\big)}\frac{\Gamma\big[\big (5-\gamma\big)/2\big]}{\Gamma\big[\big(\gamma-3\big)/2\big]},
        \end{aligned}
        \label{dFc}
    \end{equation}
where $\Gamma\big(x\big)$ is the gamma function. Note that the above expression applies only to cases $0<\gamma<1$, and $1<\gamma<2$. For $\gamma=1$ (the NFW profile), it is not valid, since $\psi\big(x\big)\propto x^{2}\mathrm{ln}x$, this case must be treated separately. Therefore, the total correction is given by the sum of the two terms, $\mathrm{d}F=\mathrm{d}F_{\mathrm{m}}+\mathrm{d}F_{\mathrm{c}}$. 

    \begin{figure}[!htb]
        \centering
        \includegraphics[width=0.45\textwidth]{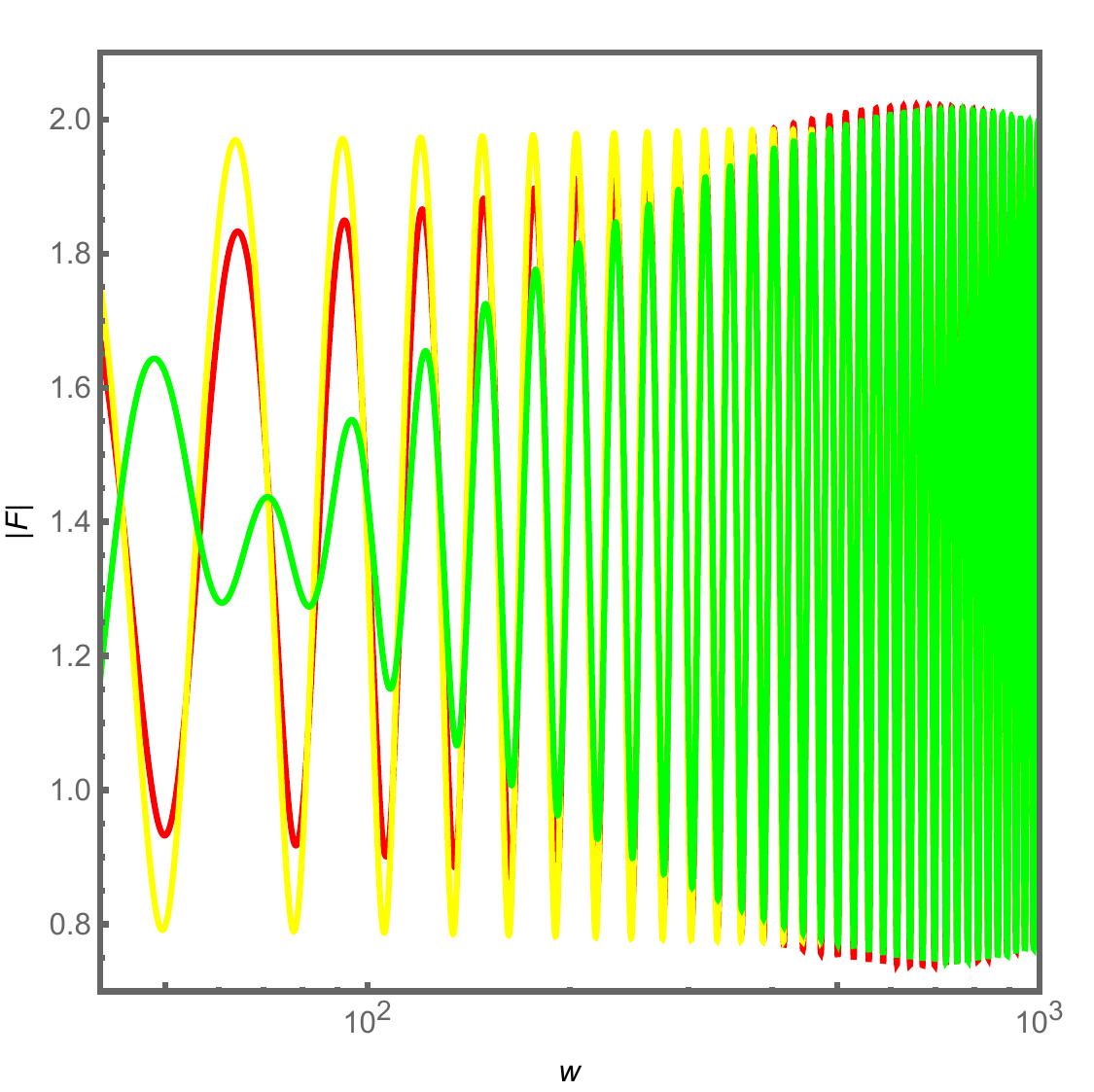}
        \caption{The amplification factor $\big|F\big|$ as a function of $w$ with $y=0.3$, $\gamma=1.89$, $b=100$, $z_{\mathrm{l}}=0.5$, $z_{\mathrm{s}}=1.5$, $\mathrm{log}\big(M_{200}/M_{\mathrm{pivot}}\big)=3$. Red: $\big|F\big|$; Yellow: $\big|F_{\mathrm{geo}}\big|$; Green: $\big|F_{\mathrm{geo}}+\mathrm{d}F\big|$. According to Eq.~(\ref{w}), the GW frequency $f_{\mathrm{obs}}$ scale corresponding to the dimensionless parameter $w$ is shown at the top of the figure. It can be clearly seen that, within the model investigated in this work, diffraction effects become relevant only for extremely low frequency.}
        \label{Fcompare}
    \end{figure}

Finally, for $y=0.3$ and $\gamma=1.89$, $\big|F\big|,\big|F_{\mathrm{geo}}\big|$ and $\big|F_{\mathrm{geo}}+\mathrm{d}F\big|$ as functions of $w$ are shown in Fig.~\ref{Fcompare}. As seen in the figure, in the quasi-geometrical optics regime $w\sim100$, the correction plays a significant role. In the geometrical optics limit $w\sim1000$, the effect of diffraction becomes negligible and all curves gradually converge. 

\begin{figure}[!htb]
        \centering
        \includegraphics[width=0.45\textwidth]{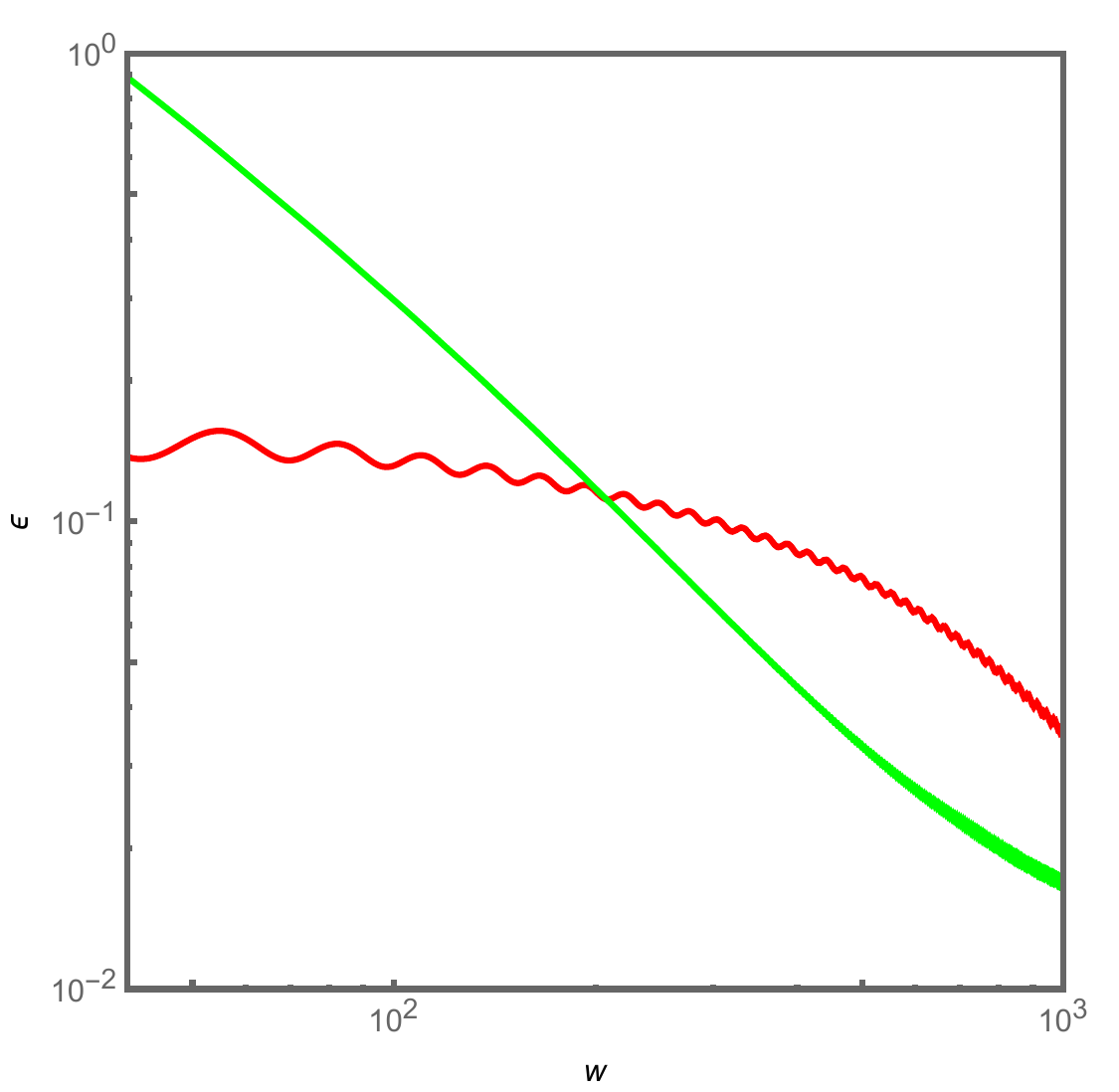}
        \caption{The errors between $\big|F\big|,\big|F_{\mathrm{geo}}\big|$ and $\big|F_{\mathrm{geo}}+\mathrm{d}F\big|$ as a function of $w$ with $y=0.3,\gamma=1.89,b=100,z_{\mathrm{l}}=0.5,z_{\mathrm{s}}=1.5,\mathrm{log}\big(M_{200}/M_{\mathrm{pivot}}\big)=3$. The amplification factor $\big|F\big|$ as a function of $w$ with $y=0.3,\gamma=1.89,b=100,z_{\mathrm{l}}=0.5,z_{\mathrm{s}}=1.5,\mathrm{log}\big(M_{200}/M_{\mathrm{pivot}}\big)=3$. Red curve denotes $\big|F-F_{\mathrm{geo}}\big|$; Green curve denotes $\big|F-\big(F_{\mathrm{geo}}+\mathrm{d}F\big)\big|$.}
        \label{error}
    \end{figure}

The detailed comparison of the modeling errors is presented in Fig.~\ref{error}. As shown in the figure, once the diffraction effects are taken into account, the description of the amplification factor $F$ by $\big|F_{\mathrm{geo}}+\mathrm{d}F\big|$ becomes more accurate and reliable than that provided by the geometrical optics approximation $\big|F_{\mathrm{geo}}\big|$. Meanwhile, for the lens model considered in this work, the observational GW frequency $f_{\mathrm{obs}}$ corresponding to Eq.~(\ref{w}) implies that diffraction effects become relevant only at extremely low frequencies ($f_{\mathrm{obs}}<10^{-9}\mathrm{Hz}$). Considering that the lowest frequency accessible to current detection technology is $\mathcal{O} \big(10^{-9}\big)$Hz, the diffraction effects can be neglected when the lens is solely attributed to a DM halo.

\bibliography{ref}

@article{10.1086/311333,
    author = "Moore, Ben and Governato, Fabio and Quinn, Thomas R. and Stadel, Joachim and Lake, George",
    title = "{Resolving the structure of cold dark matter halos}",
    eprint = "astro-ph/9709051",
    archivePrefix = "arXiv",
    doi = "10.1086/311333",
    journal = "Astrophys. J. Lett.",
    volume = "499",
    pages = "L5",
    year = "1998"
}

@article{10.1086/312287,
    author = "Moore, B. and Ghigna, S. and Governato, F. and Lake, G. and Quinn, Thomas R. and Stadel, J. and Tozzi, P.",
    title = "{Dark matter substructure within galactic halos}",
    eprint = "astro-ph/9907411",
    archivePrefix = "arXiv",
    doi = "10.1086/312287",
    journal = "Astrophys. J. Lett.",
    volume = "524",
    pages = "L19--L22",
    year = "1999"
}

@inproceedings{astro-ph/0001288,
    author = "Jing, Y. P. and Suto, Yasushi",
    title = "{Dark matter halos simulated with million particles}",
    booktitle = "{4th RESCEU International Symposium on Birth and Evolution of the Universe}",
    eprint = "astro-ph/0001288",
    archivePrefix = "arXiv",
    month = "11",
    year = "1999"
}

@article{10.1086/319136,
    author = "Keeton, Charles R. and Madau, Piero",
    title = "{Lensing constraints on the cores of massive dark matter halos}",
    eprint = "astro-ph/0101058",
    archivePrefix = "arXiv",
    doi = "10.1086/319136",
    journal = "Astrophys. J. Lett.",
    volume = "549",
    pages = "L25",
    year = "2001"
}

@article{10.1086/321437,
    author = "Wyithe, J. Stuart B. and Turner, E. L. and Spergel, D. N.",
    title = "{Gravitational lens statistics for generalized NFW profiles: parameter degeneracy and implications for self-interacting cold dark matter}",
    eprint = "astro-ph/0007354",
    archivePrefix = "arXiv",
    doi = "10.1086/321437",
    journal = "Astrophys. J.",
    volume = "555",
    pages = "504",
    year = "2001"
}

@article{10.1086/304888,
   title={A Universal Density Profile from Hierarchical Clustering},
   volume={490},
   ISSN={1538-4357},
   url={http://dx.doi.org/10.1086/304888},
   DOI={10.1086/304888},
   number={2},
   journal={The Astrophysical Journal},
   publisher={American Astronomical Society},
   author={Navarro, Julio F. and Frenk, Carlos S. and White, Simon D. M.},
   year={1997},
   month=dec, pages={493–508}
}

@article{10.1051/0004-6361/201833910,
    author = "Aghanim, N. and others",
    collaboration = "Planck",
    title = "{Planck 2018 results. VI. Cosmological parameters}",
    eprint = "1807.06209",
    archivePrefix = "arXiv",
    primaryClass = "astro-ph.CO",
    doi = "10.1051/0004-6361/201833910",
    journal = "Astron. Astrophys.",
    volume = "641",
    pages = "A6",
    year = "2020",
    note = "[Erratum: Astron.Astrophys. 652, C4 (2021)]"
}

@article{10.1111/j.1745-3933.2008.00537.x,
    author = "Duffy, Alan R. and Schaye, Joop and Kay, Scott T. and Dalla Vecchia, Claudio",
    title = "{Dark matter halo concentrations in the Wilkinson Microwave Anisotropy Probe year 5 cosmology}",
    eprint = "0804.2486",
    archivePrefix = "arXiv",
    primaryClass = "astro-ph",
    doi = "10.1111/j.1745-3933.2008.00537.x",
    journal = "Mon. Not. Roy. Astron. Soc.",
    volume = "390",
    pages = "L64",
    year = "2008",
    note = "[Erratum: Mon.Not.Roy.Astron.Soc. 415, L85 (2011)]"
}

@article{10.1146/annurev-astro-091918-104453,
    author = "Simon, Joshua D.",
    title = "{The Faintest Dwarf Galaxies}",
    eprint = "1901.05465",
    archivePrefix = "arXiv",
    primaryClass = "astro-ph.GA",
    doi = "10.1146/annurev-astro-091918-104453",
    journal = "Ann. Rev. Astron. Astrophys.",
    volume = "57",
    number = "1",
    pages = "375--415",
    year = "2019"
}

@article{10.1111/j.1365-2966.2008.13348.x,
    author = "Baldry, I. K. and Glazebrook, K. and Driver, S. P.",
    title = "{On the galaxy stellar mass function, the mass-metallicity relation, and the implied baryonic mass function}",
    eprint = "0804.2892",
    archivePrefix = "arXiv",
    primaryClass = "astro-ph",
    doi = "10.1111/j.1365-2966.2008.13348.x",
    journal = "Mon. Not. Roy. Astron. Soc.",
    volume = "388",
    pages = "945",
    year = "2008"
}

@article{10.1111/j.1365-2966.2012.21984.x,
    author = "Lidman, C. and others",
    title = "{Evidence for Significant Growth in the Stellar Mass of Brightest Cluster Galaxies over the Past 10 Billion Years}",
    eprint = "1208.5143",
    archivePrefix = "arXiv",
    primaryClass = "astro-ph.CO",
    doi = "10.1111/j.1365-2966.2012.21984.x",
    journal = "Mon. Not. Roy. Astron. Soc.",
    volume = "427",
    pages = "550",
    year = "2012"
}

@ARTICLE{2021Univ....7..139L,
       author = {{Lovisari}, Lorenzo and {Ettori}, Stefano and {Gaspari}, Massimo and {Giles}, Paul A.},
        title = "{Scaling Properties of Galaxy Groups}",
      journal = {Universe},
         year = 2021,
        month = may,
       volume = {7},
       number = {5},
          eid = {139},
        pages = {139},
          doi = {10.3390/universe7050139},
archivePrefix = {arXiv},
       eprint = {2106.13256},
 primaryClass = {astro-ph.CO},
       adsurl = {https://ui.adsabs.harvard.edu/abs/2021Univ....7..139L}
}

@article{10.1146/annurev-astro-081710-102514,
   title={Cosmological Parameters from Observations of Galaxy Clusters},
   volume={49},
   ISSN={1545-4282},
   url={http://dx.doi.org/10.1146/annurev-astro-081710-102514},
   DOI={10.1146/annurev-astro-081710-102514},
   number={1},
   journal={Annual Review of Astronomy and Astrophysics},
   publisher={Annual Reviews},
   author={Allen, Steven W. and Evrard, August E. and Mantz, Adam B.},
   year={2011},
   month=sep, pages={409–470}
}

@article{10.1111/j.1365-2966.2012.21623.x,
    author = "Ferrero, Ismael and Abadi, Mario G. and Navarro, Julio F. and Sales, Laura V. and Gurovich, Sebastian",
    title = "{The dark matter halos of dwarf galaxies: a challenge for the LCDM paradigm?}",
    eprint = "1111.6609",
    archivePrefix = "arXiv",
    primaryClass = "astro-ph.CO",
    doi = "10.1111/j.1365-2966.2012.21623.x",
    journal = "Mon. Not. Roy. Astron. Soc.",
    volume = "425",
    pages = "2817--2823",
    year = "2012"
}

@article{10.1088/0004-637X/770/1/57,
    author = "Behroozi, Peter S. and Wechsler, Risa H. and Conroy, Charlie",
    title = "{The Average Star Formation Histories of Galaxies in Dark Matter Halos from $z=$0-8}",
    eprint = "1207.6105",
    archivePrefix = "arXiv",
    primaryClass = "astro-ph.CO",
    doi = "10.1088/0004-637X/770/1/57",
    journal = "Astrophys. J.",
    volume = "770",
    pages = "57",
    year = "2013"
}

@article{10.1103/RevModPhys.77.207,
    author = "Voit, G. Mark",
    title = "{Tracing cosmic evolution with clusters of galaxies}",
    eprint = "astro-ph/0410173",
    archivePrefix = "arXiv",
    doi = "10.1103/RevModPhys.77.207",
    journal = "Rev. Mod. Phys.",
    volume = "77",
    pages = "207--258",
    year = "2005"
}

@article{astro-ph/9602053,
    author = "Bartelmann, Matthias",
    title = "{Arcs from a universal dark matter halo profile}",
    eprint = "astro-ph/9602053",
    archivePrefix = "arXiv",
    reportNumber = "MPA-923",
    journal = "Astron. Astrophys.",
    volume = "313",
    pages = "697--702",
    year = "1996"
}

@article{10.1046/j.1365-8711.2003.06276.x,
    author = "Meneghetti, Massimo and Bartelmann, Matthias and Moscardini, Lauro",
    title = "{Cluster cross-sections for strong lensing: Analytic and numerical lens models}",
    eprint = "astro-ph/0201501",
    archivePrefix = "arXiv",
    doi = "10.1046/j.1365-8711.2003.06276.x",
    journal = "Mon. Not. Roy. Astron. Soc.",
    volume = "340",
    pages = "105",
    year = "2003"
}

@article{10.1086/323961,
    author = "Takahashi, Ryuichi and Chiba, Takeshi",
    title = "{Gravitational lens statistics and the density profile of dark halos}",
    eprint = "astro-ph/0106176",
    archivePrefix = "arXiv",
    doi = "10.1086/323961",
    journal = "Astrophys. J.",
    volume = "563",
    pages = "489--496",
    year = "2001"
}

@article{10.1111/j.1365-2966.2012.21983.x,
    author = "Killedar, Madhura and Borgani, Stefano and Meneghetti, Massimo and Dolag, Klaus and Fabjan, Dunja and Tornatore, Luca",
    title = "{How Baryonic Processes affect Strong Lensing properties of Simulated Galaxy Clusters}",
    eprint = "1208.5770",
    archivePrefix = "arXiv",
    primaryClass = "astro-ph.CO",
    doi = "10.1111/j.1365-2966.2012.21983.x",
    journal = "Mon. Not. Roy. Astron. Soc.",
    volume = "427",
    pages = "533",
    year = "2012"
}

@article{10.1051/0004-6361:20040212,
    author = "Takahashi, Ryuichi",
    title = "{Quasigeometrical optics approximation in gravitational lensing}",
    eprint = "astro-ph/0402165",
    archivePrefix = "arXiv",
    doi = "10.1051/0004-6361:20040212",
    journal = "Astron. Astrophys.",
    volume = "423",
    pages = "787--792",
    year = "2004"
}

@ARTICLE{2001astro.ph..2341K,
       author = {{Keeton}, Charles R.},
        title = "{A Catalog of Mass Models for Gravitational Lensing}",
      journal = {arXiv e-prints},
         year = 2001,
        month = feb,
          eid = {astro-ph/0102341},
        pages = {astro-ph/0102341},
          doi = {10.48550/arXiv.astro-ph/0102341},
archivePrefix = {arXiv},
       eprint = {astro-ph/0102341},
 primaryClass = {astro-ph},
       adsurl = {https://ui.adsabs.harvard.edu/abs/2001astro.ph..2341K}
}

@article{10.1086/322314,
    author = "Munoz, J. A. and Kochanek, C. S. and Keeton, C. R.",
    title = "{Cusped mass models of gravitational lenses}",
    eprint = "astro-ph/0103009",
    archivePrefix = "arXiv",
    doi = "10.1086/322314",
    journal = "Astrophys. J.",
    volume = "558",
    pages = "657",
    year = "2001"
}

@article{10.1051/0004-6361/201321618,
    author = "D{\'u}met-Montoya, H. S. and Caminha, G. B. and Makler, M.",
    title = "{Analytic Solutions for Navarro--Frenk--White Lens Models for Low Characteristic Convergences}",
    eprint = "1304.0425",
    archivePrefix = "arXiv",
    primaryClass = "astro-ph.CO",
    doi = "10.1051/0004-6361/201321618",
    journal = "Astron. Astrophys.",
    volume = "560",
    pages = "A86",
    year = "2013"
}

@article{10.1051/0004-6361:20020226,
    author = "Dye, Simon and Taylor, A. N. and Greve, T. R. and Rognvaldsson, O. E. and van Kampen, E. and Jakobsson, P. and Sigmundsson, V. S. and Gudmundsson, E. H. and Hjorth., J.",
    title = "{Lens magnification by CL0024+1654 in the U and R band}",
    eprint = "astro-ph/0108399",
    archivePrefix = "arXiv",
    doi = "10.1051/0004-6361:20020226",
    journal = "Astron. Astrophys.",
    volume = "386",
    pages = "12--30",
    year = "2002"
}

@article{10.1086/589989,
    author = "Bolton, Adam S. and Treu, Tommaso and Koopmans, Leon V. E. and Gavazzi, Raphael and Moustakas, Leonidas A. and Burles, Scott and Schlegel, David J. and Wayth, Randall",
    title = "{The Sloan Lens ACS Survey. VII. Elliptical Galaxy Scaling Laws from Direct Observational Mass Measurements}",
    eprint = "0805.1932",
    archivePrefix = "arXiv",
    primaryClass = "astro-ph",
    doi = "10.1086/589989",
    journal = "Astrophys. J.",
    volume = "684",
    pages = "248--259",
    year = "2008"
}

@article{10.1088/0004-637X/748/2/129,
    author = "Anguita, T. and Barrientos, L. F. and Gladders, M. D. and Faure, C. and Yee, H. and Gilbank, D.",
    title = "{Galaxy scale lenses in the RCS2: I. First catalog of candidate strong lenses}",
    eprint = "1201.5583",
    archivePrefix = "arXiv",
    primaryClass = "astro-ph.CO",
    doi = "10.1088/0004-637X/748/2/129",
    journal = "Astrophys. J.",
    volume = "748",
    pages = "129",
    year = "2012"
}

@article{10.3847/1538-4357/aa9794,
    author = "Shu, Yiping and others",
    title = "{The Sloan Lens ACS Survey. XIII. Discovery of 40 New Galaxy-Scale Strong Lenses}",
    eprint = "1711.00072",
    archivePrefix = "arXiv",
    primaryClass = "astro-ph.GA",
    doi = "10.3847/1538-4357/aa9794",
    journal = "Astrophys. J.",
    volume = "851",
    number = "1",
    pages = "48",
    year = "2017"
}

@article{10.1051/0004-6361/202451341,
   title={The SLACS strong lens sample, debiased},
   volume={690},
   ISSN={1432-0746},
   url={http://dx.doi.org/10.1051/0004-6361/202451341},
   DOI={10.1051/0004-6361/202451341},
   journal={Astronomy \& Astrophysics},
   publisher={EDP Sciences},
   author={Sonnenfeld, Alessandro},
   year={2024},
   month=oct, pages={A325}
}

@article{10.1111/j.1365-2966.2012.21041.x,
    author = "Zitrin, Adi and Broadhurst, Tom and Bartelmann, Matthias and Rephaeli, Yoel and Oguri, Masamune and Benitez, Narciso and Hao, Jiangang and Umetsu, Keiichi",
    title = "{The Universal Einstein Radius Distribution from 10,000 SDSS Clusters}",
    eprint = "1105.2295",
    archivePrefix = "arXiv",
    primaryClass = "astro-ph.CO",
    reportNumber = "FERMILAB-PUB-12-902-A",
    doi = "10.1111/j.1365-2966.2012.21041.x",
    journal = "Mon. Not. Roy. Astron. Soc.",
    volume = "423",
    pages = "2308--2324",
    year = "2012"
}

@article{10.1103/PhysRevD.102.124076,
    author = "Guo, Xiao and Lu, Youjun",
    title = "{Convergence and Efficiency of Different Methods to Compute the Diffraction Integral for Gravitational Lensing of Gravitational Waves}",
    eprint = "2012.03474",
    archivePrefix = "arXiv",
    primaryClass = "gr-qc",
    doi = "10.1103/PhysRevD.102.124076",
    journal = "Phys. Rev. D",
    volume = "102",
    number = "12",
    pages = "124076",
    year = "2020"
}

@book{0521431085,
    author = {Press, William H. and Teukolsky, Saul A. and Vetterling, William T. and Flannery, Brian P.},
    title = {Numerical recipes in C (2nd ed.): the art of scientific computing},
    year = {1992},
    isbn = {0521431085},
    publisher = {Cambridge University Press},
    address = {USA}
}

@article{levin1982procedures,
  title={Procedures for computing one-and two-dimensional integrals of functions with rapid irregular oscillations},
  author={Levin, David},
  journal={Mathematics of Computation},
  volume={38},
  number={158},
  pages={531--538},
  year={1982}
}

@article{10.1143/ptps.133.137,
    author = "Nakamura, Takahiro T. and Deguchi, Shuji",
    title = "{Wave Optics in Gravitational Lensing}",
    doi = "10.1143/ptps.133.137",
    journal = "Prog. Theor. Phys. Suppl.",
    volume = "133",
    pages = "137--153",
    year = "1999"
}

@book{10.1007/978-3-662-03758-4,
    author = {Schneider, Peter and Ehlers, J{\"u}rgen and Falco, Emilio E.},
    title = "{Gravitational Lenses}",
    doi = "10.1007/978-3-662-03758-4",
    isbn = "978-3-540-66506-9, 978-3-662-03758-4",
    publisher = "Springer",
    series = "Astronomy and Astrophysics Library",
    year = "1992"
}

@article{10.1103/PhysRevLett.116.061102,
    author = "Abbott, B. P. and others",
    collaboration = "LIGO Scientific, Virgo",
    title = "{Observation of Gravitational Waves from a Binary Black Hole Merger}",
    eprint = "1602.03837",
    archivePrefix = "arXiv",
    primaryClass = "gr-qc",
    reportNumber = "LIGO-P150914",
    doi = "10.1103/PhysRevLett.116.061102",
    journal = "Phys. Rev. Lett.",
    volume = "116",
    number = "6",
    pages = "061102",
    year = "2016"
}

@article{10.1086/177173,
    author = "Navarro, Julio F. and Frenk, Carlos S. and White, Simon D. M.",
    title = "{The Structure of cold dark matter halos}",
    eprint = "astro-ph/9508025",
    archivePrefix = "arXiv",
    doi = "10.1086/177173",
    journal = "Astrophys. J.",
    volume = "462",
    pages = "563--575",
    year = "1996"
}

@article{10.1007/BF00654034,
    author = "Bontz, Robert J. and Haugan, Mark P.",
    title = "{A diffraction limit on the gravitational lens effect}",
    doi = "10.1007/BF00654034",
    journal = "Astrophys. Space Sci.",
    volume = "78",
    number = "1",
    pages = "199--210",
    year = "1981"
}

@article{10.1086/164389,
       author = {{Deguchi}, S. and {Watson}, W.~D.},
        title = "{Diffraction in Gravitational Lensing for Compact Objects of Low Mass}",
      journal = {\apj},
         year = 1986,
        month = aug,
       volume = {307},
        pages = {30},
          doi = {10.1086/164389},
       adsurl = {https://ui.adsabs.harvard.edu/abs/1986ApJ...307...30D}
}

@article{10.1086/377430,
    author = "Takahashi, Ryuichi and Nakamura, Takashi",
    title = "{Wave effects in gravitational lensing of gravitational waves from chirping binaries}",
    eprint = "astro-ph/0305055",
    archivePrefix = "arXiv",
    doi = "10.1086/377430",
    journal = "Astrophys. J.",
    volume = "595",
    pages = "1039--1051",
    year = "2003"
}

@ARTICLE{2020ScPC....3....7B,
       author = {{Bertone}, Gianfranco and {Croon}, Djuna and {Amin}, Mustafa and {Boddy}, Kimberly K. and {Kavanagh}, Bradley and {Mack}, Katherine J. and {Natarajan}, Priyamvada and {Opferkuch}, Toby and {Schutz}, Katelin and {Takhistov}, Volodymyr and {Weniger}, Christoph and {Yu}, Tien-Tien},
        title = "{Gravitational wave probes of dark matter: challenges and opportunities}",
      journal = {SciPost Physics Core},
         year = 2020,
        month = oct,
       volume = {3},
       number = {2},
          eid = {007},
        pages = {007},
          doi = {10.21468/SciPostPhysCore.3.2.007},
archivePrefix = {arXiv},
       eprint = {1907.10610},
 primaryClass = {astro-ph.CO},
       adsurl = {https://ui.adsabs.harvard.edu/abs/2020ScPC....3....7B}
}

@ARTICLE{2021MNRAS.502L..16C,
       author = {{Cao}, Shuo and {Qi}, Jingzhao and {Biesiada}, Marek and {Liu}, Tonghua and {Li}, Jin and {Zhu}, Zong-Hong},
        title = "{Measuring the viscosity of dark matter with strongly lensed gravitational waves}",
      journal = {\mnras},
         year = 2021,
        month = mar,
       volume = {502},
       number = {1},
        pages = {L16-L20},
          doi = {10.1093/mnrasl/slaa205},
archivePrefix = {arXiv},
       eprint = {2012.12462},
 primaryClass = {astro-ph.CO},
       adsurl = {https://ui.adsabs.harvard.edu/abs/2021MNRAS.502L..16C}
}

@ARTICLE{2022A&A...659L...5C,
       author = {{Cao}, Shuo and {Qi}, Jingzhao and {Cao}, Zhoujian and {Biesiada}, Marek and {Cheng}, Wei and {Zhu}, Zong-Hong},
        title = "{Direct measurement of the distribution of dark matter with strongly lensed gravitational waves}",
      journal = {\aap},
         year = 2022,
        month = mar,
       volume = {659},
          eid = {L5},
        pages = {L5},
          doi = {10.1051/0004-6361/202142694},
archivePrefix = {arXiv},
       eprint = {2202.08714},
 primaryClass = {astro-ph.CO},
       adsurl = {https://ui.adsabs.harvard.edu/abs/2022A&A...659L...5C}
}

@ARTICLE{2021PhRvD.104f3001C,
       author = {{Choi}, Han Gil and {Park}, Chanung and {Jung}, Sunghoon},
        title = "{Small-scale shear: Peeling off diffuse subhalos with gravitational waves}",
      journal = {\prd},
         year = 2021,
        month = sep,
       volume = {104},
       number = {6},
          eid = {063001},
        pages = {063001},
          doi = {10.1103/PhysRevD.104.063001},
       adsurl = {https://ui.adsabs.harvard.edu/abs/2021PhRvD.104f3001C}
}

@ARTICLE{2018PhRvD..98j4029D,
       author = {{Dai}, Liang and {Li}, Shun-Sheng and {Zackay}, Barak and {Mao}, Shude and {Lu}, Youjun},
        title = "{Detecting lensing-induced diffraction in astrophysical gravitational waves}",
      journal = {\prd},
         year = 2018,
        month = nov,
       volume = {98},
       number = {10},
          eid = {104029},
        pages = {104029},
          doi = {10.1103/PhysRevD.98.104029},
archivePrefix = {arXiv},
       eprint = {1810.00003},
 primaryClass = {gr-qc},
       adsurl = {https://ui.adsabs.harvard.edu/abs/2018PhRvD..98j4029D}
}

@ARTICLE{2022PhRvD.106b3018G,
       author = {{Guo}, Xiao and {Lu}, Youjun},
        title = "{Probing the nature of dark matter via gravitational waves lensed by small dark matter halos}",
      journal = {\prd},
         year = 2022,
        month = jul,
       volume = {106},
       number = {2},
          eid = {023018},
        pages = {023018},
          doi = {10.1103/PhysRevD.106.023018},
archivePrefix = {arXiv},
       eprint = {2207.00325},
 primaryClass = {astro-ph.CO},
       adsurl = {https://ui.adsabs.harvard.edu/abs/2022PhRvD.106b3018G}
}

@ARTICLE{2025PhRvL.135k1402J,
       author = {{Jana}, Souvik and {Kapadia}, Shasvath J. and {Venumadhav}, Tejaswi and {More}, Surhud and {Ajith}, Parameswaran},
        title = "{Probing the Nature of Dark Matter Using Strongly Lensed Gravitational Waves from Binary Black Holes}",
      journal = {\prl},
         year = 2025,
        month = sep,
       volume = {135},
       number = {11},
          eid = {111402},
        pages = {111402},
          doi = {10.1103/7q31-3qwz},
archivePrefix = {arXiv},
       eprint = {2408.05290},
 primaryClass = {astro-ph.CO},
       adsurl = {https://ui.adsabs.harvard.edu/abs/2025PhRvL.135k1402J}
}

@ARTICLE{2019PhRvL.122d1103J,
       author = {{Jung}, Sunghoon and {Shin}, Chang Sub},
        title = "{Gravitational-Wave Fringes at LIGO: Detecting Compact Dark Matter by Gravitational Lensing}",
      journal = {\prl},
         year = 2019,
        month = feb,
       volume = {122},
       number = {4},
          eid = {041103},
        pages = {041103},
          doi = {10.1103/PhysRevLett.122.041103},
archivePrefix = {arXiv},
       eprint = {1712.01396},
 primaryClass = {astro-ph.CO},
       adsurl = {https://ui.adsabs.harvard.edu/abs/2019PhRvL.122d1103J}
}

@ARTICLE{2025PhRvD.112f3055L,
       author = {{Li}, Zhijin and {Guo}, Xiao and {Cao}, Zhoujian and {Zhang}, Yun-Long},
        title = "{Detectability of dark matter density distribution via gravitational waves from binary black holes in the Galactic Center}",
      journal = {\prd},
         year = 2025,
        month = sep,
       volume = {112},
       number = {6},
          eid = {063055},
        pages = {063055},
          doi = {10.1103/zr7l-7y5c},
archivePrefix = {arXiv},
       eprint = {2506.19327},
 primaryClass = {astro-ph.HE},
       adsurl = {https://ui.adsabs.harvard.edu/abs/2025PhRvD.112f3055L}
}

@ARTICLE{2020MNRAS.495.2002L,
       author = {{Liao}, Kai and {Tian}, Shuxun and {Ding}, Xuheng},
        title = "{Probing compact dark matter with gravitational wave fringes detected by the Einstein Telescope}",
      journal = {\mnras},
         year = 2020,
        month = jun,
       volume = {495},
       number = {2},
        pages = {2002-2006},
          doi = {10.1093/mnras/staa1388},
archivePrefix = {arXiv},
       eprint = {2001.07891},
 primaryClass = {astro-ph.CO},
       adsurl = {https://ui.adsabs.harvard.edu/abs/2020MNRAS.495.2002L}
}

@ARTICLE{2018PhR...730....1T,
       author = {{Tulin}, Sean and {Yu}, Hai-Bo},
        title = "{Dark matter self-interactions and small scale structure}",
      journal = {\physrep},
         year = 2018,
        month = feb,
       volume = {730},
        pages = {1-57},
          doi = {10.1016/j.physrep.2017.11.004},
archivePrefix = {arXiv},
       eprint = {1705.02358},
 primaryClass = {hep-ph},
       adsurl = {https://ui.adsabs.harvard.edu/abs/2018PhR...730....1T}
}

@article{10.1038/s41586-020-2642-9,
    author = "Wang, Jie and Bose, Sownak and Frenk, Carlos S. and Gao, Liang and Jenkins, Adrian and Springel, Volker and White, Simon D. M.",
    title = "{Universal structure of dark matter haloes over a mass range of 20 orders of magnitude}",
    eprint = "1911.09720",
    archivePrefix = "arXiv",
    primaryClass = "astro-ph.CO",
    doi = "10.1038/s41586-020-2642-9",
    journal = "Nature",
    volume = "585",
    number = "7823",
    pages = "39--42",
    year = "2020"
}

@article{10.1051/0004-6361/201219539,
    author = "Retana-Montenegro, E. and Frutos-Alfaro, F. and Baes, M.",
    title = "{Analytical shear and flexion of Einasto dark matter haloes}",
    eprint = "1207.4281",
    archivePrefix = "arXiv",
    primaryClass = "astro-ph.CO",
    doi = "10.1051/0004-6361/201219539",
    journal = "Astron. Astrophys.",
    volume = "546",
    pages = "A32",
    year = "2012"
}

@article{10.1051/0004-6361/201118543,
    author = "Retana-Montenegro, E. and Van Hese, E. and Gentile, G. and Baes, M. and Frutos-Alfaro, F.",
    title = "{Analytical properties of Einasto dark matter haloes}",
    eprint = "1202.5242",
    archivePrefix = "arXiv",
    primaryClass = "astro-ph.CO",
    doi = "10.1051/0004-6361/201118543",
    journal = "Astron. Astrophys.",
    volume = "540",
    pages = "A70",
    year = "2012"
}

@article{10.1093/mnras/283.3.L72,
    author = "Navarro, Julio F. and Eke, Vincent R. and Frenk, Carlos S.",
    title = "{The cores of dwarf galaxy halos}",
    eprint = "astro-ph/9610187",
    archivePrefix = "arXiv",
    doi = "10.1093/mnras/283.3.L72",
    journal = "Mon. Not. Roy. Astron. Soc.",
    volume = "283",
    pages = "L72--L78",
    year = "1996"
}

@article{10.1093/mnras/stz1890,
    author = "Ben{\'\i}tez-Llambay, Alejandro and Frenk, Carlos S. and Ludlow, Aaron D. and Navarro, Julio F.",
    title = "{Baryon-induced dark matter cores in the eagle simulations}",
    eprint = "1810.04186",
    archivePrefix = "arXiv",
    doi = "10.1093/mnras/stz1890",
    journal = "Mon. Not. Roy. Astron. Soc.",
    volume = "488",
    number = "2",
    pages = "2387--2404",
    year = "2019"
}

@article{10.1086/377489,
   title={Gravitational Lensing by a Compound Population of Halos: Standard Models},
   volume={595},
   ISSN={1538-4357},
   url={http://dx.doi.org/10.1086/377489},
   DOI={10.1086/377489},
   number={2},
   journal={The Astrophysical Journal},
   publisher={American Astronomical Society},
   author={Li, Li‐Xin and Ostriker, Jeremiah P.},
   year={2003},
   month=oct, pages={603–613} }

@article{10.1093/mnras/stw1707,
    author = "Mantz, Adam B. and Allen, Steven W. and Morris, R. Glenn",
    title = "{Cosmology and astrophysics from relaxed galaxy clusters {\textendash} V. Consistency with cold dark matter structure formation}",
    eprint = "1607.04686",
    archivePrefix = "arXiv",
    primaryClass = "astro-ph.CO",
    doi = "10.1093/mnras/stw1707",
    journal = "Mon. Not. Roy. Astron. Soc.",
    volume = "462",
    number = "1",
    pages = "681--688",
    year = "2016"
}

@article{10.1093/mnras/stu1284,
    author = "Vegetti, Simona and Vogelsberger, Mark",
    title = "{On the density profile of dark matter substructure in gravitational lens galaxies}",
    eprint = "1406.1170",
    archivePrefix = "arXiv",
    primaryClass = "astro-ph.CO",
    doi = "10.1093/mnras/stu1284",
    journal = "Mon. Not. Roy. Astron. Soc.",
    volume = "442",
    number = "4",
    pages = "3598--3603",
    year = "2014"
}

\end{document}